\documentclass{aa}  

\usepackage{graphicx}
\usepackage[caption=false]{subfig}
\usepackage{txfonts}
\usepackage{tabto}
\usepackage{natbib}
\bibpunct{(}{)}{;}{a}{}{,} % to follow the A&A style

\usepackage{xcolor}

\definecolor{correction-blue}{RGB}{85,72,250}
\definecolor{correction-magenta}{RGB}{165, 0, 124}
\definecolor{comment-red}{RGB}{190,50,0}
\definecolor{edit-green}{RGB}{37, 132, 78}%{47,172,102}

\begin{document}

   \title{Black Mirror: The impact of rotational broadening on the search for reflected light from 51 Pegasi b with high resolution spectroscopy}

   \titlerunning{Impact of rotational broadening on reflected light from 51 Pegasi b}
    
   \author{E. F. Spring  \inst{1}
          \and
          J. L. Birkby \inst{2,1,3}
          \and
          L. Pino \inst{4}
          \and
          R. Alonso \inst{5,6}
          \and
          S. Hoyer \inst{7}
          \and
          M. E. Young \inst{2}
          \and
          P. R. T. Coelho \inst{8}
         \and
          D. Nespral \inst{5}
          \and
          M. L\'opez-Morales \inst{3}
          }

    \institute{Anton Pannekoek Instituut (API), Universiteit van Amsterdam,
          Science Park 904, 1098 XH Amsterdam, Netherlands\\
          \email{e.f.spring@uva.nl}
          \and
          Astrophysics, Department of Physics, University of Oxford,
          Denys Wilkinson Building, Keble Road, Oxford, OX1 3RH, UK  
          \and
          Center for Astrophysics | Harvard and Smithsonian,
          60 Garden Street, Cambridge MA 02138, USA
          \and
          INAF - Osservatorio Astrofisico di Arcetri, Largo E. Fermi 5, I-50125, Florence, Italy
          \and
          Instituto de Astrofísica de Canarias, E-38200 La Laguna, Tenerife, Spain
          \and
          Departamento de Astrofísica, Universidad de La Laguna, E-38206 La Laguna, Tenerife, Spain
          \and
          Aix Marseille Univ, CNRS, CNES, LAM, 38 rue Frédéric Joliot-Curie, 13388 Marseille, France
          \and
          Universidade de São Paulo, Instituto de Astronomia, Geofísica e Ciencias Atmosféricas, Rua do Matão 1226, 05508-090, São Paulo, Brazil
}
   \date{Received Month Date, Year; accepted Month Date, Year}

% \abstract{}{}{}{}{} 
% 5 {} token are mandatory
 
  \abstract
   {The extreme contrast ratios between stars and their planets at optical wavelengths make it challenging to isolate the light reflected by exoplanet atmospheres. Yet, these reflective properties reveal key processes occurring in the atmospheres, and they also span wavelengths that include the potential O$_2$ biosignature. High resolution cross-correlation spectroscopy (HRCCS) offers a robust avenue for developing techniques to extract exoplanet reflection spectra.}
   {We aimed to extract the optical reflected light spectrum of the non-transiting hot Jupiter 51 Pegasi b by adapting techniques designed to remove tellurics in infrared HRCCS to instead remove optical stellar lines. Importantly, we investigated the as of yet neglected impact of the broadening of the reflected host star spectrum due to the difference between the stellar rotation and the planet's orbital velocity.}
   {We used 484, $R=115,000$ optical spectra of 51 Pegasi b from HARPS-N and HARPS, which we aligned to the exact stellar rest frame, in order to effectively remove the contaminating host star. However, some stellar residuals remained, likely due to stellar activity.  We cross-correlated with an appropriately broadened synthetic stellar model to search for the planet's Doppler-shifting spectrum.}
   {We detect no significant reflected light from 51 Pegasi b, and report a signal-to-noise (S/N) $=3$ upper limit on the contrast ratio of 76.0 ppm ($7.60\times10^{-5}$) when including broadening, and 24.0 ppm ($2.40\times10^{-5}$) without.  These upper limits rule out radius and albedo combinations of previously claimed detections.}
   {Broadening can significantly impact the ability of HRCCS to extract reflected light spectra and it must be considered when determining the contrast ratio, radius, and albedo of the planet. Asynchronous systems ($P_{rot,\star}\ne P_{orb}$) are most affected, including most hot Jupiters as well as Earth-size planets in the traditional habitable zones of some M-dwarfs.}


   \maketitle
%
%---------------------------------------------

\section{Introduction} \label{intro}

The use of high resolution cross-correlation spectroscopy (HRCCS, resolving power $R >$ 100\,000) has proven to be a powerful technique in the characterisation of giant exoplanet atmospheres using ground-based telescopes. The technique can be applied to transiting and non-transiting planets alike, and has already delivered studies of both transmission and thermal emission spectroscopy, revealing compositions, atmospheric structures, rotation, and day-to-night winds~\citep[e.g.][]{Snellen2010, Brogi2016}. Moreover, it holds great promise for identifying biosignatures in the nearest rocky exoplanet atmospheres using the large photon collecting power of the upcoming extremely large telescopes (ELTs) \citep[see e.g.][]{Snellen2015, Hawker2019, Serindag2019}. However, HRCCS relies on the noise being photon-dominated, which limits its effectiveness from the ground at wavelengths longer than $\gtrsim5~\mu$m due to the thermal background \citep{Snellen2015}. Thus, any search for biosignatures with HRCCS at the ELTs will likely need to focus on optical and near-infrared wavelengths. 

One of the key, although not definitive, biosignatures is oxygen \citep{MeadowsBiosig2018}. It has strong optical spectral features ($\sim760$ nm) that are accessible to HRCCS in transmission \citep{Snellen2013}. However, the nearest rocky exoplanets, such as Proxima b, are non-transiting and have very low thermal emission at optical wavelengths. This leaves only the reflected optical spectrum as an avenue to search for oxygen in their atmospheres with HRCCS. 

Shortly after the discovery of 51 Pegasi b \citep{MayorQueloz95}, attempts to use HRCCS to detect reflected light began \citep{Charbonneau1999, CollierCameron1999}. However, despite multiple attempts, only upper limits or contested detections have been made, and the success of the method remains inconclusive \citep[see e.g.][]{Charbonneau1999, CollierCameron1999, Leigh2003a, Rodler2010, Rodler2013, Hoeijmakers2018, GAPS2021, Martins15, Borra2018ACF}. Although hot Jupiters have advantageous inflated radii and short orbital periods, their close proximity to their host stars is thought to leave many of them bereft of a reflective cloud deck \citep[e.g.][]{Rowe2008, Cowan_2011, Heng2013}, resulting in low expected geometric albedo ($A_g\lesssim0.1$), making them very dark with low planet-to-star contrast ratios that are difficult to detect~\citep[e.g.][]{Hoeijmakers2018}. This is supported by observations of photometric optical phase curves from Kepler, K2, TESS and CHEOPS \citep{Coughlin2012, Angerhausen2015, Esteves2015, Wong2020, Wong2021, Hooton2021}. There are, however, a few notable exceptions, including Kepler-7 b with $A_g \sim 0.3$ suggesting the presence of a reflective cloud deck over at least some of the planet \citep{Kipping2011,Demory2011,Demory2013,Heng2021}, and HD 189733 b with an increasing albedo towards bluer wavelengths as observed with low resolution Hubble Space Telescope spectra \citep{Evans2013, Barstow2014}. Thus, the literature shows that, whilst the majority of hot Jupiters have a very low $A_g$, there are sufficient exceptions that the diversity in albedo warrants investigation. In addition to the challenge of low albedo, for reflection HRCCS, the stellar and planetary spectra are highly correlated because the targeted planets reflect the spectrum of their host star. Thus, reflection HRCCS must overcome the substantial additional challenge of successfully disentangling two very similar signals. 

With these challenges in mind, in this paper we present our study of optical reflected light from 51 Pegasi b using HRCCS, where we remove stellar and telluric contaminants directly from the spectra rather than in cross-correlation space, adapting processes used previously for infrared HRCCS analysis techniques (see Section~\ref{intro_cc}). In this respect, we were aided by the superior stability of the HARPS-N spectrograph at the TNG used to obtain our data. In Section~\ref{intro_cc} we describe the adaptations necessary to use HRCCS in reflected light. This includes the important consideration of rotational broadening due to the apparent rotation of the star as seen from the planet during its orbit. This has so far been neglected in previous studies of 51 Pegasi b, but has the potential to impact how well a planet spectrum can be recovered by HRCCS and the interpretation of its contrast ratio (see Section~\ref{intro_rot_broad}). We describe our observations in Section~\ref{obsv}, and Section~\ref{method} details our post-processing methods, including how we remove the host star spectrum contaminant. Our results and upper limits are shown in Section~\ref{results} and we assess our adaptations for reflection HRCCS in Section~\ref{discuss} along with an in depth comparison to previous studies of 51 Pegasi b. We conclude our study in Section~\ref{conclude}.

\section{Reflected light high resolution cross-correlation spectroscopy (HRCCS)}

\subsection{Previous studies of reflected light with HRCCS}\label{sec:intro_compare}
Several previous works have attempted reflection HRCCS, with some studies resulting in planet-to-star contrast upper limits, for example at the $1.5\times10^{-5}$ level for $\tau$ Bo\"o b \citep{Hoeijmakers2018}. Given its bright host star magnitude, four previous works have sought the reflected light from 51 Pegasi b \citep{Martins15, Borra2018ACF, Marcantonio2019ICA, GAPS2021}. We briefly summarise the status of these 51 Pegasi b investigations here as motivation for this work, as they have presented conflicting conclusions.

The first,~\citet[][hereafter M15]{Martins15}, used a set of sparse and randomly sampled archival HARPS-N spectra, and followed the method outlined in~\cite{Martins2013}, which removes the contaminating host star spectrum in cross-correlation space. They generated cross-correlation functions (CCFs) using the weighted G2 binary mask in the HARPS-N data reduction pipeline \citep{Pepe2002}, first for star-only spectra when the planet day side is hidden, and then for each star+planet spectrum. They divide out the star-only CCF from the star+planet CCFs to remove the host star spectrum, in order to obtain a planet-only CCF. They then use the height of the peak of the planet-only CCF as a metric for the strength of the planet detection, which they refer to as the signal amplitude. M15 used their measured signal amplitude to suggest evidence of a detection of the reflected light from 51 Pegasi b at 3$\sigma$ level, corresponding to a very bright planet-to-star contrast ratio of $1.2\times10^{-4}$. This would correspond to a very inflated planet radius of $R_p=1.9R_J$ with a bright geometric albedo ($A_g=0.5$). They also noted significant broadening of the signal, but demonstrate that this is introduced by their processing methods in the low signal-to-noise (S/N) regime.

\citet{Borra2018ACF} use the same archival data as M15, and a similar method to obtain a planet-only CCF. However, they generate their CCFs by using the observed spectra themselves as the cross-correlation templates, rather than using the G2 weighted mask. These so-called autocorrelation functions (ACFs) have the advantage that they are always perfectly aligned to the stellar rest frame\footnote{As noted by~\citet[][hereafter S21]{GAPS2021}, the general definition of the autocorrelation function is of a signal cross-correlated with a copy of itself. The autocorrelation function (ACF) as defined by~\cite{Borra2018ACF}, however, refers to a spectrum that is cross-correlated with a stellar template. To avoid confusion in this work, we use the acronym `ACF' to refer to the~\cite{Borra2018ACF} method, and the term `autocorrelation function' retains its standard meaning.}.

\citet{Borra2018ACF} find an even deeper signal amplitude, which they interpret as a detection at the $5.5\sigma$ level. The width of their planet-only ACF is broader than their stellar ACF, but to a lesser extent than seen in the CCFs of M15. \citet{Borra2018ACF} propose that 51 Pegasi b is tidally locked, and that the apparent broadening is due to noise. They do not make inferences about the resultant planet properties, such as the contrast ratio or $A_g$ and $R_p$, but the deeper amplitude suggests an even brighter or larger planet.

\citet{Marcantonio2019ICA} again used the same archival data as M15, but use instead the method of Independent Component Analysis (ICA). In contrast to M15 and~\cite{Borra2018ACF}, they do not recover a reflected light signal from 51 Pegasi b, which they ascribe to the insufficient S/N of the dataset.

Finally,~\citet[][hereafter S21]{GAPS2021} presented an analysis following the method of \citet{Borra2018ACF}, but used an augmented archival dataset which only included long dedicated observing sequences from individual nights. S21 do not recover a reflected light signal from 51 Pegasi b, reaching instead a $3\sigma$ upper limit on the contrast ratio of $10^{-5}$, which they use to predict that 51 Pegasi b is an average-sized hot Jupiter ($0.9\leq R_p\leq1.5R_J$) with low albedo ($A_g<0.1$). They also show that no significant broadening was imparted by their processing techniques. Their result is in stark contrast to the previous analyses presented in M15 and~\cite{Borra2018ACF}. S21 suggest these previous claims of reflected light from 51 Pegasi b may be affected by an unlucky combination of false positives.

\subsection{Adaptations for reflection HRCCS} \label{intro_cc}

First, let us consider the theory behind HRCSS, as applies for transmission, emission and reflection HRCSS. High-resolution spectroscopy resolves molecular bands into a dense forest of thousands of spectral lines, which each adhere to a unique pattern in wavelength space that is difficult to reproduce with random noise. For short orbit exoplanets, the success of HRCCS relies on using the large Doppler shift  ($\sim\,\rm km\,s^{-1}$) of the exoplanet during its orbit to separate its spectrum from its host star. The spectral lines of the exoplanet change considerably in wavelength during even a fraction of its orbit, while its host star and Earth's telluric lines appear quasi-static. Consequently, any spectral feature that does not change in wavelength over time can be identified and removed, typically using algorithms such as Principle Component Analysis (PCA) or \textsc{Sysrem} \citep{sysrem2005, sysrem2007}, leaving behind the Doppler-shifting exoplanet spectrum. However, there remains the challenge of the very high contrast ratio between the exoplanet and its host star, which varies from $10^{-3}$ for some ultra hot Jupiters in the near infrared to only $10^{-10}$ for an Earth-analogue orbiting a Sun-like star in the optical. In order to amplify any signal from the exoplanet, its individual spectral lines are combined by cross-correlating the residuals (i.e. the observed spectra after removing the time invariant features) with a forward model of the exoplanet atmosphere. In an idealised case, where each line is assumed to have equal depth and the data is photon-noise dominated, this increases the S/N of the planet by \citep[following][]{Snellen2015, Birkby2018exoplanet}:

\begin{equation}
S/N_p = \left(\frac{F_{p}}{F_\star}\right)\ S/N_\star\sqrt{N_{lines}}
     \label{eq_SN}
\end{equation}

\noindent where $F_p$ is the flux from the planet, $F_\star$ is the flux from the star, $\rm S/N_\star$ is the total S/N of the observed spectra and $N_{lines}$ is the number of spectral lines.

HRCCS was developed in the near infrared, in order to study exoplanets in transmission and thermal emission ~\citep[e.g.][]{Snellen2010, Brogi2012, Birkby2013, Brogi2016, Birkby2017, CasasayasBarris2019, Webb2020, Boucher2021}, and it has only recently been applied in the optical for ultra-hot Jupiters such as Kelt-9 b~\citep[e.g.][]{Yan2018, Hoeijmakers2018Nature, Hoeijmakers2019, Pino2020} and Wasp-76 b~\citep[e.g.][]{Ehrenreich2020, Deibert2021}. In all these cases, the exoplanet spectrum is typically very distinct from its host star. A host star may contain e.g. CO or H$_{2}$O lines but these are typically modified in strength compared to those in the exoplanet spectrum due to the different abundances, structures, and temperatures of their respective atmospheres.

In reflected light, however, the exoplanet reflection spectrum (hereafter planet spectrum, $F_{p}(\lambda)$) is a replica of  the host star's spectrum (hereafter stellar spectrum, $F_{\star}(\lambda)$), with the exoplanet's geometric albedo\footnote{i.e. its brightness at full illumination compared to a Lambertian disk.} as function of wavelength ($A_{g}(\lambda)$) imprinted on top, as follows:

\begin{equation}
F_{p}(\lambda) = F_{\star}(\lambda)\ g(\alpha)\ A_{g}(\lambda)\ \left(\frac{R_p}{a}\right)^2
     \label{eq_Fp}
\end{equation}

\noindent where $g(\alpha)$ is the phase function, $\alpha$ is the phase angle, $R_{p}$ is the planet radius and $a$ is the semi-major axis.  For more detail on the phase function, see Section~\ref{model_inject}.

Thus, for reflection HRCCS the cross-correlation template will match with not only the exoplanet's spectrum but also with any residual host star spectrum that was not removed by the cleaning algorithms, potentially to a level that prohibits reaching the required contrast ratios to detect the planet. Unlike near infrared spectra, where removing telluric contamination is prioritised, optical spectra are comparatively free of strong tellurics, and removal of the host star spectrum is instead the priority.

From the reference frame of the Earth, the exoplanet's velocity $V_{p}$ varies with the orbital phase, $\phi$, according to:

\begin{equation}
V_{p}(\phi)= K_{p}\sin(2\pi\phi) + v_{bary} + v_{sys}
\label{eq_RVp}
\end{equation}

\noindent where the first term is its radial velocity ($RV_p$) and $K_{p}$ is the $RV$ semi-amplitude of the exoplanet, the second term $v_{bary}$ is its apparent motion due to the Earth orbiting the barycentre of the Solar system, and the last term $v_{sys}$ is the constant systemic velocity of the star-planet system through the galaxy with respect to the barycentre of the Solar System. The same equation applies to the host star to find $V_{\star}(\phi$), by substituting $K_{p}$ with $-K_{\star}$. The intra-night variation in $v_{bary}$ can be sufficiently large that an ultra stable spectrograph will measure the host star spectrum moving during the half-night of observations, offsetting it from the less dominant, but not completely negligible telluric lines, and negating the time-invariant assumptions made in the cleaning algorithms. Consequently, we need to operate in the rest frame of the star rather than the barycentric frame to use the cleaning algorithms designed for HRCCS, albeit at the expense of the quality of telluric removal. We discuss and illustrate this issue further in our data analysis in Section~\ref{align}.

As the cross-correlation template needs to be predominantly a replica of the host star, it could be created from the observations themselves, by making a high S/N reference spectrum of the host star. However, there are potential issues with this approach. At the contrast ratios we aim to achieve, even small instrumental systematics and tellurics could act to obscure the planet signal in the cross-correlation (see Section~\ref{discuss_interp}). But importantly, as described below in Section~\ref{intro_rot_broad}, the planet may broaden the stellar lines it reflects, which necessitates changing the shape of the stellar lines in the cross-correlation template to achieve maximum correlation. Broadening the observed stellar spectrum also broadens any telluric contamination or instrumental systematics, which could then obscure any planet signal in the cross-correlation. The lines in the planet's albedo function are also neglected when using only an observed stellar spectrum. Consequently, high resolution spectral libraries offer a useful alternative for providing a cross-correlation template.

%Begin figure
   \begin{figure*} %  TEMP here to sort spacing for co-authors
    \centering
       {%
          \includegraphics[width=\textwidth]{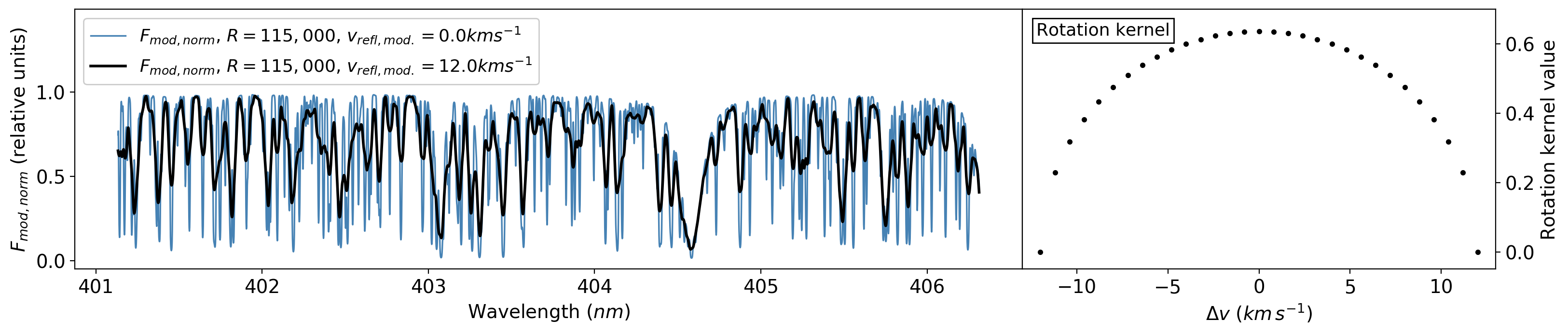}%
        }

   \caption{Model spectrum of reflected light for 51 Pegasi b based on a continuum normalised synthetic spectrum of its host star, at the spectral resolution of HARPS-N ($F_{mod,\,norm}$, $R=115,000$, blue line in left-hand panel). The black solid line shows the effect of broadening the model to $v_{refl}=12.0$ km\,s$^{-1}$ via convolution with the rotation kernel shown in the right-hand panel. It is approximately equivalent to $R\sim25,000$. Many of the narrow spectral features are blended and shallower in the broadened model, which reduces the last term of Equation~\ref{eq_SN} and lowers $\rm S/N_\textit{p}$.}
    \label{fig:model_broaden}         
    \end{figure*}
  %End figure 

\subsection{The impact of rotational broadening on reflected light} \label{intro_rot_broad}

The light emitted from and reflected by an exoplanet is subject to rotational broadening. From the perspective of an observer in the Earth or barycentric reference frame (Observer A), the light emitted and reflected from an exoplanet is Doppler broadened due to that exoplanet's rotation on its own axis. This is the exoplanet's projected rotational velocity, as seen from the Earth's line of sight, $v_{proj,p}$. In the case of reflected light, however, there is an additional source of broadening to consider due to the rotation of the exoplanet's host star and the planet's orbital period. Let us imagine another observer (Observer B) situated on the exoplanet itself. In some systems, such as $\tau$ Bo\"o b~\citep{Rodler2010, Hoeijmakers2018}, the planet's orbit and the stellar rotation are fully synchronised, such that it would appear to Observer B that the star was not rotating. However, for most hot Jupiters -- such as 51 Pegasi b and other tidally locked planets -- this is not the case. The planet keeps the same hemisphere facing the star, but the star does not maintain the same hemisphere facing the planet throughout its orbit. Thus, from the perspective of Observer B, the star will appear to be rotating at a velocity that depends on the difference between the true stellar rotation period $P_{rot,\star}$ and planet orbital period $P_{orb}$ (see Equation~\ref{eq_srefp}). Let's call this the star's rotational velocity in the reference frame of the exoplanet $v_{\star,p}$. In order to understand intuitively why $v_{\star,p}$ has an impact on the rotational broadening, consider that due to $v_{\star,p}$, each line in the stellar spectrum originated at different velocities across the stellar disk as observed by Observer B. However, Observer B cannot observe parts of the stellar disk individually, and thus the star appears as an integrated disc. Therefore, from Observer B's perspective, each stellar line will be a composite of the Doppler shifts from the integrated velocities, causing each line to appear broadened.

Thus, from the perspective of Observer A in the Earth or barycentric reference frame, the disk-integrated light reflected from the exoplanet will be rotationally broadened according to two factors: $v_{proj,p}$ and $v_{\star,p}$. In order to quantify the extent of the rotational broadening of the light reflected from 51 Pegasi b, we referred to equations 9 - 11 in~\cite{Rodler2010}. 

    We first assumed that 51 Pegasi b is tidally locked on a circular orbit, and thus $P_{rot,p} = P_{orb}=4.230784\pm0.000004$ days \citep{GAPS2021}. Then,  $v_{proj,p}$ follows as:

\begin{equation}
    v_{proj,p} = 2 \pi \sin{i} \frac{R_p}{P_{rot,p}} = 1.46 \pm0.12 \,\rm km\,s^{-1}
    \label{eq_pproj}
\end{equation}

\noindent where $i$ is the inclination of the planet's orbit relative to the Earth, which we set to $i = 80.9\pm1.3\,^{\circ}$~\citep{Brogi2013}, and $R_p$ is the exoplanet radius, which for this non-transiting planet, for the purposes of constraining the rotational broadening, we have set to be $1.2\pm0.1\,R_J$.\footnote{Based on the data currently available in the NASA Exoplanet Archive, $1.2\pm0.1\,R_J$ is the approximate mean radius value for short-orbit transiting exoplanets ($1<P_{orb}<10$ days) with similar masses to 51 Pegasi b ($0.3<M_{p}<0.6\,M_{J}$). The true mass of 51 Pegasi b has been measured directly to be $0.46\pm0.02\,M_J$ \citep{Brogi2013}.}

The $v_{\star,p}$ is calculated as (assuming no misalignment of the star-planet spin-orbit axes): 

\begin{equation}
    v_{\star,p} = 2 \pi R_\star \left(\frac{1}{P_{rot,\star}} - \frac{1}{P_{orb}}\right) = -11.93\pm0.45\,\rm km\,s^{-1}
        \label{eq_srefp}
\end{equation}

\noindent where $R_\star=1.237\pm0.047\,R_\sun$ is the stellar radius~\citep{vanBelle2009}, and $P_{rot,\star}$ is the stellar rotation period, for which we used $P_{rot,\star} = 21.9\pm0.4$ days~\citep{Simpson2010}. The negative value of $v_{\star,p}$ here indicates that to Observer B, the star would appear to be rotating counter to its true direction of orbit.

Finally, the two components of the rotational broadening, $v_{proj,p}$ and $v_{\star,p}$, are added in quadrature to give:

\begin{equation}
    v_{refl} = \sqrt{v_{proj,p}^2 + v_{\star,p}^2} = 12.02\pm0.45\,\rm km\,s^{-1}
    \label{eq_vrefl}
\end{equation}

  %Begin figure
   \begin{figure*} 
    \centering
       {%
          \includegraphics[width=18.2cm]{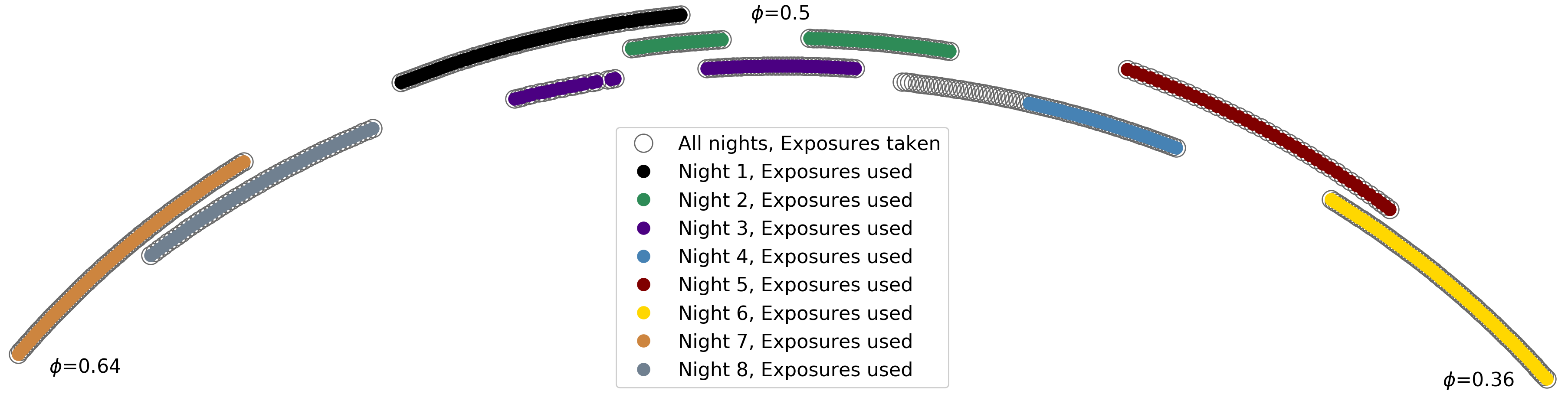}%
        }

   \caption{Phases, $\phi$, covered during the eight nights. The open grey circles show all observations and the solid coloured circles show the frames used in the final analysis (see Table~\ref{tab:obvs}) from each night.  Discarding unusable data from Night 4 left a gap in the phase coverage. The smallest phase observed ($\phi = 0.36$), the largest phase observed ($\phi = 0.64$), and the point of inferior conjunction ($\phi = 0.5$), are marked.}
    \label{fig:phase}         
    \end{figure*}
 %End figure

\noindent where $v_{refl}$ is the total rotational broadening. Thus, whenever we accounted for rotational broadening in this work, we set $v_{refl} = 12.0\, \rm km\,s^{-1}$. This important value is based on an approximation that $R_p = 1.2\pm0.1\,R_J$. For 51 Pegasi b the radius value has a minimal impact on the value of $v_{refl}$, as the largest contribution to $v_{refl}$ comes from $v_{\star,p}$. For example, a value of $R_p = 0.6\,R_J$ gives a $v_{refl} =\,11.95\,\rm km\,s^{-1}$, and a value of $R_p = 1.8\,R_J$ gives a $v_{refl} =\,12.13\,\rm km\,s^{-1}$.

Figure~\ref{fig:model_broaden} shows the impact of broadening on the depth of the spectral lines, making them shallower and more challenging to detect. These broadening effects are discussed extensively in \citet{Strachan2020}, who show that the amplitude of the spectral lines is deepest in a fully synchronised system ($P_{rot,\star}=P_{orb}$), and gets shallower as the two periods differ. The effect plateaus for slower rotating host stars, but the line depth amplitude rapidly decreases when the star is rotating faster than the planetary orbit. However, in cases where the cross-correlation template for the planet is significantly broader than the lines in the host star spectrum, this can in fact help to mitigate contamination of the CCF by residual host star features as the line shapes do not match (see Section~\ref{cc_method}). 

The star's projected velocity ($v_{proj,\star}$, calculated using an adapted Equation~\ref{eq_pproj}, such that $v_{proj,\star} = 2 \pi \sin{i} (R_\star/P_{rot,\star})$) could affect the discrepancy between the broadening of the stellar and planet lines for some systems, as if a star rotates sufficiently rapidly its stellar lines will also be be broadened. However, we do not account for $v_{proj,\star}$ in this work, as for $v_{proj,\star} = 2.86\,\rm km\,s^{-1}$ for 51 Pegasi, which is only just above the velocity per resolution elements of HARPS-N ($\Delta v = 2.6$ km\,s$^{-1}$) and HARPS ($\Delta v = 2.5$ km\,s$^{-1}$). Thus, the impact of $v_{proj,\star}$ will be negligible, and the stellar lines will be effectively unbroadened. See Section~\ref{obsv} for more detail on the velocity per resolution element of HARPS-N and HARPS.
\\

\section{Observations} \label{obsv}

% Observations table
\begin{table*}
    \centering
       \caption{Summary of the eight nights of observations.} 
    \begin{tabular}{c|c|c|c|c|c|c|c|c}
          \textbf{Night \tablefootmark{$\ast$}}&\textbf{Programme}&\textbf{Programme PI}&\textbf{Start Date}&\textbf{Exp.} \tablefootmark{$\dag$}&\textbf{Number \tablefootmark{$\S$}}&\textbf{Mean S/N\tablefootmark{$\P$}}&\textbf{Orbital}\tablefootmark{$\ast\ast$}&\textbf{$\Delta RV_p\,$}\tablefootmark{$\ddag$}\\
          &&&\textbf{in UTC}&\textbf{Time \emph{(s)}}&\textbf{of Frames}&\textbf{per Frame}&\textbf{Phase, $\phi$}&\textbf{(km s$^{-1}$)}\\
         \hline
         \hline
        1 & CAT15B\_146 & S. Hoyer &  27/10/2015 & 200 & 75 (75) & 147.6 & 0.517 - 0.564 & $-37.9$ \\
        2 & CAT16B\_146 & S. Hoyer &  12/10/2016 & 200 & 63 (63) & 165.9 & 0.474 - 0.523 & $-40.8$ \\
        3 & CAT16B\_43  & R. Alonso &  29/10/2016 & 200  & 60 (59) & 112.6 & 0.489 - 0.544 & $-45.5$ \\
        4 & CAT16B\_43  & R. Alonso &  02/11/2016 & 200  & 76 (41) & 173.4 & 0.435 - 0.460 & $-19.7$\\
         \hline
        5 & 091.C-0271 & N.C. Santos & 29/09/2013 & 450 & 39 (39) & 193.8 & 0.395 - 0.444 & $-35.7$\\
        6 & GAPS & G. Micela & 26/07/2017 & 200 & 78 (78) & 121.3 & 0.359 - 0.407 & $-29.6$\\
        7 & GAPS & G. Micela & 27/07/2017 & 200 & 81 (81) & 146.3 & 0.593 - 0.642 & $-30.2$\\ 
        8 & 0101.C-0106(A) & J.H.C. Martins & 20/08/2018 & 300 & 48 (48) & 166.7 & 0.569 - 0.612 & $-30.2$\\
        
    \end{tabular}
    
       \tablefoot{The eight nights of observations include the dedicated HARPS-N observing programme (top four rows) and the archival data (bottom four rows). Both sets of observations are ordered by their start dates (in UTC). \\
       \tablefoottext{$\ast$}{The number with which each night of observations is referred to throughout this paper, for example, data collected from programme CAT15B\_146 are referred to as Night 1.}
       \tablefoottext{$\dag$}{The exposure time for each frame}
       \tablefoottext{$\S$}{The total number of frames ($N_f$) collected per night, with the $N_f$ actually used in our analysis in brackets (see Section 3).}
       \tablefoottext{$\P$}{The mean S/N per frame, for the frames that we used in our analysis. Discarding one frame from Night 3 and 35 frames from Night 4 resulted in an increase in the mean S/N from 110.8 to 112.6 and 119.9 to 173.4 respectively.}
       \tablefoottext{$\ast\ast$}{The orbital phase ($\phi$) at the start and the end of the used frames.}
       \tablefoottext{$\ddag$}{The change in 51 Pegasi b's $RV_p$ over the course of the used frames. }} 

    \label{tab:obvs}

\end{table*}

In this work, made use of both dedicated and archival observations of the main-sequence star 51 Pegasi, a bright star of spectral type G2 IV and apparent magnitude V = 5.46 mag. We ran our dedicated observation programme at the TNG using HARPS-N, while the archival data was collected at both HARPS-N and HARPS. Together, these datasets almost precisely replicate the data analysed by S21, enabling comparison between our results. The data is summarised in Table~\ref{tab:obvs}. We used the time of inferior conjunction ($T_{c}$) in the orbital solution of S21 to calculate the orbital phases ($\phi$) of these time-critical observations, which in total spanned $ 0.395 < \phi < 0.612 $. The total change in the radial velocity of the planet ($RV_p$) during the observations was $\sim\Delta RV_p= -206.5 \,\rm km\,s^{-1} $. We note that while Nights 2, 4, 7 and 8 have the highest average mean S/N in our dataset (after the removal of poor quality frames), the overlapping $RV$ of the planet and stellar lines during the phases obtained in Night 2, the fewer usable frames obtained in Night 4, and the relative distance from $\phi = 0.5$ for Nights 7 and 8 result in Night 1 having the best sensitivity for extracting the planet spectrum. Details of our dedicated observing programme and the archival data we used are presented below in Sections~\ref{obs_HARPSN} and \ref{obs_arc}.  

\subsection{Dedicated HARPS-N observing programme} \label{obs_HARPSN}
We observed 51 Pegasi b continuously during four half-nights between 2015-2016, under Programmes CAT15B\_146, CAT16B\_146 and CAT16B\_143 (hereafter Night 1, 2, 3 and 4, see Table~\ref{tab:obvs}).  We collected a total of 274 exposures spanning $ 0.43 < \phi < 0.56 $, chosen to coincide with the passage of 51 Pegasi b when its illuminated day side hemisphere was orientated towards the Earth (see Figure~\ref{fig:phase}) for maximum signal. However, some of the exposures were collected when the radial velocity of the planet ($RV_p$) overlapped with the radial velocity of the star ($RV_\star$) that is at $\phi$ = 0.5. While not ideal due to increased contamination by the host star (see Section~\ref{cc_stel_cont}), this only affected a small number of frames.

The observations were carried out in visitor mode using HARPS-N \citep{HARPS}, a fibre-fed echelle spectrograph on the Nasmyth B Focus of the 3.6-TNG telescope, located at the Observatorio del Roque de Los Muchachos, La Palma, Spain. There are two HARPS-N fibres, one for the object and one for calibration (sky or Th-Ar lamp), each with an aperture of 1 arcsecond.  Each fibre's input is re-imaged by the spectrograph optics onto the two $4096\times2048$ pixel CCDs, designed for enhanced response in the optical regime, specifically $385<\lambda<691$ nm. This results in a resolving power $R \sim 115\,000$, with a velocity per resolution element of  $c/R = 300 000/115 000 = 2.6$ km\,s$^{-1}$, which is sampled by the HARPS-N full width at half maximum FWHM = 3.2 pixels, resulting in a velocity resolution per pixel of $2.6/3.2 = 0.8$ km\,s$^{-1}$. For each frame, a spectrum comprising 69 echelle orders is formed per fibre.

The lamp calibration frames were all taken prior to the start of science observations, and during science observations the reference fibre was put on the sky. Simultaneous reference lamps were not used concurrent with science observations in order to minimise the possibility of cross-contamination. Thus, only lamp calibration frames were available to determine the wavelength solution. We note that, given the high stability of the temperature and pressure conditions in the chamber containing HARPS-N, the wavelength solution obtained from the calibration frames at the beginning of the night will be sufficiently precise for our purposes \citep{Nugroho2020a}. We do not need to flux calibrate the spectra and thus did not observe standard stars.

The observing conditions across the nights were variable. Despite this, the length of each exposure was kept at 200 seconds, in order to maintain consistent instrumental noise properties.  Nights 2 and 3 included periods of dome closure due to poor weather, and Night 3 was further disrupted by tracking problems, resulting in fewer frames collected on these nights. Adverse observing conditions at the end of Night 4 also substantially reduced the S/N of these spectra, hence we have discarded 35 frames from our analysis, which left a short gap in the phase coverage of the dataset (see Figure~\ref{fig:phase}). S21 similarly cut 31 frames from this night of data. The additional four frames that we cut from Night 4 is the only difference between the datasets used in this work and ~S21.

\subsection{Archival HARPS-N and HARPS data} \label{obs_arc}
Nights 6 and 7 are archival HARPS-N spectra from GAPS programme (see Table~\ref{tab:obvs}). We used only the optical data associated with this program as reflected light from 51 Pegasi b is strongest at these wavelengths. This programme used the same 200 s exposure times as our dedicated program, and we used all 159 spectra collected across the two nights.

Nights 5 and 8 are archival HARPS spectra collected with the ESO La Silla 3.6m telescope under programmes 091.C-0271 and 0101.C-0106(A), respectively. While these programmes contain spectra obtain sporadically and sparsely across multiple nights each, we only use spectra obtained during long continuous observing sequences during any single night in order to match the observing strategy in our dedicated observing program, and the analysis of S21. Night 5 corresponds to the data used in the previous analyses of reflected light from 51 Pegasi b (M15;~\citealp{Borra2018ACF,Marcantonio2019ICA}). Both nights have different exposure times to the HARPS-N datasets (see Table~\ref{tab:obvs}), but have photon-dominated noise, and so we opted to use them to remain consistent with S21.

\section{Method} \label{method}
The light reflected from 51 Pegasi b is much fainter than its host star. Therefore, to reveal the planet spectrum, we needed to remove stellar contamination from the observed spectra, ideally to the photon noise level, before we could search for the planet signal using the cross-correlation method. Because we rely on the Doppler-shift of the planet spectrum to isolate it from the star, we aimed to remove all features that were stationary in wavelength over time.

For ease of understanding and replication, in this Section we have first summarised the function of the HARPS-N data reduction software (DRS, \cite{LovisPepe2007}, see Section~\ref{process}) that is used by the TNG archive from which we obtained the extracted spectra. The further post-processing that we carried out is detailed in Section~\ref{post-process}. Section~\ref{model_inject} describes: i) the process for injecting a synthetic stellar spectrum into the data, which we used to test the efficacy of the pipeline and to obtain an upper limit on the planet-to-star contrast ratio; and ii) how we accounted for the rotational broadening of the planet signal in the pipeline. In Section~\ref{stel_remove} we describe our method for removing stellar lines from the dataset, including the use of the \textsc{Sysrem} algorithm. Section~\ref{cc_method} shows how we performed the cross-correlation. Each night of data was processed separately until after the cross-correlation was performed.

\subsection{HARPS-N DRS and the e2ds format} \label{process}
 %Begin figure 
\begin{figure*}
    \centering
       {%
          \includegraphics[width=\textwidth]{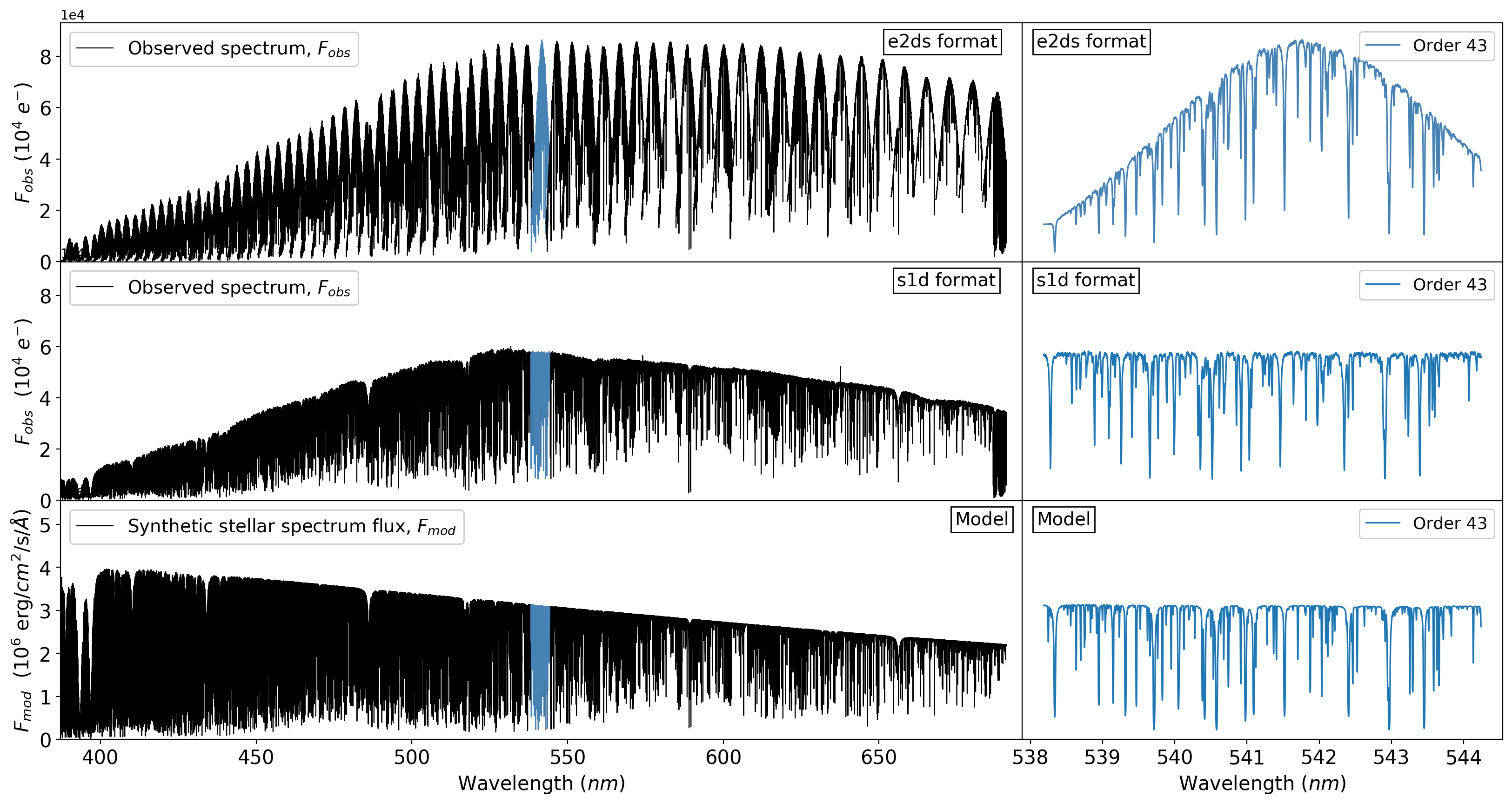}%
        }
   \caption{Extracted spectra ($F_{obs}$) in both e2ds (top panel) and s1d (middle panel) formats, and the synthetic stellar spectrum ($F_{mod}$, bottom panel) used in the cross-correlation. The units $e^{-}$ are photoelectrons. A zoom around wavelengths $538.2 - 533.2$ nm (order 43) is shown in blue and the right-hand panels. These plots show the dense forest of spectral lines in our dataset. The continuum variation in the e2ds format is the uncorrected blaze function, while in the s1d format we see the residual low order instrument response in our non-flux calibrated data.}
    \label{fig:spectrum}         
    \end{figure*}
 %End figure  

Every exposure collected by HARPS-N is processed by the HARPS-N DRS, taking the data from raw $4096\times4096$ pixel detector format to extracted spectra and radial velocity drift ($RV_{drift}$) measurements. For our dataset, the DRS corrected for bad columns on the CCD and the bias, then extracted the 69 individual orders in each frame using the Horne optimum extraction method~\citep{Horne1986}, and flat-fielded.

The DRS provides data in two formats: extracted two-dimensional spectra (e2ds) and extracted one-dimensional spectra (s1d, see top two panels, Figure~\ref{fig:spectrum}). The e2ds format provides a two-dimensional matrix for each frame, where each row represents a single order (i.e. $4096\times69$). Thus each row starts at a different wavelength value and spans a different wavelength range.  The unique wavelength solution for each order is calculated by the DRS. There is a significant overlap in wavelength range between each order. The e2ds spectra remain in the Earth rest frame. Conversely, s1d format provides a single one-dimensional `stitched' spectrum for each frame. To make this, the 69 orders are first blaze-corrected, and then merged together and re-binned onto a regularly spaced wavelength grid by the DRS, and are then interpolated onto the barycentric reference frame. The s1d format has been the chosen format for several publications (see e.g. \citealp{Hoeijmakers2018}; S21); however, the additional steps of merging the orders and rebinning the data risks increasing the systematics~\citep{Pino2020}, hence we elected to use the less processed e2ds data.
 
The DRS  provides a wavelength solution (a third order polynomial) per order for every spectrum, which is in pixel spacing for the e2ds format (i.e. an irregular wavelength grid). For each night, we used the wavelength solutions obtained from the lamp calibration frames (see Section~\ref{obsv}). This wavelength solution was sufficiently precise to determine the $\sim$km\,s$^{-1}$ variations of the planet radial velocity.

\subsection{Post-processing the data}  \label{post-process}
\subsubsection{Separating the orders}
To facilitate data handling and the removal of stellar lines, we first reorganised the extracted spectral matrices. For each night, we rearranged the e2ds spectra into 69 two-dimensional matrices, one for each spectral order, where each column corresponded to a single wavelength channel (detector pixel) and each row showed the spectrum at a given time (i.e. a 69 matrices of $4096\times N_f$). This served to highlight variations in time and wavelength, such as changes in flux due to clouds coverage or seeing effects, as well as barycentric motion and long term radial velocity drift. Accounting for the latter helps especially in the removal of the stellar lines (see Section~\ref{align}).

\subsubsection{Outlier removal}
Despite bad pixel correction by the DRS, some outliers still remain that propagate into the extracted spectra. These outliers can adversely affect the removal of stellar lines \citep[see e.g.][]{Birkby2013}. To highlight them, for each matrix, we first continuum normalised each row (see Section~\ref{stel_remove} for a description of our continuum normalisation method), then subtracted the mean column value from each wavelength channel. To determine if a pixel was an outlier, we calculated the median and standard deviation of its six nearest neighbour pixels in the same row and if it was more than six standard deviations above or below this median, it was replaced by the linear interpolation of these neighbours in the e2ds spectra. The choice of the $6 \sigma$  threshold led to approximately 0.3\% of pixels being identified as an outlier across the dataset, but was sufficiently high that variation due to any potential planet signal would not be mistakenly removed. 

\subsubsection{Alignment to the stellar rest frame} \label{align}

\begin{figure}
    \centering
       {%
          \includegraphics[width=9cm]{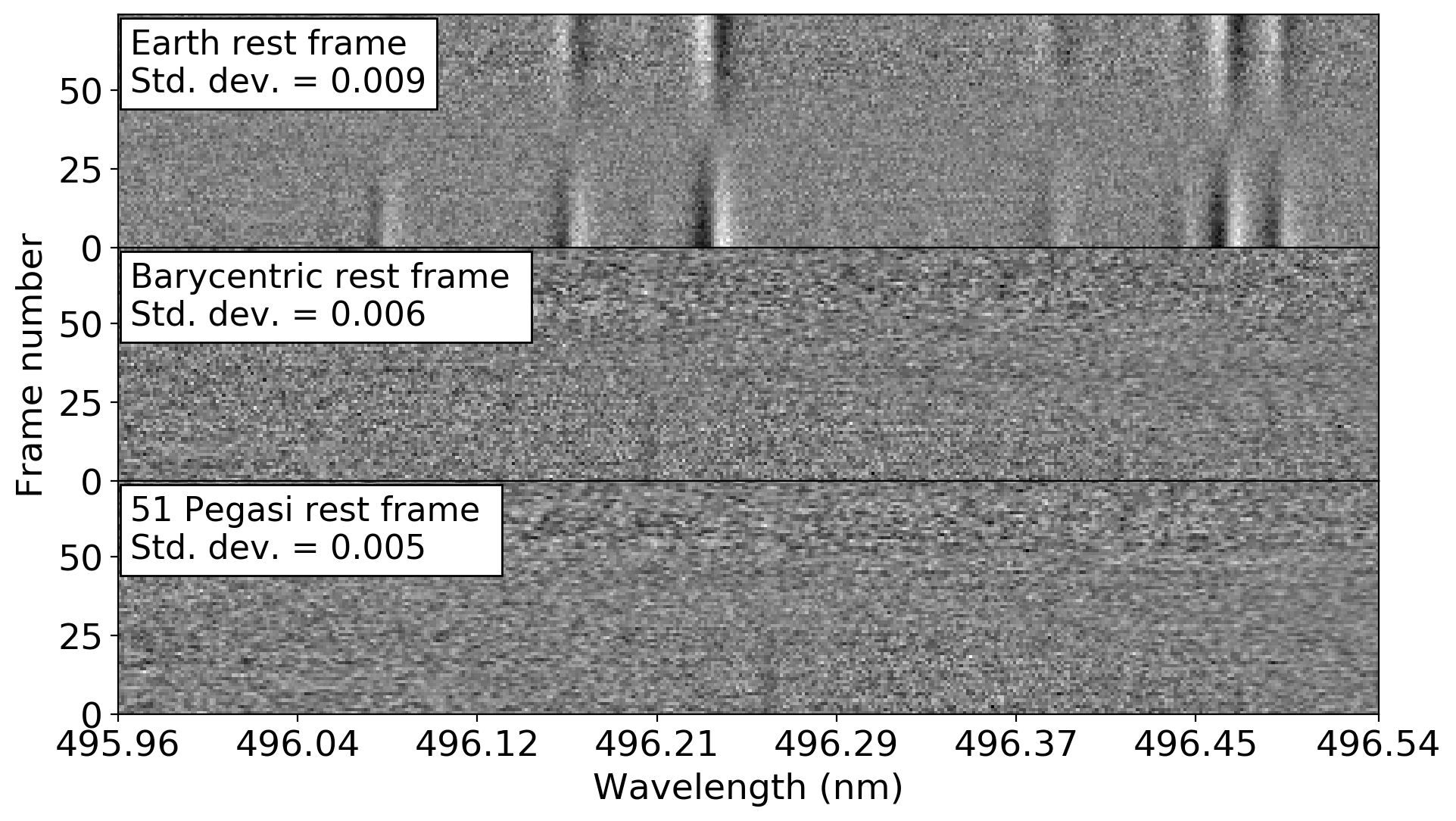}%
        }

   \caption{Section of order 35 from Night 1, aligned to different rest frames, with the corresponding standard deviation shown in the top left of each panel.  For each alignment, the matrix has been continuum normalised and then the mean of each column has been subtracted. This exposes the characteristic `X' pattern caused by misalignment of the stellar lines. The noise is reduced closest to the photon noise in the host star's true rest frame, reducing the standard deviation by a factor of 1.8.}
    \label{fig:aligned}         
    \end{figure}

The optical spectra of the 51 Pegasi system were dominated by the stellar lines, and were far less significantly contaminated by the telluric lines that plague high resolution spectra in the infrared regime \citep[e.g.][]{Birkby2017}. Here, we demonstrate that the ideal rest frame to remove this major stellar contaminant is the rest frame of the star, which departs from previous infrared studies with HRCCS that aligned to the telluric rest frame \citep[e.g.][]{Birkby2013}. This is visualised in Figure~\ref{fig:aligned}, where each extracted spectrum has been continuum normalised (see Section~\ref{stel_remove}), and then the mean of each wavelength channel has been subtracted to highlight any spectral features that are not stationary in wavelength over time. Any under- or over-subtraction by the mean column values results in a characteristic `X' pattern in the residuals, indicating misalignment of the stellar lines over time. Previous works have shifted the spectra to the barycentric reference frame to mitigate this effect for stellar lines. The middle panel of Figure~\ref{fig:aligned} highlights the $\sim33\%$ reduction in the standard deviation of the residuals when the spectra are first aligned to the barycentric reference frame. The value for the Barycentric Earth Radial Velocity ($BERV$) was taken from the DRS. We then further aligned our spectra to the true rest frame of the host star using the $RV_{drift}$ values calculated by the DRS. We note that the $RV_{drift}$ exceeded the expected values due to the influence of 51 Pegasi b alone.  In this true stellar rest frame, we achieve an additional $\sim11\%$ reduction in the standard deviation of the residuals, resulting in a $\sim44\%$ total reduction in comparison with the residuals in the Earth's rest frame, indicating that the stellar rest frame is the best rest frame to work in for removing stellar contamination (see the bottom panel of Figure~\ref{fig:aligned}).

We used linear interpolation to shift the extracted spectra into the true stellar rest frame. A side effect was that any telluric lines were then misaligned, and were therefore removed less effectively (see Section~\ref{stel_remove}). We prioritised the optimal removal of the host star's spectrum rather than the tellurics as discussed in Section~\ref{intro}, noting also that only the reddest spectral orders of HARPS-N were significantly affected by telluric lines. However, in the case of seeking the reflected spectrum from an Earth-like planet, one would need to carefully consider both sources of contamination, including micro-tellurics \citep{Cunha2014}.

 %Begin figure
 %End figure  

\subsection{Injecting a model}   \label{model_inject}
We consider four separate cases throughout this work: 
 
 \begin{enumerate}
     \item No model injected, assume any planet signal is rotationally broadened according to $v_{refl} = 12.0\,\rm km\,s^{-1}$
     \item A model injected that is broadened to match HARPS-N's spectral resolution, and rotationally broadened according to $v_{refl} = 12.0\,\rm km\,s^{-1}$
     \item No model injected
     \item A model injected that is broadened to HARPS-N's spectral resolution only
 \end{enumerate}

\noindent We inject a model planet into our data in order to determine an upper limit on the star-planet contrast ratio (see Section~\ref{res_upp_lim}). Case 2 enables us to determine the impact of rotational broadening on our ability to recover a realistically broadened signal from 51 Pegasi b, whilst case 4 allows us to find the maximum sensitivity of the observations in the case where we are limited by the instrument resolution. Figure~\ref{fig:model_broaden} shows the model planet spectrum in both cases. As discussed in Section~\ref{intro}, we use a synthetic spectrum of the host star as our reflected light model in order to avoid any systematics and tellurics that might otherwise contaminate a reference made from the observed data, and to correctly broaden the spectral lines. In this work, for simplicity we assume that $A_{g}(\lambda)=A_{g}$ is constant with wavelength. This approximation is typical in previous HRCCS studies, whilst more detailed models of $A_{g}(\lambda$) are reserved for high S/N detections. 

\subsubsection{Synthetic model planet spectra}
We obtained the synthetic stellar spectrum from the \citet{Coelho_2014} stellar library, with $T_{eff}=5750$ K, $\log(g)=4.5$, $[Fe/H]=0.2$, and $[\alpha/H]=0.0$ at a wavelength resolution of $R=300,000$. We use the continuum normalised spectrum as the model ($F_{mod,\,norm}(\lambda^\prime)$, where $\lambda^\prime$ is in the laboratory rest frame). 

In case 4, where we only consider broadening due to the instrument resolution of HARPS-N, we simply convolve the spectrum with a Gaussian to match the HARPS-N spectral resolution ($R=115,000$). In case 2, to account for the rotational broadening in the reflected light from 51 Pegasi b, we convolve the model with a rotation kernel (see right panel, Figure~\ref{fig:model_broaden}) using $v_{refl} = 12.0\,\rm km\,s^{-1}$ (as calculated for 51 Pegasi b in Section~\ref{intro_rot_broad}) with spacing, $v_{step}$ = 0.8$\,\rm km\,s^{-1}$ corresponding to HARPS-N's velocity resolution per pixel\footnote{To ensure that our results from all eight nights were possible to combine, we used HARPS-N's velocity resolution throughout our analysis.}

 \begin{figure*} 
    \centering
        \vspace{-1.5mm}
          \includegraphics[width=18.4cm]{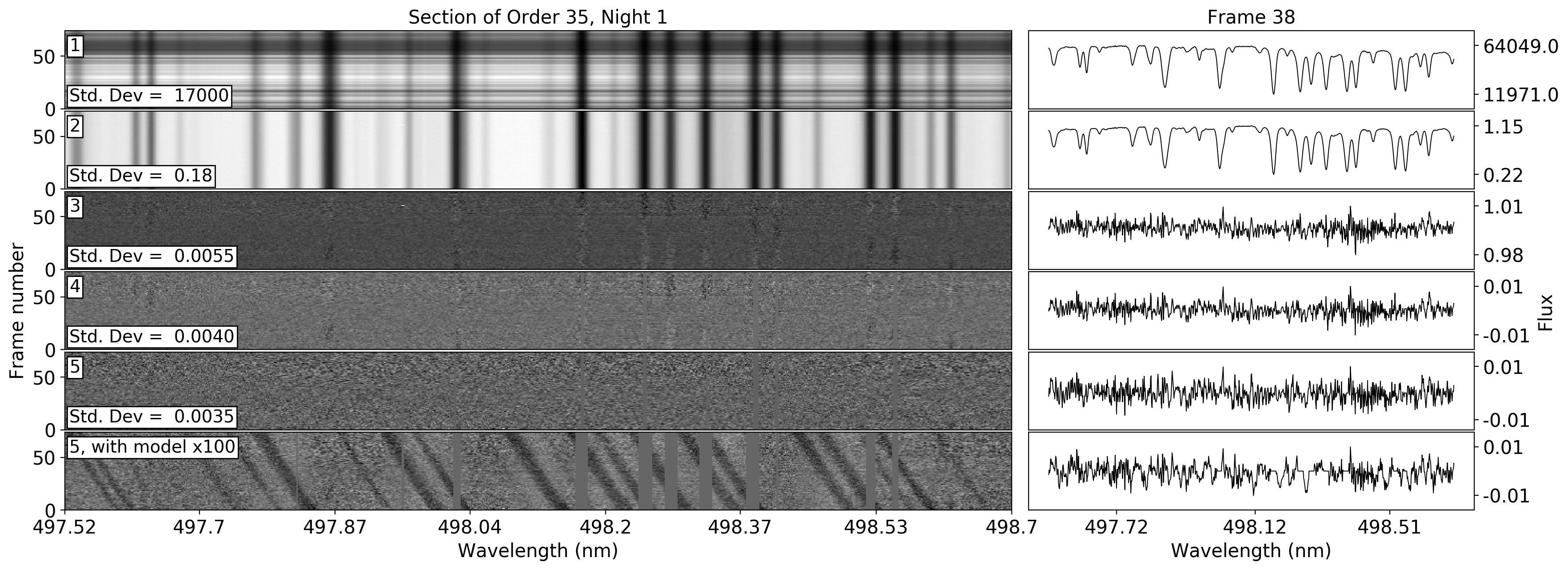}%
        \vspace{-3mm}

    \caption{\textbf{Left panels: }sections of the order 35, from Night 1, shown at various stages of post-processing, demonstrating the removal of the stellar residuals. \textbf{Right panels: }sections of frame 38 from order 35, demonstrating the removal of the stellar residuals from a single spectrum. \textbf{Row 1:} observed spectra after bad pixel correction and alignment to the true stellar rest frame; \textbf{Row 2:} after the removal of each order's continuum; \textbf{Row 3:} after division by a master spectrum; \textbf{Row 4:} after masking the columns with over 1.4x the mean column variance, followed by one iteration of the \textsc{Sysrem} algorithm; \textbf{Row 5:} after weighting columns by their variance; \textbf{Bottom row:} same stage as for Row 5, but with a model planet ($R_p = 1.9 R_J$, $A_g = 0.5$, $v_{refl,\,mod} = 0.0\,\rm km\,s^{-1}$) injected at 100x nominal strength after the alignment to the stellar rest frame. At each stage of post-processing, the standard deviation of the entire order decreases.}
    \label{fig:waterfall}
    \end{figure*} 

\subsubsection{Adding the model to the data}
 When injecting a model, we assume that the observed spectra $F_{obs}(\lambda)$ contain only light from the host star, that is $F_{obs}(\lambda)=F_\star(\lambda)$. The sequence of injecting a model planet into the observed spectra is as follows: i) broaden the model according to case 2 or 4; ii) scale the model according to $g(\alpha)$, $A_g$, $R_p$ and $a$ (see Equations~\ref{eq_Fp}, \ref{eq_phase_angle} and \ref{eq_phase_function}); iii) Doppler-shift the model according to the orbital phase at the time of observation; iv) multiply the Doppler-shifted model, $F_{mod,\,norm}(\lambda)$, by the continuum of the observed spectra, $F_{obs,\,cont}(\lambda)\,$; and finally v) add the model to the data. We include step iv because our observed spectra are not flux calibrated, so multiplying by $F_{obs,\,cont}$ ensures that the model is injected at the correct contrast ratio. 

The $A_{g}$ factor assumes that all the light reflected from 51 Pegasi b's dayside is directed towards Earth, but of course this is not the case. In order to account for this, the phase function, $g(\alpha)$, outputs a factor between 0 and 1 that indicates the proportion of the light that is reflected towards Earth's line of sight. The phase function is determined by the phase angle $\alpha$, which is calculated by:

\begin{equation}
    \cos{\alpha}\,=\,-\sin{i}\,\cos{(2\pi\phi)}
    \label{eq_phase_angle}
\end{equation}

\noindent Following~S21, if we assume that 51 Pegasi b follows Lambert's scattering law, then the phase function is calculated by:

\begin{equation}
    g{(\alpha)}\,=\,\frac{\sin{\alpha}\,+\,(\pi - \alpha)\cos{\alpha}}{\pi}
    \label{eq_phase_function}
\end{equation}

\noindent The phase function is maximised to 1 in the case where an exoplanet has $i = 90^{\circ}$ and if at inferior conjunction ($\phi\,=\,0.5$, although, in the case of 51 Pegasi b,  the planet would then be blocked from the Earth's line of sight by its host star). The phase function is minimised when the exoplanet is in superior conjunction ($\phi\,=\,0.0$).

To account for the Doppler shift associated with each observed spectrum, we offset $F_{mod,\,norm}(\lambda^\prime)$ by $\Delta \lambda = RV_{p}(K_p,\phi)/c$, where the planet $RV_p$ semi-amplitude $K_{p}=\,133 \rm km\,s^{-1}$ was taken from literature (e.g. \citealt{Brogi2013, Birkby2017}) and the phase was calculated as in Section~\ref{post-process}. This resulted in a new wavelength grid for each frame of $\lambda= \lambda^\prime\ (1 + \Delta\lambda)$, and we interpolated $F_{mod,\,norm}(\lambda^\prime)$ onto each new wavelength grid, generating $F_{mod,\,norm}(\lambda)$ in Equation~\ref{eq_inj} below at each time step. In all instances, the model is injected after the alignment of the observed spectra to the rest frame of the host star, but prior to the removal of the stellar lines.

The sequence of model injection is described by the following equation, where $F_{mod,\,norm}(\lambda,v_{refl})$ is the appropriately broadened and Doppler-shifted model:

\begin{multline}
     \label{eq_inj}
F_\star(\lambda)+F_{p}(\lambda)=\\
 F_{obs}(\lambda)+\left[F_{mod,\,norm}(\lambda,v_{refl})\,F_{obs,\,cont}(\lambda)\,g(\alpha)\,A_g\left(\frac{R_p}{a}\right)^2\right]
\end{multline}

\subsection{Removal of stellar and telluric lines} \label{stel_remove}

51 Pegasi b orbits a G2 IV star that has abundant spectral features at optical wavelengths. As discussed in Section~\ref{intro}, these same lines appear in the planet spectrum, but Doppler-shifted by the planet's radial velocity, and thus their multiplicity aides the recovery of the planet spectrum in the cross-correlation (boosting the S/N by $\sqrt{N_{lines}}$). However, it also means the cross-correlation template will match with any residual host star spectrum left in the data. Hence, the removal of the host star spectrum is key to extracting the faint planet spectrum. In infrared HRCCS where tellurics dominate, stellar lines have been removed by modelling them directly \citep[][]{Brogi2016,Chiavassa2019}. However, given that they are numerous and the main source of contamination in our optical spectra, and that we expect systematic effects (e.g. air mass variations) to impact the spectra as well, we opt instead for a data-driven approach. With the spectra now in the true stellar rest frame, the stellar lines are stationary in wavelength over time and we can remove the host star spectrum, as described below and illustrated for Night 1 in Figure~\ref{fig:waterfall} (similar for the other nights can be found in Appendix~\ref{appen_additional_plots} in Figure~\ref{fig:waterfall_appen}).

We first removed the stellar continuum from each spectrum. To model the continuum, we divided each residual spectrum into 20 sections and assigned the median of each as the continuum value at the central wavelength of each section. We interpolated these values onto the  wavelength grid for their order, then divided each residual spectrum by the continuum model. 

We then created a master spectrum for each individual order by taking the mean of all the frames. We then divided each  spectrum in the order by the master spectrum. This successfully removed the weaker stellar lines, however, at this stage, there were still traces of the deeper stellar lines in the data and other high order systematic effects (see Figure~\ref{fig:waterfall}). We opted to mask the strongest residuals before any further cleaning. To make the mask, we first calculated the variance of the residuals in each column, and then calculated the median column variance. We divided each column's variance by this value and applied a sigma clip. For cases 1 and 2 (see Section~\ref{model_inject}) we used a lower 1.4 sigma clip, resulting in a more aggressive stellar mask. This minimised the number of times that the cleaning algorithm \textsc{Sysrem} (see below) needed to be run, as we found that broadened planet signals are susceptible to removal by \textsc{Sysrem} (see Section~\ref{sec:sysrem}). For cases 3 and 4 we used a higher 1.6 sigma clip, resulting in a less of the data being masked and available for use in the cross correlation. A full investigation of the optimal masking approach for HRCCS is beyond the scope of this work, and deferred to future study.

\subsubsection{\textsc{Sysrem\label{sec:sysrem}}}
 
    %Begin Figure
    \begin{figure*}%[t!]
    \centering
        \subfloat{%
          \includegraphics[width=18.4cm]{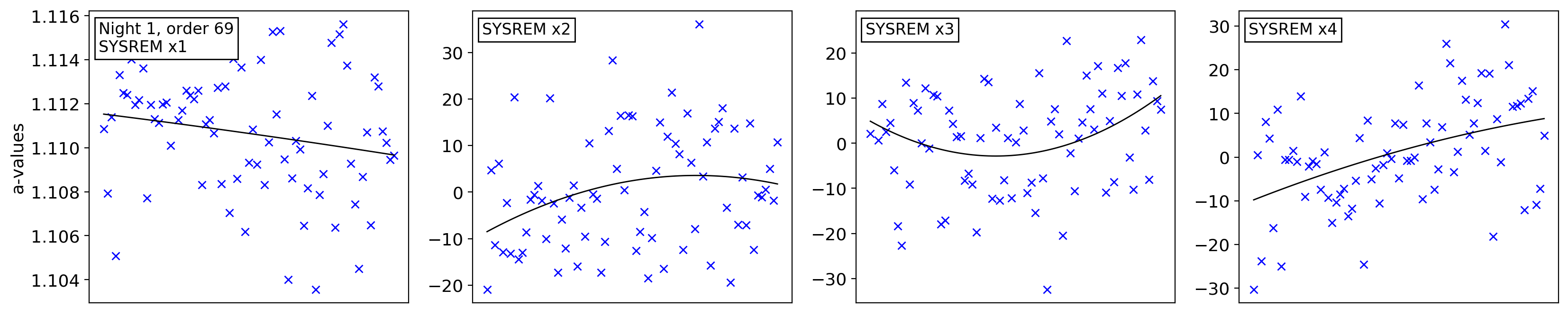}%
        }\vspace{-3mm}

        \subfloat{%
          \includegraphics[width=18.4cm]{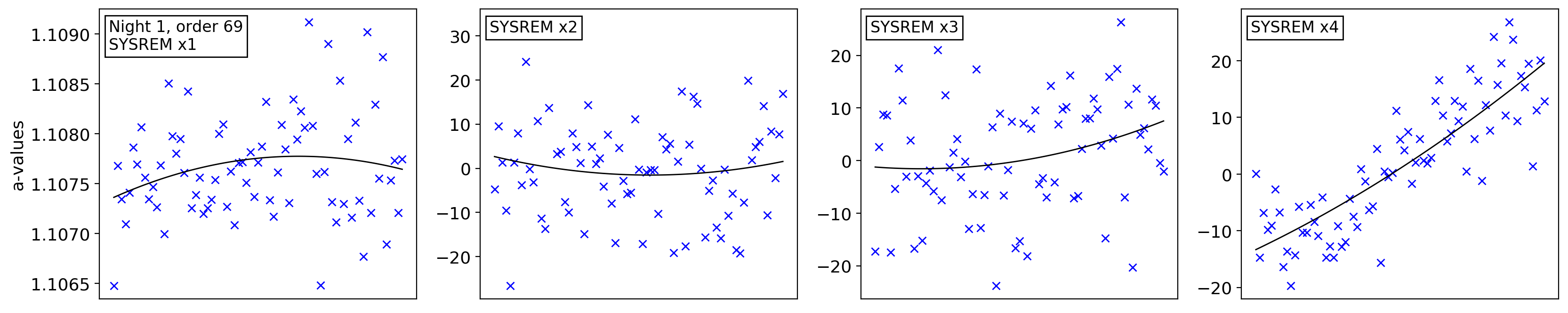}%
        }\vspace{-3mm}
        
        \subfloat{%
          \includegraphics[width=18.4cm]{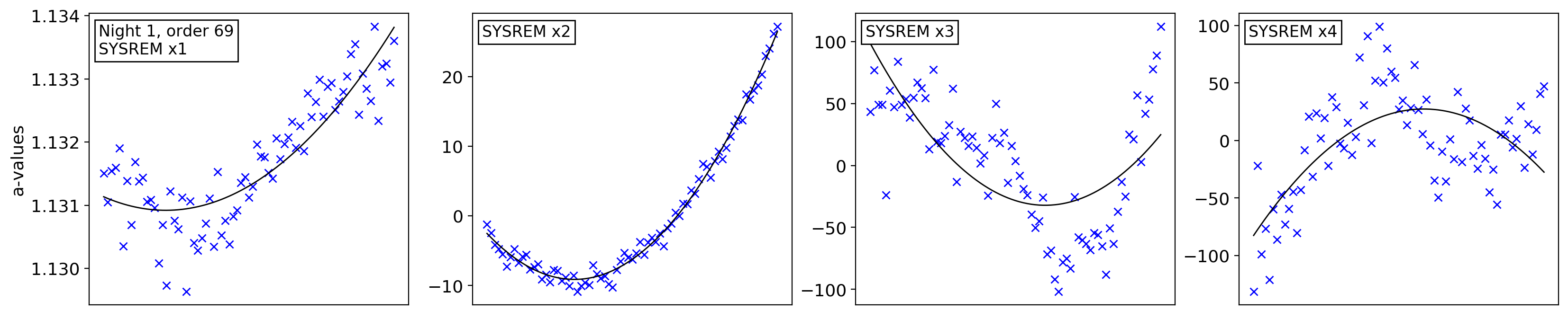}%
        }\vspace{-3mm}
         
        \subfloat{%
          \includegraphics[width=18.4cm]{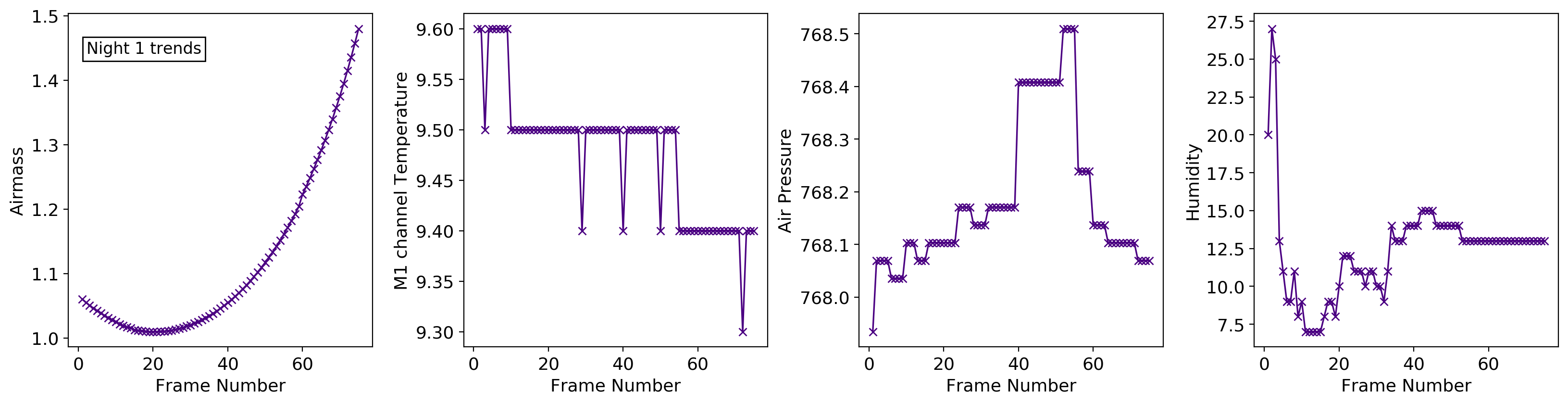}%
        }
     \caption{Sets of \textsc{Sysrem} a-values from Night 1 of data that has been masked with a 1.4$\sigma$ clip (blue crosses, top three rows, see Section~\ref{stel_remove}), $2^{nd}$ degree polynomials fitted to detect any trends (black lines, top three rows), and possible causes of the observed \textsc{Sysrem} trends (purple crosses, bottom row). \textbf{Top three rows:} The rows show the a-values subtracted during the first, second, third and fourth iteration of \textsc{Sysrem} for orders 1, 35 and 69, respectively. There is a large variety in the shape of the trend removed during each iteration, and little similarity between the trends, indicating a strong wavelength dependence. \textbf{Bottom row:} the air mass, primary mirror channel temperature, local air pressure and humidity during Night 1 of observations. The air mass values tightly correlate with the a-values for order 69 where tellurics were strongest. The other physical trends do not correlate with any of the sets of a-values.}
    \label{fig:sys_rem}        
    \end{figure*}
      %End Figure  

We next used the \textsc{Sysrem} algorithm~\citep{sysrem2005, sysrem2007} to further remove remaining high order variance. \textsc{Sysrem} is similar to PCA but allows unique uncertainties to be assigned to the data points. It was developed to identify and subtract systematic trends from multiple light curves. For a set of $M$ light curves, $\{l_0, l_1,...,l_x,...l_M\}$, each with $N$ points, each iteration of \textsc{Sysrem} will converge on two trends: the a-values, $\{a_0, a_1, ..., a_x, ..., a_N\}$ (see Figure~\ref{fig:sys_rem}), which correspond to the number of points in each light curve, (i.e. $N_f$); and the c-values, $\{c_0, c_1, ..., c_x, ..., c_M\}$, which correspond to the number of light curves, $M$ (i.e. the wavelengths channels). Once these trends have converged, each light curve $l_x$ is de-trended by subtracting its c-value multiplied by the a-values, $\{c_x a_0, c_x a_1, ..., c_x a_x, ..., c_x a_N\}$. This provides a new set of de-trended light curves, and completes a single iteration of \textsc{Sysrem}. By treating each spectral order as a set of multiple light curves, and thus each individual wavelength channel of an order as a single light curve, \textsc{Sysrem} is readily adaptable to high-resolution spectroscopic datasets. Because the planet is Doppler-shifting, it in principle does not present as a trend common to any of the light curves. \textsc{Sysrem} has been used previously, with an emphasis on the removal of telluric features from high-resolution spectroscopic data \citep[see e.g.][]{Birkby2013,Birkby2017,Merritt2020, Gibson2020, Yan2020, Kesseli2020, Nugroho2020a, Nugroho2020b, Nugroho2021}. For the individual uncertainties, we took the inverse square-root of each data-point prior to any removal of the stellar lines. This gave data-points with a higher S/N a lower associated uncertainty, and vice versa. We set the uncertainties for the masked regions to $1\times10^{8}$, namely a very high number so that \textsc{Sysrem} down weights them in its calculations.

\textsc{Sysrem} cannot be allowed to run indefinitely, as it will begin to converge on random patterns in the noise (see Figure~\ref{fig:sys_rem}). Subtracting these would simply distort the noise and run the risk of obscuring any planet signal. Thus, the optimal stopping point for the algorithm must be determined, which will vary per dataset. In principle, the systematic trends could vary per order per night, as conditions are not uniform, and therefore optimal results could be obtained by using different numbers of \textsc{Sysrem} iterations per order, per night. However, some methods for optimising per order \textsc{Sysrem} have the potential to constructively add noise at the planet's velocity, and in the limit of large datasets, this constructively compounded noise could create the appearance of a planet detection. We defer a full investigation into the optimal application of \textsc{Sysrem} beyond that of \citet{Cabot19} to future work, as it merits its own study. In this work, we cautiously adopt the same approach that others have in recent literature of applying the same number of \textsc{Sysrem} iterations for each order across all the nights to avoid possible biasing \citep{Cabot19, Merritt2020, Gibson2020, Yan2020, Nugroho2020a, Nugroho2020b, Nugroho2021}.

To determine when to stop \textsc{Sysrem}, we used a variation of the $\Delta$CCF method that is described in detail below in Section~\ref{ccf_weights}, and further justified in Section~\ref{discuss_sys}. In brief, the $\Delta$CCF is simply the difference of the cross correlation functions (CCFs) created by Cases 1 and 2, or by Cases 3 and 4, that is the model-injected minus observed CCFs, where the injected model corresponded to the $R_p$ and $A_g$ as suggested by M15. These $\Delta$CCFs are created for successive iterations of \textsc{Sysrem} until it is clear that \textsc{Sysrem} begins to remove the planet signal.

Some of the trends that \textsc{Sysrem} identified reflected a physical quantity, and other trends did not (see Figure~\ref{fig:sys_rem}). For the reddest spectral orders, the trends identified by \textsc{Sysrem} are highly correlated with the variation in air mass, likely due to the higher concentration of now misaligned telluric lines in this region.

\begin{figure}
    \centering
       \subfloat{%
          \includegraphics[width=9cm]{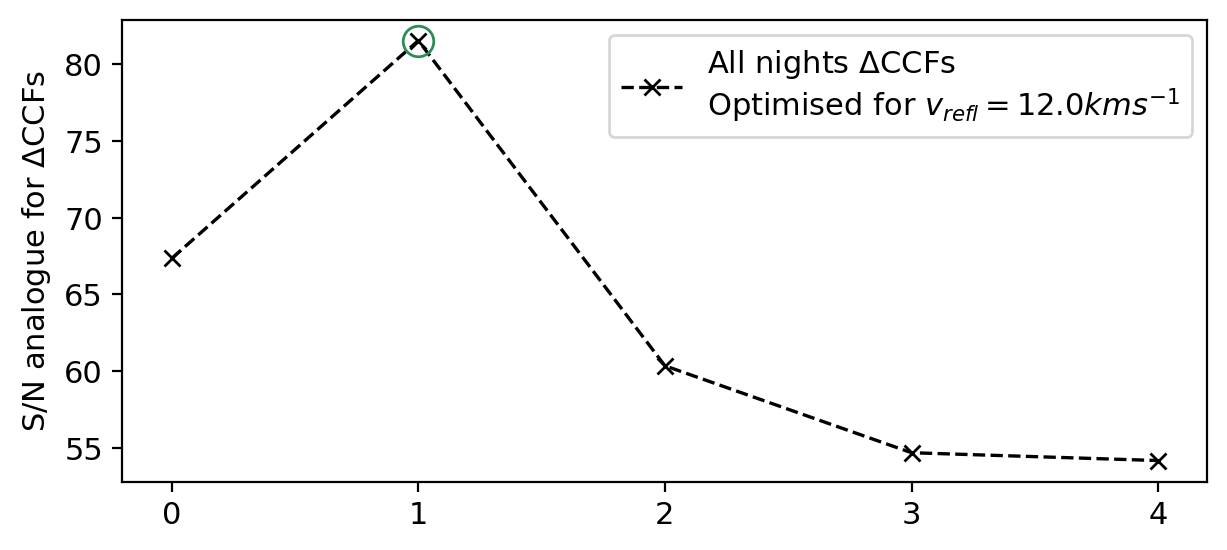}%
        }\vspace{-1.5mm}
        \subfloat{%
          \includegraphics[width=9.05cm]{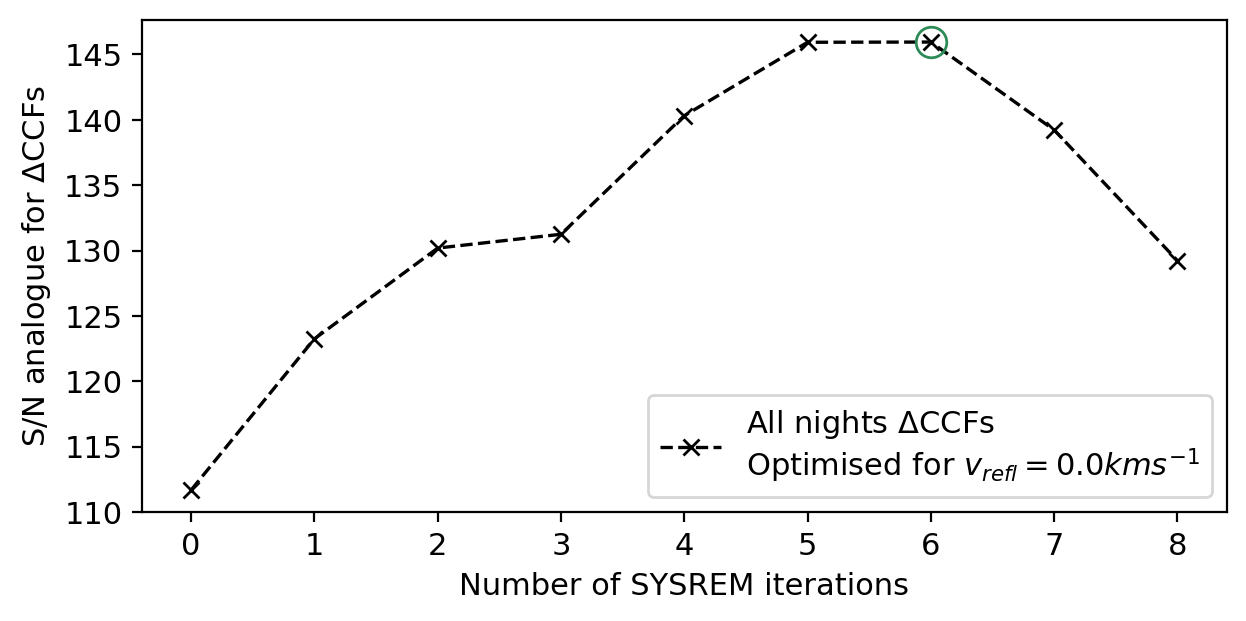}%
        }
   \caption{Optimising the S/N of the $\Delta$CCFs at $\Delta v$ = 0$\rm km\,s^{-1}$ in the planet rest frame,  for all orders and all the nights combined, optimised for cases 1 and 2 (top panel) and for cases 3 and 4 (bottom panel). More \textsc{Sysrem} iterations improves the model recovery when rotational broadening is not included, indicating that \textsc{Sysrem} is likely removing the wings of broadened signals and thus should be used with less iterations.}
    \label{fig:opt_sys}         
    \end{figure}

For the realistic cases, with a rotational broadening (cases 1 and 2, see Section~\ref{model_inject}), we assessed the results of using zero, one, two, three, and four iterations of \textsc{Sysrem}. We found that wings of the broadened model spectral lines were susceptible to being removed by \textsc{Sysrem}. One iteration of \textsc{Sysrem} provided the optimal balance between removing stellar and telluric contamination, without removing the injected broadened model (see top panel, Figure~\ref{fig:opt_sys}). Thus, when running our pipeline for cases 1 or 2 we applied one iteration of \textsc{Sysrem} to each order, across all the nights. This did not adequately remove the stellar contamination, and thus we decided to use a more aggressive stellar mask prior to running \textsc{Sysrem} for cases 1 and 2.

For the most sensitive case, without rotational broadening (cases 3 and 4, see Section~\ref{model_inject}), we assessed the results of using up to eight iterations of \textsc{Sysrem}. Six iteration of \textsc{Sysrem} provided the optimal result (see bottom banel, Figure~\ref{fig:opt_sys}). Thus, when running our pipeline we applied six iterations of \textsc{Sysrem} to each order, across all the nights. This meant we were able to use a less aggressive stellar mask prior to running \textsc{Sysrem} for cases 3 and 4.

    %Begin Figure      
    \begin{figure}
    \centering
        {
          \includegraphics[width=9cm]{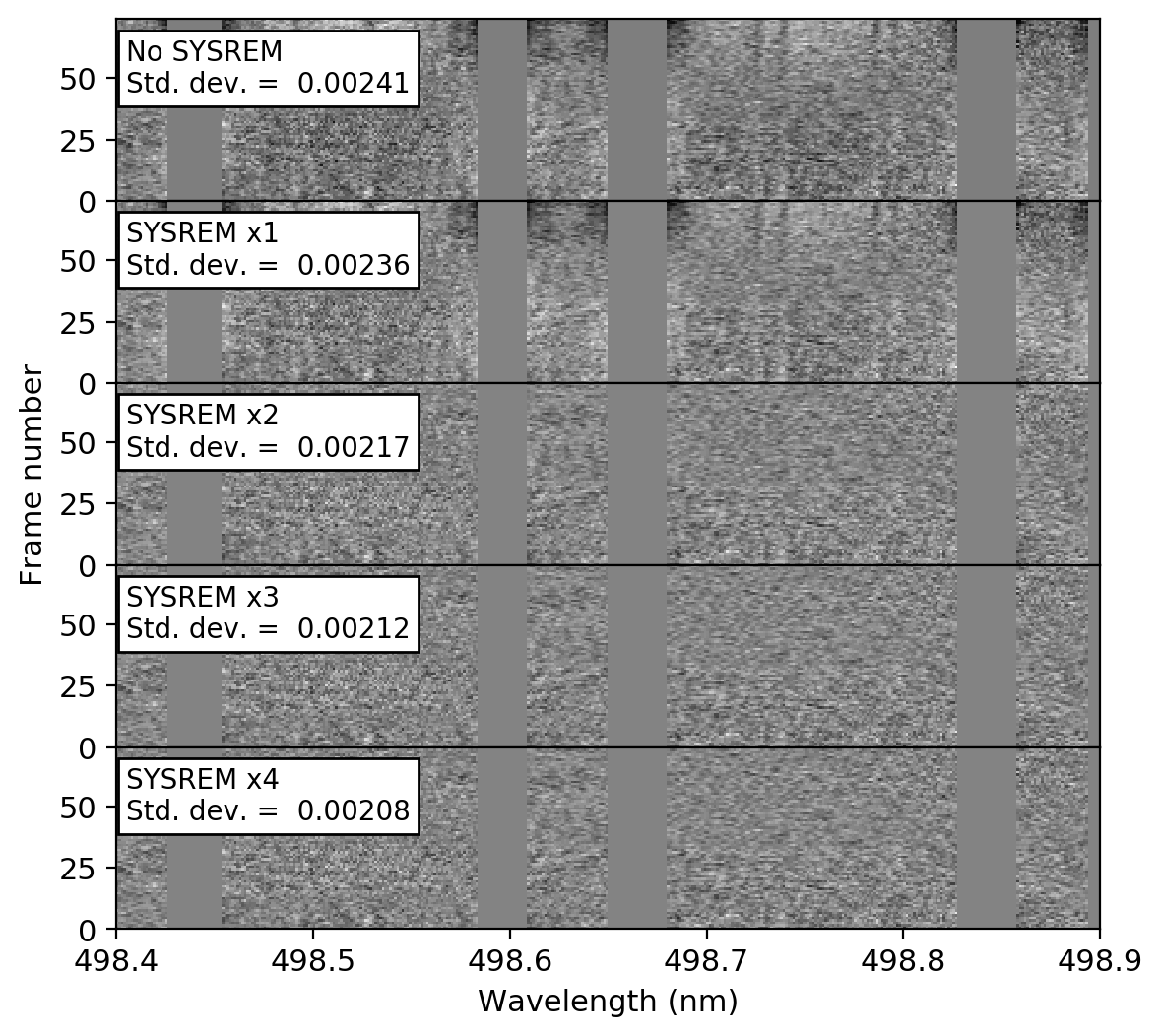}%
        } 
   \caption{Section of order 69, from Night 1 of observations, after zero, one, two, three and four iterations of the \textsc{Sysrem} algorithm respectively. After each iteration, the standard deviation of the order is reduced}.
    \label{fig:sys_rem_retrieved}
    \end{figure}
     %End Figure  

     %Begin figure
  %End figure  

\subsubsection{Weighting residuals by variance}
Following the literature \citep[see e.g.][]{Snellen2010, Birkby2017, Hoeijmakers2018}, we normalised our data after running \textsc{Sysrem} by its S/N. Each column was divided by its variance, and multiplied by the average variance per order. This served to down weight noisy wavelength channels (columns), such as any remaining stellar or telluric contamination, and the noisy fringes as the edges of each order.

 %Begin Figure 
    \begin{figure}
    \centering
        {
          \includegraphics[width=9cm]{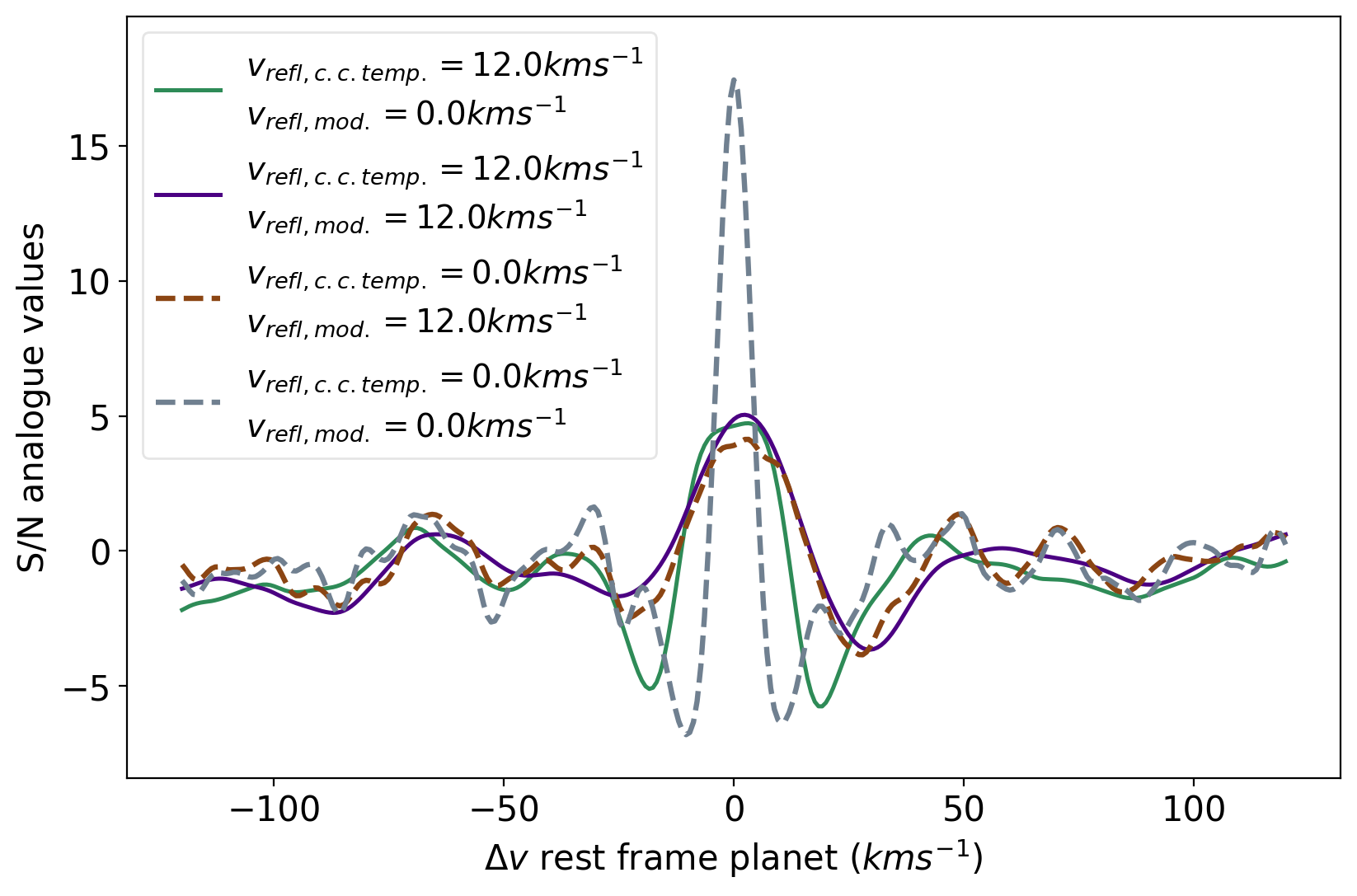}%
        } 
   \caption{Comparison of the summed (all orders combined) CCFs for Night 1 with a model (\emph{mod}) injected with parameters $R_p=1.9R_J$ and $A_g=0.5$, when the cross-correlation template (\emph{c. c. temp}) is broadened (solid lines) and is not broadened (dashed lines). Mismatched broadening (green and brown lines) results in lower S/N in the CCF peak compared to matched broadening (purple and grey lines).}

    \label{fig:cc_extract}        
    \end{figure}
    %End Figure   

%Begin Figure      
    \begin{figure}
    \centering
        {
          \includegraphics[width=9cm]{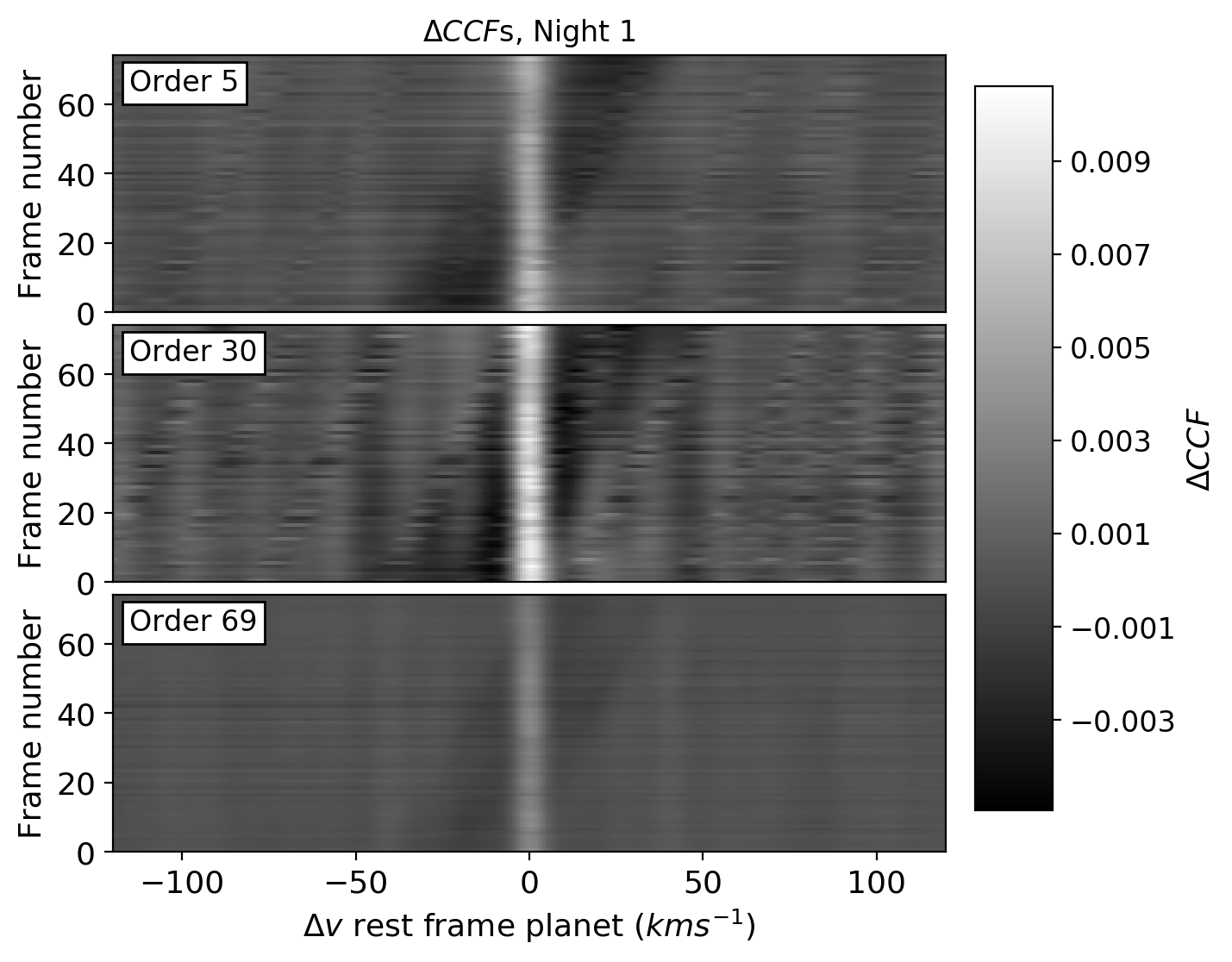}%
        } 
   \caption{$\Delta$CCFs from different orders from Night 1 (see Section~\ref{ccf_weights}) demonstrating the sensitivity of each order in recovering an injected model .}
    \label{fig:signal_only}
    \end{figure}
%End Figure  

    %Begin Figure      
    \begin{figure}
    \centering
        \subfloat{%
          \includegraphics[width=9cm]{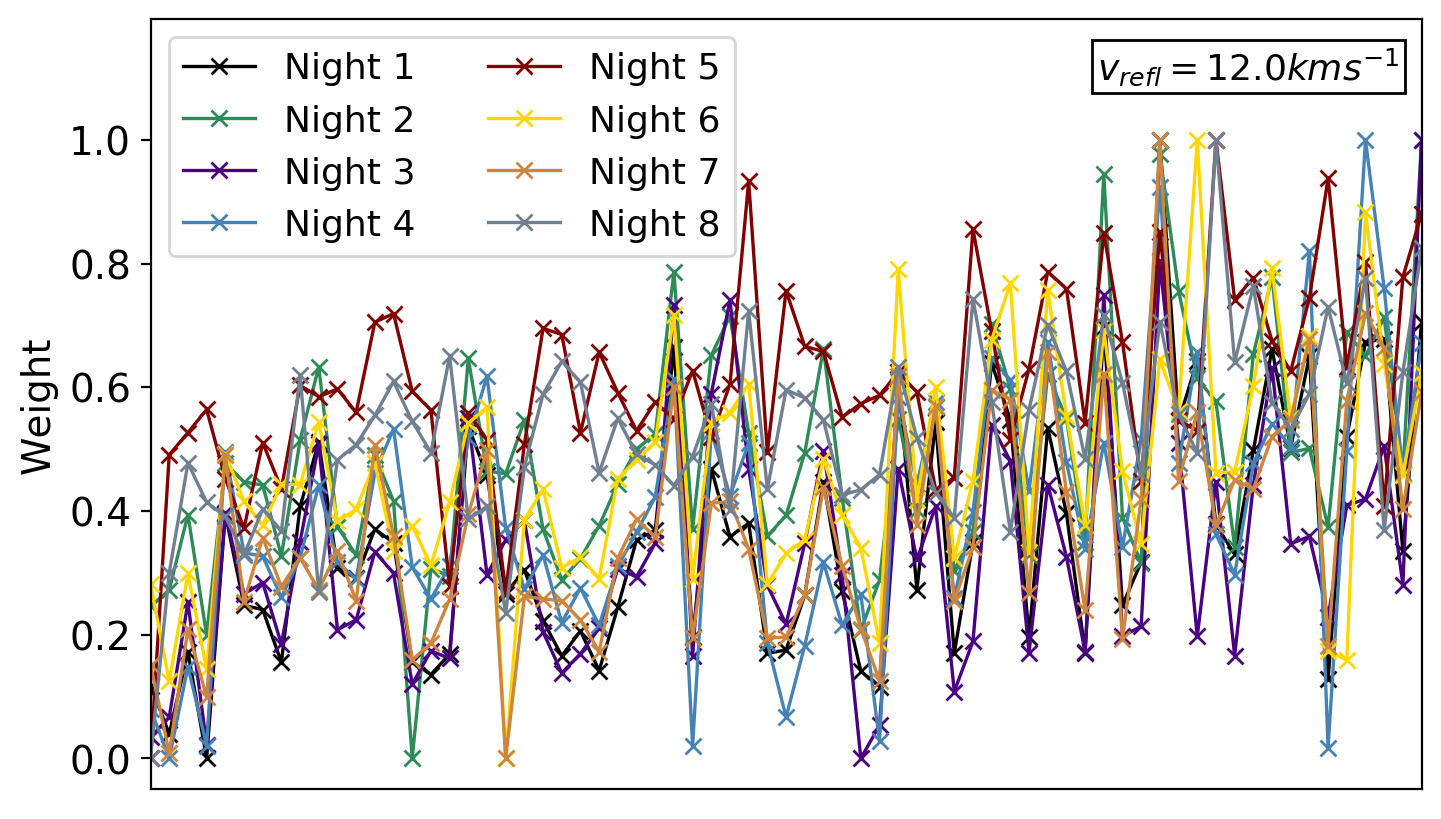}%
        }\vspace{-1.5mm}
         \subfloat{%
          \includegraphics[width=9cm]{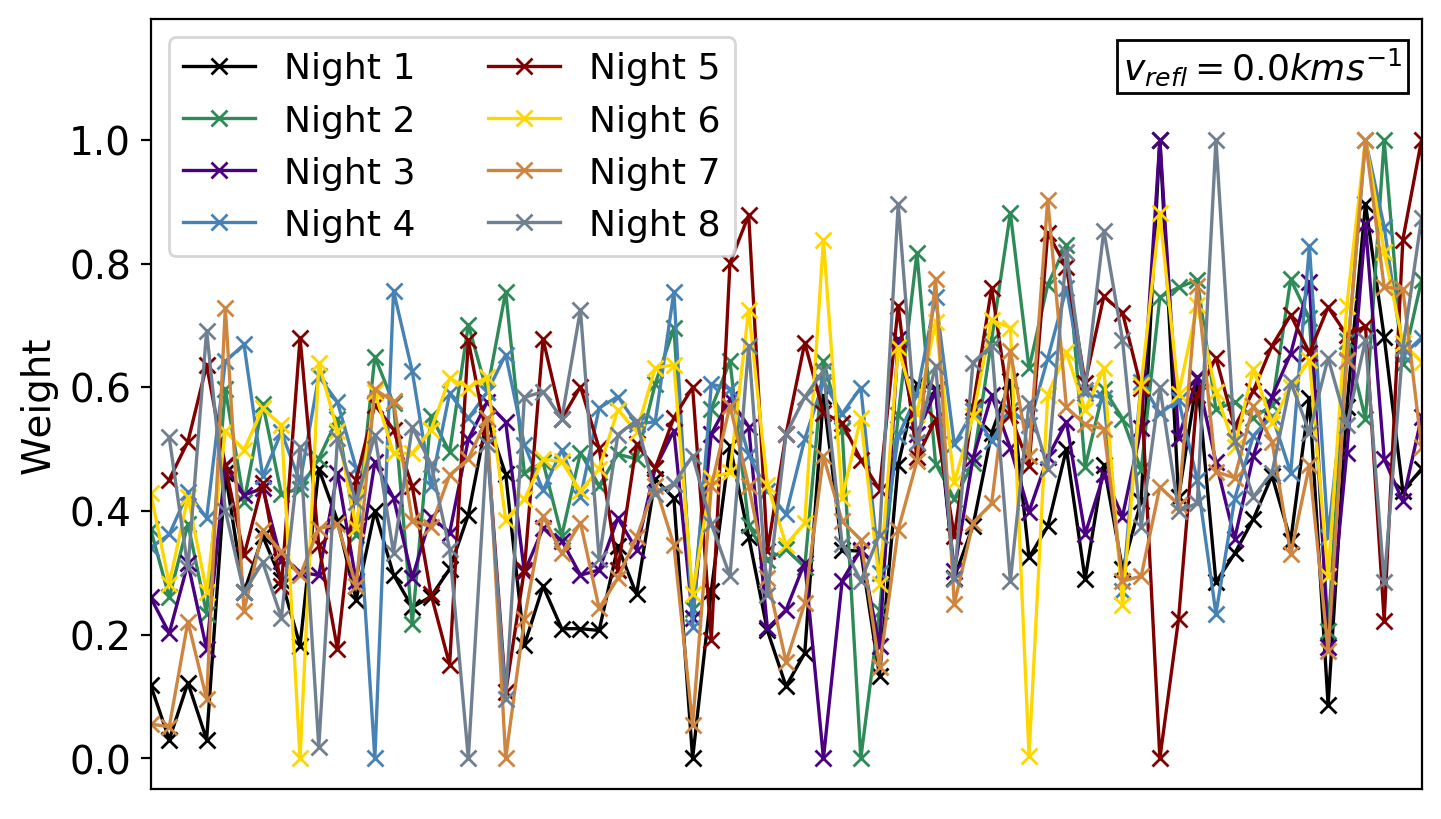}%
        }\vspace{-1.5mm}
        \subfloat{%
          \includegraphics[width=9.05cm]{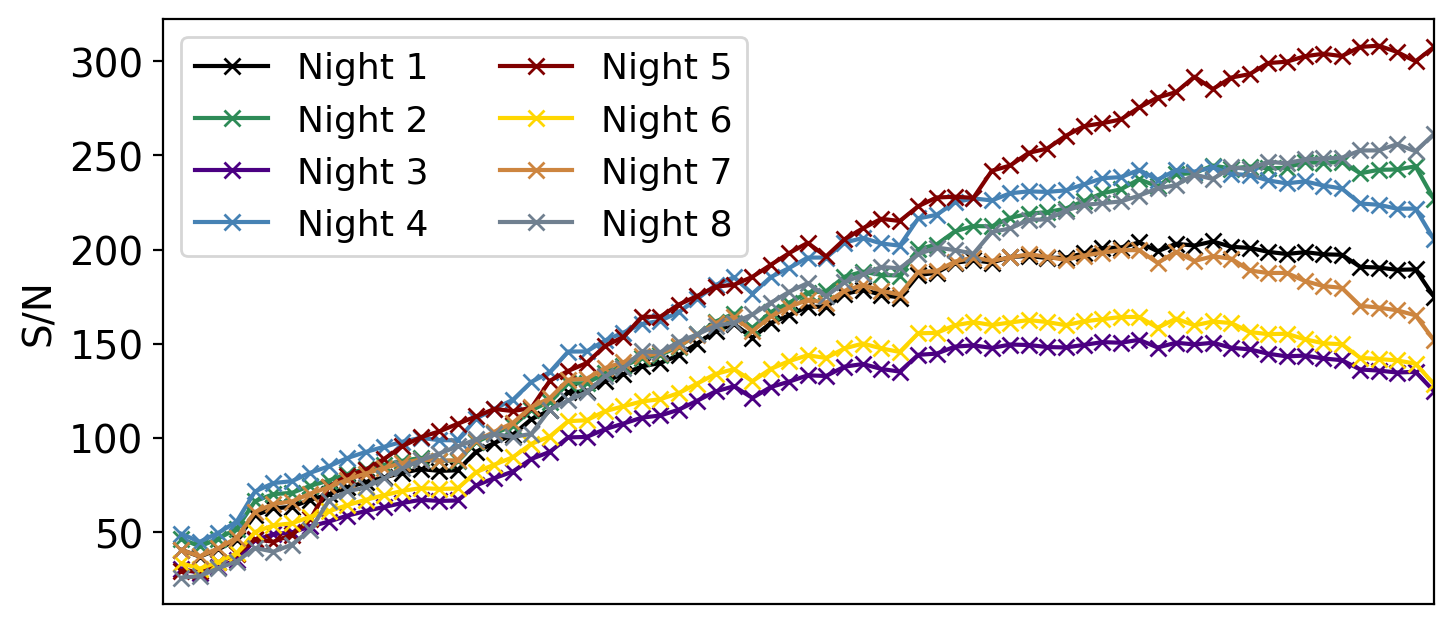}%
        }\vspace{-1.2mm}
        \subfloat{%
          \includegraphics[width=9cm]{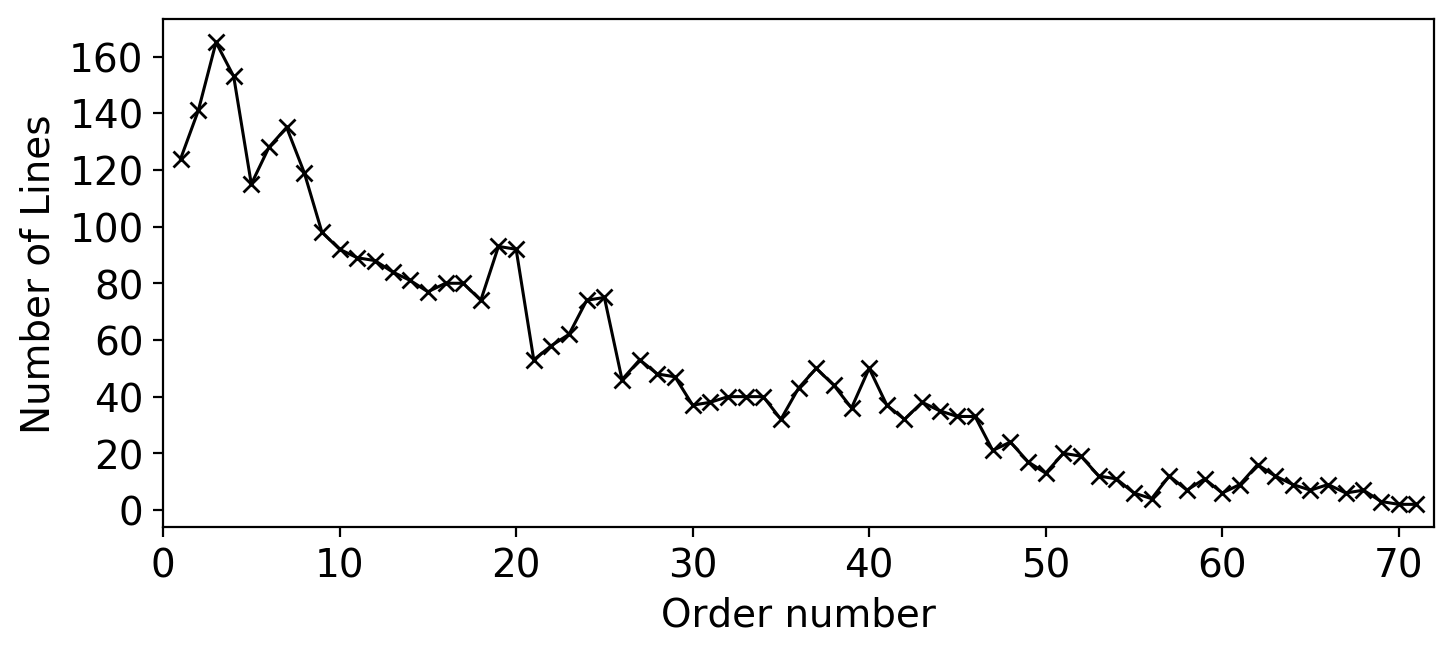}%
        }
   \caption{\textbf{Top two panels: }CCF weights for each night of data, for cases 1 and 2 (top panel) and for cases 3 and 4 (second panel). \textbf{Third panel: }S/N per order, for each night of data. The S/N increases towards the red showing the sensitivity of both HARPS detectors. \textbf{Bottom panel: }number of stellar lines per order, which is the same for each night of data.}
    \label{fig:weights}
    \end{figure}
     %End Figure  

    %Begin Figure
    \begin{figure*}
    \centering
    {%
          \includegraphics[width=18cm]{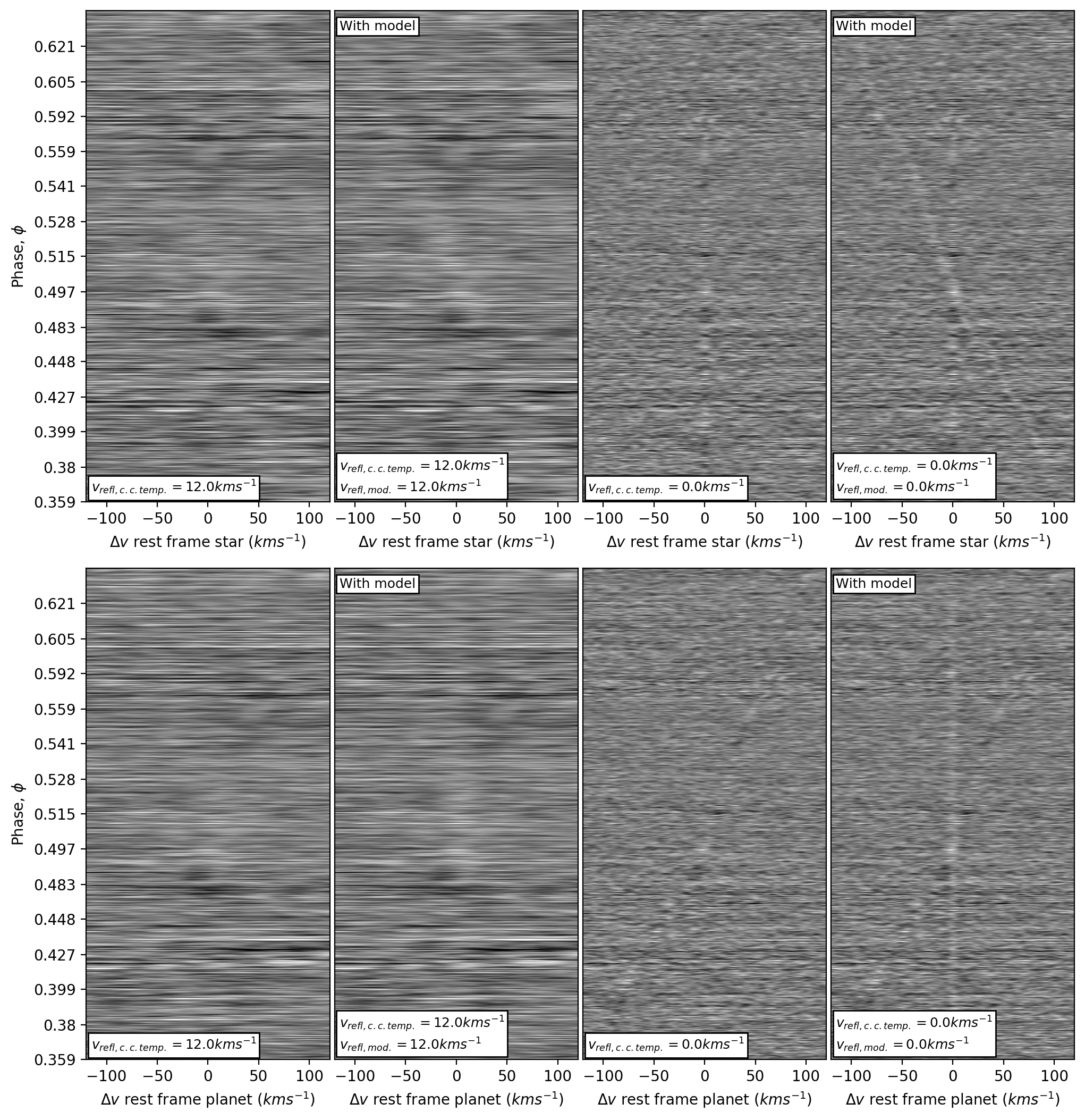}%
        } 
       \caption{All-nights CCF matrices for cases 1, 2, 3 and 4, with the CCFs from all eight nights ordered by phase. The top panels are in the rest frame of the star and the bottom panels are in the rest frame of the planet. The cases where the cross-correlation template is rotationally broadened (cases 1 and 2, $v_{refl,\, c.c.\,temp} = 12.0 \,\rm km\,s^{-1}$, left two columns) and is not broadened (cases 3 and 4, $v_{refl,\, c.c.\,temp} = 0.0 \,\rm km\,s^{-1}$, right two columns) are shown. Cases 2 and 4, where a model ($mod$) following that proposed by M15 ($R_p=1.9, A_g=0.5$) has been injected are indicated by `With model'. For case 2, the injected model returns a very weak trail, as the CCF peak is spread out over more $\Delta v$ increments. For case 4, the injected model is clearly visible to the eye as a blue-shifting trail in the stellar rest frame and as a vertical stationary trail in the planet rest frame. The phase increments are not evenly spaced due to uneven phase coverage (see Figure~\ref{fig:phase}). The faint vertical trails at $\Delta v = 0.0\,km\,s^{-1}$ in the top panels is due to residual stellar lines. The stellar residuals are much more pronounced for cases 3 and 4, because the stellar residuals match with the non-broadened cross-correlation template.}
       
   \label{fig:cc_plot}
    \end{figure*}
   %End Figure

\begin{figure*} 
\centering
        \subfloat{%
          \includegraphics[width=8.5cm]{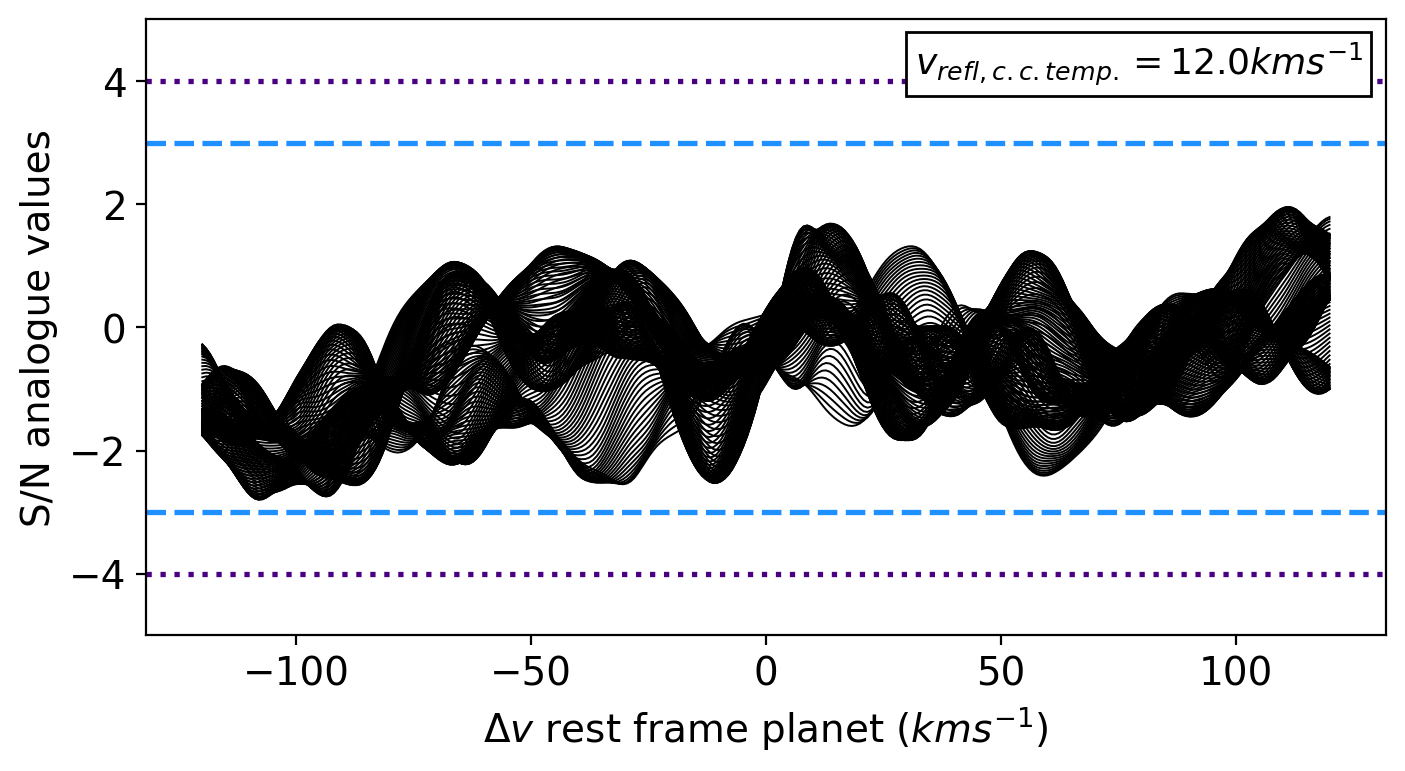}%
        }\hspace{7mm}
     \subfloat{%
          \includegraphics[width=8.5cm]{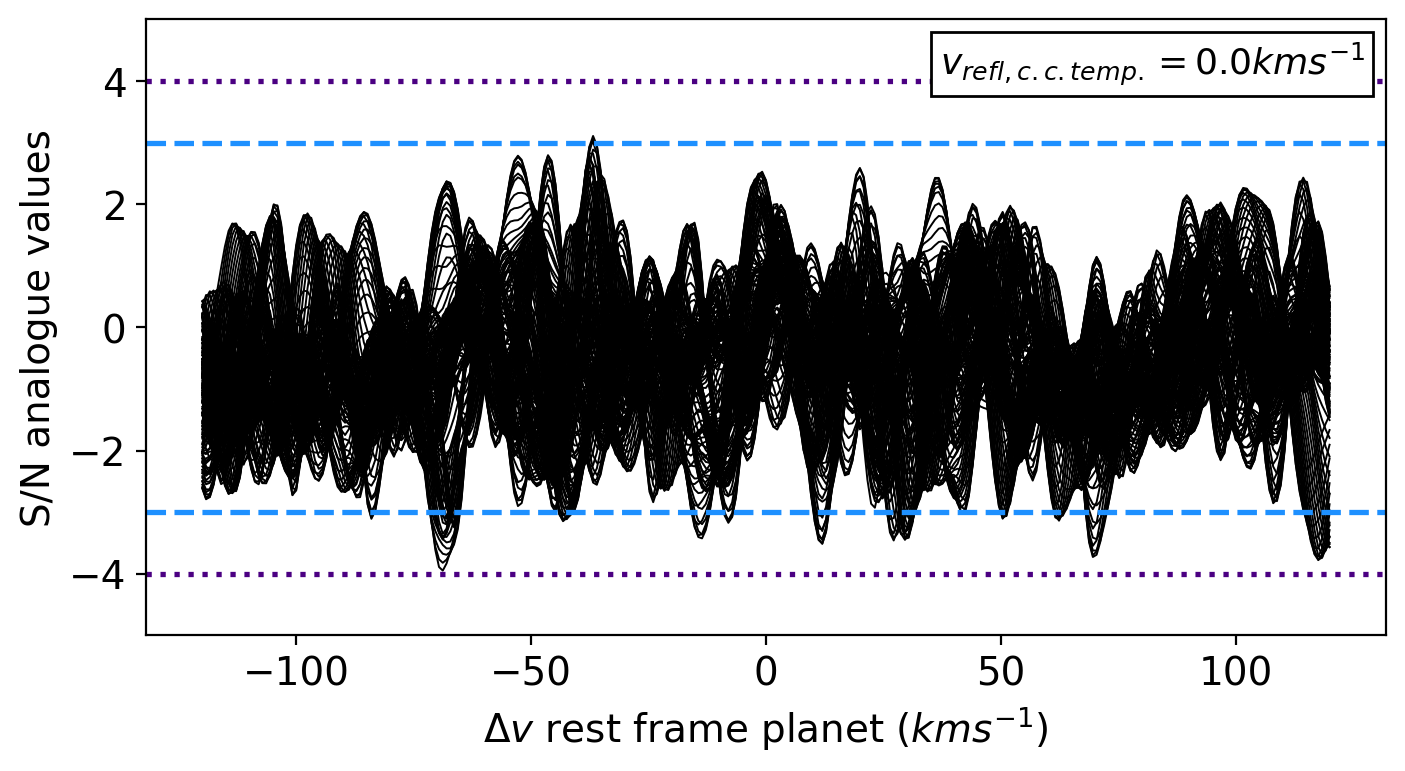}%
        }
\caption{Summed 1D CCFs of the observed data in terms of S/N, where each lines corresponds to a different $K_p$ in $40 \le K_p \le 200$ km\,s$^{-1}$, for cases 1 (left) and 3 (right), with cross-correlation templates with and without broadening ($v_{refl,\,c.c.\,temp} = 12.0$ km\,s$^{-1}$ and $0.0$ km\,s$^{-1}$ respectively). The dashed light blue line and the dotted purple line mark the S/N $=\pm3$ and S/N $=\pm4$ thresholds. The cross-correlation template without rotational broadening results in stronger noise spikes due to matching with narrower lines in the residual host star spectrum.}

\label{fig:sigma_noise_plot}
\end{figure*}

\subsection{Cross-correlation} \label{cc_method}
With the host star spectrum removed as optimally as possible, we can now proceed to extract the faint planet spectrum by cross-correlating with a model template. In each frame, the stellar light reflected by 51 Pegasi b is Doppler-shifted according to Equation~\ref{eq_RVp}. Although we expect 51 Pegasi b to appear at $K_p = 133 \,\rm km\,s^{-1}$ and $\Delta v = 0$ in the planet rest frame \citep{Brogi2013, Birkby2017}, we explored a range of $\Delta v \pm \, 180$ km\,s$^{-1}$ in steps of the HARPS-N velocity resolution per pixel ($v_{step}=0.8$ km\,s$^{-1}$), and $40\le K_p$ (km\,s$^{-1})\le200$ in increments of $1$ km\,s$^{-1}$. This was to explore the noise properties of the surrounding velocity space and to search for velocity offsets caused by astrophysical effects such as bright spots. For our cross-correlation template, we used the same synthetic spectrum of 51 Pegasi, $F_{mod,\,norm}$, that we used in the model injection as described in Section~\ref{model_inject}.
 
Before each cross-correlation, the wavelength grid of the model was offset to the required Doppler-shift in wavelength space by $\Delta v$ using $\lambda= \lambda^\prime (1+(\Delta v/c))$, then interpolated onto the wavelength grid of the data. We then cross-correlated the residuals with $F_{mod,\,norm}(\lambda)$ by calculating the Pearson correlation coefficient, $\rho$, as:

\begin{equation}
    \rho_{X,Y} = \frac{C_{X,Y}}{\sqrt{C_{X,X} \cdot C_{Y,Y}}}
\end{equation}

\noindent where $\rho_{X,Y}$ is the Pearson correlation coefficient between two matrices $X$ and $Y$, and $C$ is the covariance\footnote{$\rho_{X,Y}=$1 indicates perfect correlation, 0 no correlation, and -1 perfect anti-correlation.}. In our case, $X$ was the residuals spectrum and $Y$ was the cross-correlation template $F_{mod,\,norm}(\lambda)$. This created a cross-correlation function (CCF) for each frame of each individual order, which we arranged as a CCF matrix for each order with one CCF per row and one $v_{step}$ per column. 

\subsubsection{Matching the cross-correlation template to the planet's rotational broadening}
 Cross-correlation is insensitive to absolute values hence no scaling was applied to the continuum normalised synthetic spectrum when making the cross-correlation templates. However, it is sensitive to the shape of the line, and thus to the rotational broadening. As we were searching for a planet signal that was considerably rotationally broadened, we needed to establish if our cross-correlation template also needed to be broadened to achieve maximum S/N. In precision radial velocity surveys, a non-broadened template (or even a binary mask) is used as this results in a CCF with a very precisely known peak position, but as we are interested in the S/N of the peak of the CCF, we may wish to optimise CCF peak strength instead. To test this, using Night 1 of our data, we calculated the CCFs for the four different scenarios arising when injecting a model with and without broadening, and cross-correlating using a template with or without broadening. The results are shown in Figure~\ref{fig:cc_extract}, and indicate that like-for-like scenarios result in the best cross correlation, namely when the cross-correlation template matches the broadening of the injected model. Therefore, we match our cross-correlation templates to the expected broadening of the planet spectrum in question. For 51 Pegasi b, we calculated this in Section~\ref{intro_rot_broad} to be $v_{refl} = 12.0$ km\,s$^{-1}$.

\subsubsection{Weighting the cross-correlation functions}\label{ccf_weights}
Once we had a CCF matrix for each order per night, we proceeded to weight and sum the orders from each night to maximise the strength of the planet signal. Each order spans a unique number of reflected stellar spectral lines, and the HARPS-N detector is not uniformly sensitive in wavelength. Consequently, certain orders contributed more to the planet signal than others. To account for this, we weighted every order of each night between 0 and 1. To determine the weights, we follow a similar approach to \citet{Hoeijmakers2018} and calculated a $\Delta$ CCF for each order on each night. $\Delta$CCF = $\Delta$CCF$_{mod,\,inj} - \Delta$CCF$_{obs}$ is the difference between the CCF obtained after injecting a model into every spectrum\footnote{The model had HARPS-N resolution with $R_{p} = 1.9R_J$, $A_g = 0.5$ (M15)}., and the CCF obtained from the observed spectra. The S/N of the peak of the $\Delta$ CCF demonstrates the effectiveness of each order on each night in recovering the injected planet spectrum (see Figure~\ref{fig:signal_only}). The weight ($w$) for each $i$-th order was then calculated as:

    \begin{equation}
       w_i = \frac{\rm S/N_\textit{i} - \rm S/N_{min}}{\rm S/N_{max} - \rm S/N_{min}} 
    \end{equation}

\noindent where $\rm S/N_{min}$ is the minimum peak $\rm S/N$ obtained in an order on that night, and similarly for $\rm S/N_{max}$. Each  $i$-th order was multiplied by its weight $w_i$. There was a close similarity between the weights across all eight nights (see Figure~\ref{fig:weights}). The weights are primarily driven by the S/N of each order, but countered at the bluer ends by the number of stellar lines per order.

Once weighted, the orders were summed to create a single CCF matrix per night. The CCFs in these nightly matrices were then collected and ordered by phase to create the final all-nights CCF matrix shown in Figure~\ref{fig:cc_plot}. The phase increments are not evenly spaced due to gaps in the phase coverage (see Figure~\ref{fig:phase}).

\subsubsection{Stellar residual features in the CCFs} \label{cc_stel_cont}

For cases 3 and 4, when the cross-correlation template is not rotationally broadened ($v_{refl,\,c.c.\,temp.} = 0.0$ km\,s$^{-1}$), there is a noticeable vertical feature at $\Delta v = 0 $ km\,s$^{-1}$ in the stellar rest frame (for example, see the top right panels of Figure~\ref{fig:cc_plot}).

Because these CCFs were created without rotationally broadening the cross-correlation template, the template will match well with the narrower lines of any residual host star spectrum. Thus, this vertical feature is most likely a stellar residual feature. This stellar residual feature will add noise to the Doppler-shifting trail of the planet, which overlaps with it when $RV_p\sim RV_\star$ at $\phi\sim~0.5$. This could potentially provide a false boost to a recovered planet signal. However, for cases 1 and 2, where the cross-correlation template is rotationally broadened ($v_{refl,\,c.c.\,temp.} = 12.0 \,\rm km\,s^{-1}$), the vertical feature is substantially reduced. We postulate that this robustness against the stellar residuals largely stems from the poorer match between the broadened lines of the planet's cross-correlation template and the narrower lines of the host star spectrum (see Figure~\ref{fig:cc_extract}). 

We note that, in the special case of $\tau$ Bo\"o b and other fully synchronised systems where the planet spectrum is not broadened by $v_{\star,p}$ (see Equation~\ref{eq_srefp}), it would be pertinent to mask these stellar residual features in the CCFs (as was done by \citealt{Hoeijmakers2018}). However, in the case of 51 Pegasi b and most other hot Jupiters where only the planet is tidally-locked, we do not need to mask the CCFs.

\section{Results} \label{results}

    %Begin Figure
   \begin{figure*}
    \centering
          \includegraphics[width=18.4cm]{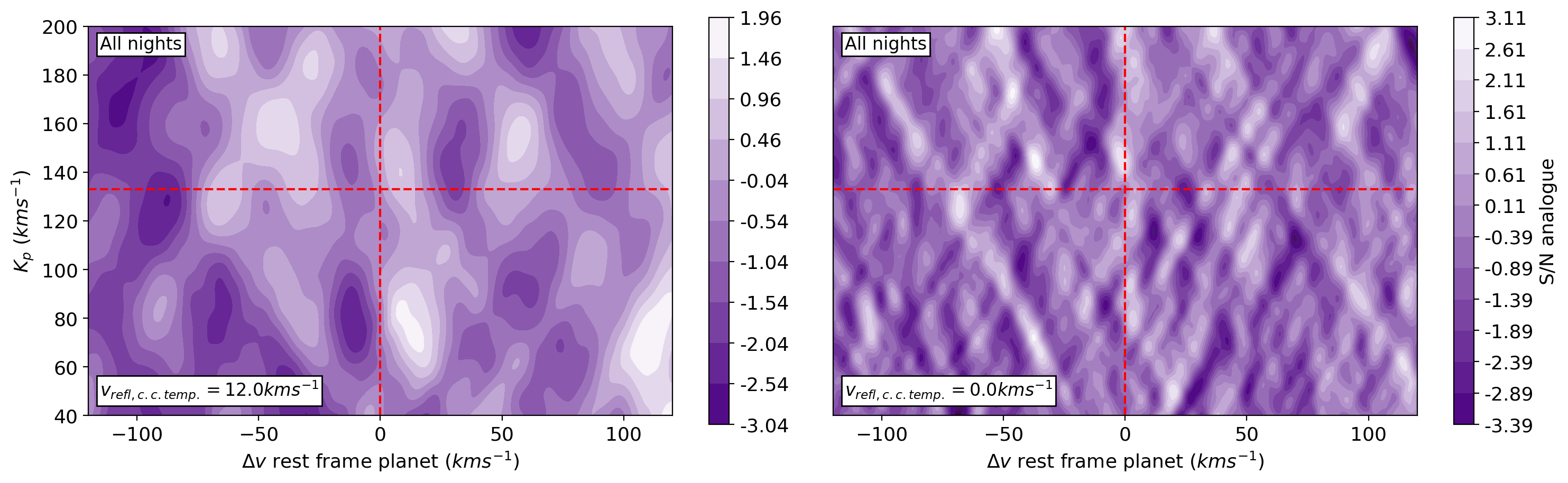}%
    
   \caption{$K_p$ - $\Delta v$ maps showing the S/N of the CCFs for a range of $K_{p}$, for all the nights combined. The red dashed lines mark the expected location of any signal from 51 Pegasi b. The plots shows case 1 (left) and case 3 (right), with and without a rotationally broadened cross-correlation template ($v_{refl,\,c.c.\,temp} = 12.0 \,\rm km\,s^{-1}$ and $0.0 \,\rm km\,s^{-1}$ respectively). No planet signal is recovered, with $\rm S/N = -0.16$ and $\rm S/N = -0.30$ at $K_{p}=133 \rm km\,s^{-1}$ and $\Delta v=0 \rm km\,s^{-1}$ for the left and right plots respectively. The colour bars are calibrated to peak at the highest value in each $K_p$ - $\Delta v$ map, and descend in increments of 0.5.}
    \label{fig:ripple_ccsig_nomod}      
    \end{figure*}
     %End Figure   

%Begin Figure
   \begin{figure*}
    \centering
          \includegraphics[width=18.4cm]{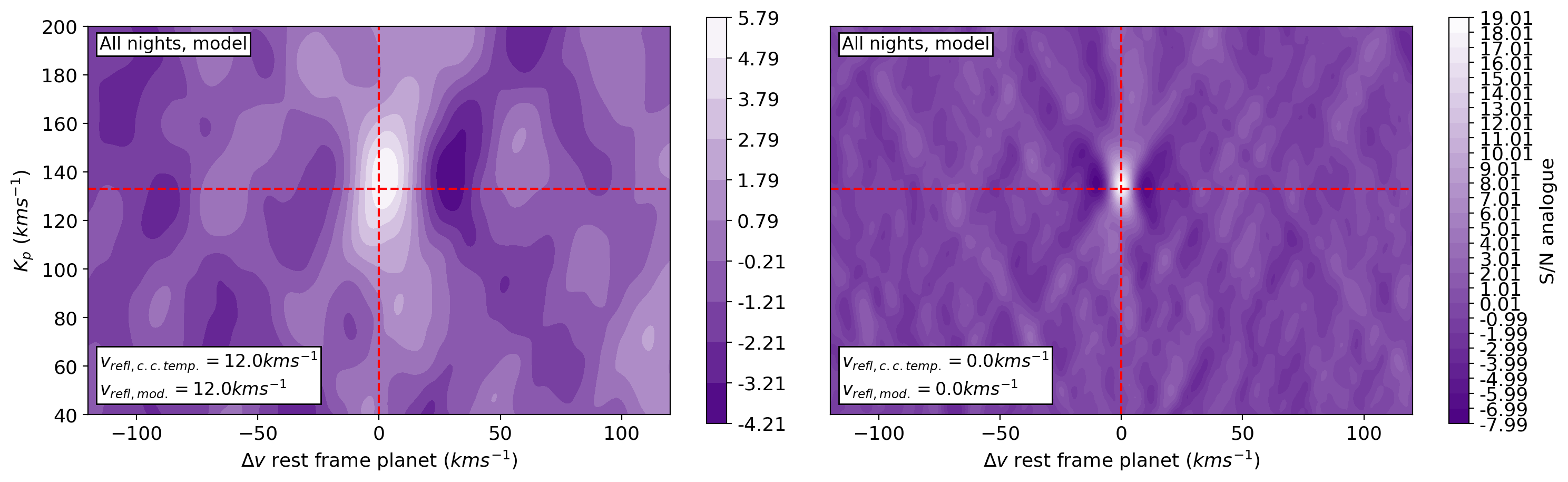}%
          
   \caption{As for Figure~\ref{fig:ripple_ccsig_nomod}, $K_p$ - $\Delta v$ maps showing the S/N of the CCFs for all the nights combined, but with a model injected with parameters $A_g = 0.5, R_p = 1.9R_J$ (M15). The plots show case 2 (left) and case 4 (right), with and without rotational broadening ($v_{refl,\,mod} = 12.0 \,\rm km\,s^{-1}$ and $0.0 \,\rm km\,s^{-1}$ respectively). The cross-correlation templates match the broadening of the model in each case. The broadened model is weakly recovered with $\rm S/N = 5.50$ at $K_{p}=133 \rm km\,s^{-1}$ and $\Delta v=0 \rm km\,s^{-1}$. Without broadening the model is recovered very strongly with $\rm S/N = 18.82$ at $K_{p}=133 \rm km\,s^{-1}$ and $\Delta v=0 \rm km\,s^{-1}$. The colour bars are calibrated to peak at the highest value in each $K_p$ - $\Delta v$ map, and descend in increments of 1.0.}
    \label{fig:ripple_ccsig_mod}      
    \end{figure*}
%End Figure   

\subsection{$K_p$ - $\Delta v$ S/N analogue maps} \label{SNA}

To maximise the recovery of the planet signal, we interpolated the CCF matrices into the planet's rest frame, which aligns any planet signal as a vertical trail at $\Delta v = 0 \,\rm km\,s^{-1}$ (the vertical trail due to the model injection is visible in the bottom rightmost panel of Figure~\ref{fig:cc_plot}). We then summed the CCF matrices into a single 1D CCF. However, as discussed in Section~\ref{cc_method}, we assume that we do not know the $K_{p}$ \emph{a priori} that would align the planet signal, and we also want to explore any offset in the planet signal from $\Delta v=0$ km\,s$^{-1}$. Thus, prior to summing, we align the CCF matrices for a range of $K_p$ values, $40 < K_p < 200 \,\rm km\,s^{-1}$, and then create a summed 1D CCF for each $K_{p}$. As a final step, we calculate the S/N of each data point in each summed 1D CCF. To do this, we first calculated the standard deviation in each summed 1D CCF, (excluding data $\Delta v = \pm 40 \,\rm km\,s^{-1}$ to avoid contamination from any planet signal, then divided the entire summed 1D CCF by this value to get the S/N (see Figure~\ref{fig:sigma_noise_plot}). This metric is sometimes referred to in literature as the cross-correlation significance ($\sigma$), the S/N, or the S/N ratio (sometimes abbreviated as SNR) \citep[see e.g.][where occasionally the standard deviation is calculated instead using all summed 1D CCFs]{Snellen2010, Birkby2013,Yan2020, Kesseli2020}. We will refer to our approach as the S/N analogue, or S/N for brevity. The $K_p$ - $\Delta v$ map format serves to highlight the tight contours that form when $K_{p}$ and $\Delta v$ approach the values that where a planet signal is present. \\
\\

\begin{figure*}
\centering
        \subfloat{%
          \includegraphics[width=8.8cm]{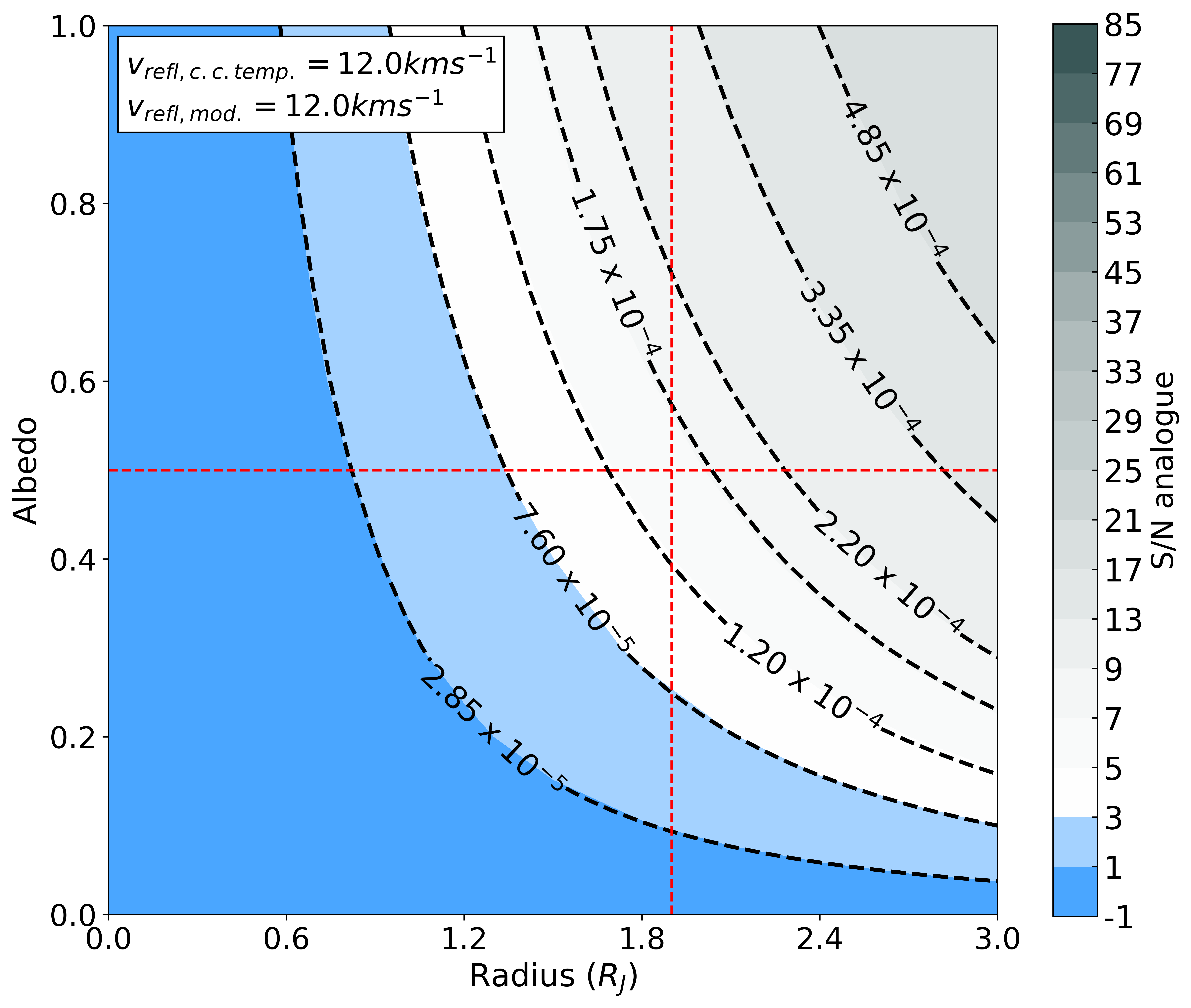}%
        }\hspace{4mm}
        \subfloat{%
          \includegraphics[width=8.8cm]{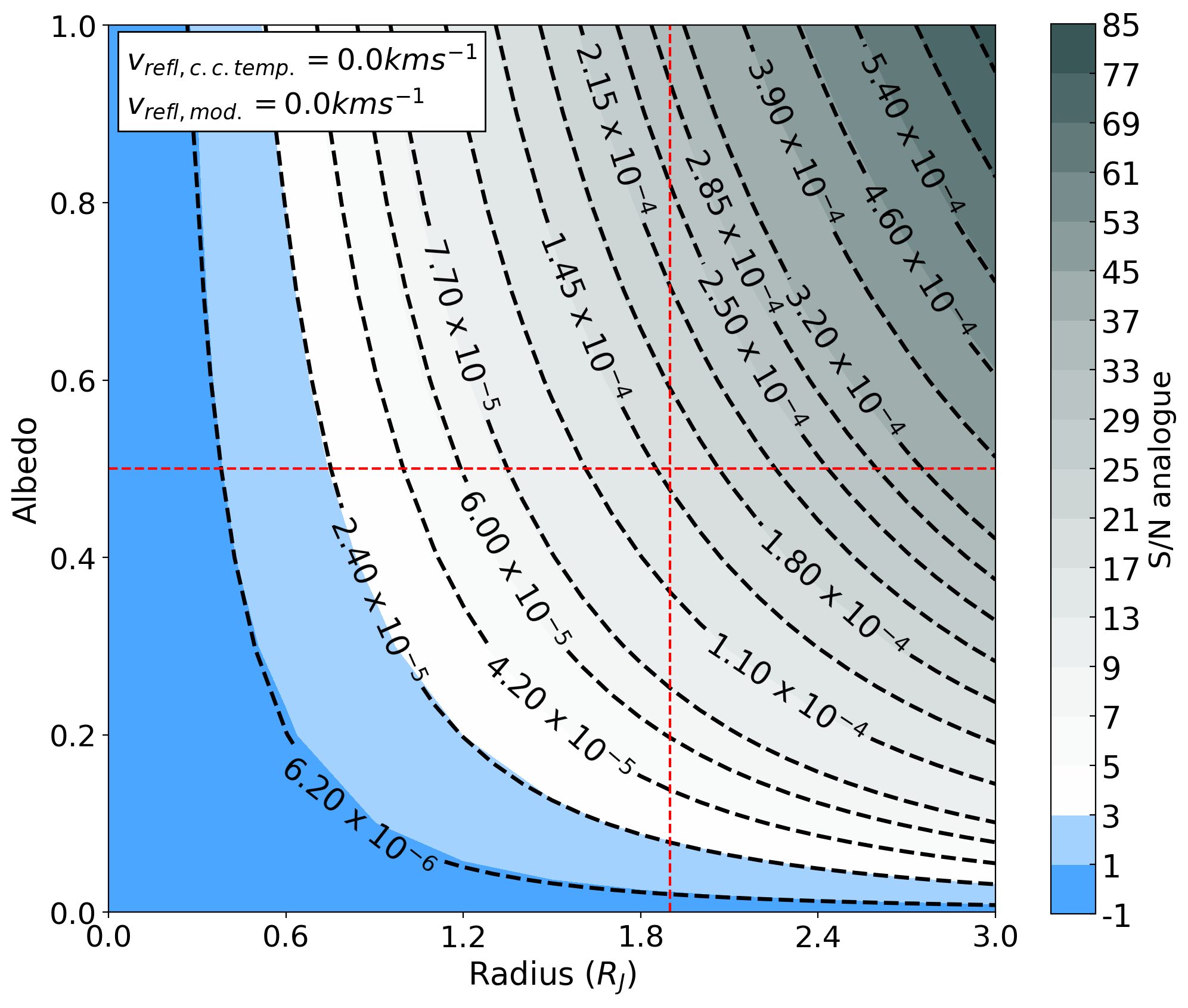}%
        }
        
\caption{S/N recovered by our pipeline for a suite of injected reflected light models at $K_p = 133$ km\,s$^{-1}$ and $\Delta v = 0$ km\,s$^{-1}$, as a function of  geometric albedo ($A_g$) and planet radius ($R_p$), for models with (left) and without (right) rotational broadening. The white/grey contours are above the threshold for detection, while blue regions cover statistically viable scenarios for 51 Pegasi b, given the upper limits on the $F_p/F_\star$ contrast that we derive in Section~\ref{res_upp_lim}. The black dashed contours mark out the $F_p/F_\star$ contrast values. The red dashed lines mark the $A_g = 0.5$ and $R_p = 1.9R_J$ parameters proposed by  M15 for 51 Pegasi b. The blue to grey colour scale is not evenly spaced, but increases in increments of two up to S/N $\le$ ten, in increments of four when ten $<$ S/N $\le$ 40 and in increments of eight when S/N $>$ 40. The colourbars begin at -1 because the S/N value where no model is injected is less than zero for the cases with and without broadening. The colour scale is the same for both plots.}

\label{fig:contour_plots}
\end{figure*} 

\subsection{Non-detection of 51 Pegasi b}\label{non_detect}
As described at the beginning of Section~\ref{model_inject}, we analysed our observed data for cases 1 and 3, where no model was injected. This allowed us to determine whether we had detected reflected light from 51 Pegasi b. Figure~\ref{fig:sigma_noise_plot} shows the distribution of S/N values in the observed summed 1D CCFs, which gives deviations of S/N$\sim\pm3$. Similar structure is seen by \citealt{Hoeijmakers2018, Cabot19} and elsewhere in the literature.  Thus, we adopt S/N$=3$ as our threshold for detection.

We ordered the 1D summed CCFs by $K_p$ to create a 2D $K_p$ - $\Delta v$ map in terms of S/N, as described in Section~\ref{SNA}. For case 1 (with rotational broadening), at 51 Pegasi b's expected $K_{p}=133 \rm km\,s^{-1}$ and $\Delta v=0 \rm km\,s^{-1}$ the S/N $= -0.16$ in the all-nights matrix. For case 3 (without rotational broadening), the S/N $= -0.30$ at $K_{p}=133 \rm km\,s^{-1}$ and $\Delta v=0 \rm km\,s^{-1}$. There is also no value $\rm S/N>3$ anywhere in the region of $K_{p}=133 \rm km\,s^{-1}$ and $\Delta v=0 \rm km\,s^{-1}$, for the case with or without the broadening of the cross-correlation template. Consequently, we do not claim a detection of 51 Pegasi b in reflected light. The $K_p$ - $\Delta v$ maps for the results from all the nights combined are displayed in Figure~\ref{fig:ripple_ccsig_nomod}. The $K_p$ - $\Delta v$ maps for the individual nights for the broadened case can be seen in Appendix~\ref{appen_additional_plots} in Figure~\ref{fig:ripple_appen}.

\subsection{Impact of rotational broadening} \label{impact_rot}

In order both to demonstrate the impact of rotational broadening on our ability to recover the reflected light spectrum of the planet, and to determine the maximum sensitivity of our data, we injected two different models into our dataset (case 2 and case 4, see Section~\ref{model_inject}). Both models were injected with parameters $A_g = 0.5, R_p = 1.9R_J$ at $K_p=133 \rm km\,s^{-1}$ and $\Delta v=0 \rm km\,s^{-1}$ to simulate the scenario proposed by M15. The $K_p$ - $\Delta v$ maps for the results from all the nights combined are displayed in Figure~\ref{fig:ripple_ccsig_mod}.

For case 2, the physically realistic case with rotational broadening, the model is recovered with the $\rm S/N = 5.50$ at the injected $K_{p}=133 \rm km\,s^{-1}$ and $\Delta v=0 \rm km\,s^{-1}$ location, while for case 4, without rotational broadening, the model is recovered with very high S/N $= 18.82$. The negative values of the noise distribution reach approximately -4 and -8 for the two cases, respectively. This is mainly due to the negative wings or `shadows', either side of the detections. These shadows appear due to the column-wise nature of the removal of the stellar and telluric contamination, which can subtract the wings of the planet spectral lines as they cross each column, resulting in anti-correlation with the template at this final stage.

The discrepancy between our recovery of the models with and without broadening make clear that rotational broadening has a substantial impact on the ability of HRCCS to recover the planet's reflected light spectrum. Our recovery of the non-broadened model at high S/N is consistent with S21, as discussed further in Section~\ref{discuss_interp}.

\subsection{Upper limit on the radius and albedo of 51 Pegasi b} \label{res_upp_lim}

As there was no statistically significant recovery of light reflected from 51 Pegasi b in the observed spectra, we proceeded to calculate upper limits on its radius and albedo. We considered both the physical scenario with rotational broadening (cases 1 and 2), and the scenario without rotational broadening (cases 3 and 4), with the latter being a test of the maximum sensitivity of our data and pipeline. We did this by injecting a series of models for $0.0 \le A_g \le 1.0$ (increment of 0.1 between each model injection) and $0.0 \le R_p \le 3.0 R_J$ (increment of 0.3 between each model injection) values using Equation~\ref{eq_inj}. As for cases 2 and 4 the model was injected at 51 Pegasi b's expected $K_p = 133\,\rm km\,s^{-1}$ and $\Delta v = 0.0\,\rm km\,s^{-1}$, and we calculated the S/N of the recovery. The full results of our model injections are shown in Figure~\ref{fig:contour_plots} with (left panel) and without (right panel) rotational broadening, where the S/N value is taken at $K_p = 133\,\rm km\,s^{-1}$ and $\Delta v =0\rm km\,s^{-1}$ . 

As shown in Section~\ref{non_detect}, our threshold for detection is $\rm S/N>3$. We adopt this S/N $=3$ upper limit cut\footnote{The literature sometimes calls this cut a $3\sigma$ upper limit see e.g. \citet{Hoeijmakers2018}, which would be the case in pure Gaussian noise.} as the threshold for when our pipeline can sufficiently recover a model and thus reject its parameters as a possible physical scenario for 51 Pegasi b. Any model recovered with $\rm S/N <3$ was kept as a statistically plausible model for 51 Pegasi b.

For the physical scenario with rotational broadening (case 1), we find a $\rm S/N = 3$ upper limit on the contrast ratio of $\frac{F_p}{F_\star}<7.60\times10^{-5}$, resulting in relatively loose constraints $A_g$ and $R_p$. For example, a typical hot Jupiter with the same mass of 51 Pegasi b\footnote{$1.2R_J$ is the approximate mean radius value for short-orbit transiting exoplanets with similar masses to 51 Pegasi b in the NASA Exoplanet Archive.} has $R_p\sim1.2R_J$, which corresponds to $A_g < 0.62$. 

For the scenario without rotational broadening (case 3), which represents the limit of the sensitivity of our data and pipeline for a planet orbiting 51 Pegasi b in a fully synchronised orbit at $a=0.052$ au, we find a $\rm S/N = 3$ upper limit of $\frac{F_p}{F_\star}<2.40\times10^{-5}$, corresponding to a much more tightly constrained $A_g < 0.20$ for $R_p\sim1.2R_J$. This is comparable to the upper limit on the contrast ratio found by \citet{Hoeijmakers2018} with archival data for the fully synchronised $\tau$ Bo\"o b hot Jupiter ($\frac{F_p}{F_\star}<1.5\times10^{-5}$), and the upper limit of S21 for 51 Pegasi b ($\frac{F_p}{F_\star}\sim10^{-5}$). If 51 Pegasi b had similar albedo to HD 189733b (i.e. $A_g=0.40$ \citep{Evans2013}), our results would give constraints of $R_p < {1.50\,R_J}$ with rotational broadening, and $R_p < {0.84\,R_J}$ without rotational broadening. 

It is evident that our analysis very strongly rules out the radius and albedo parameters suggested by M15 for 51 Pegasi b ($A_g=0.5$,$R_p=1.9R_J$). For the case they consider (our case 4), we recover the corresponding injected model at $\rm S/N = 18.82$, so it is clearly recovered. Even adding rotational broadening (our case 2), we still recover the model at  $\rm S/N = 5.50$, while the observed data alone reveal no signals greater than $\rm S/N = 3$.

\subsection{Confirming the source of the broadening}

%Begin Figure 
    \begin{figure}
    \centering
        {
          \includegraphics[width=9cm]{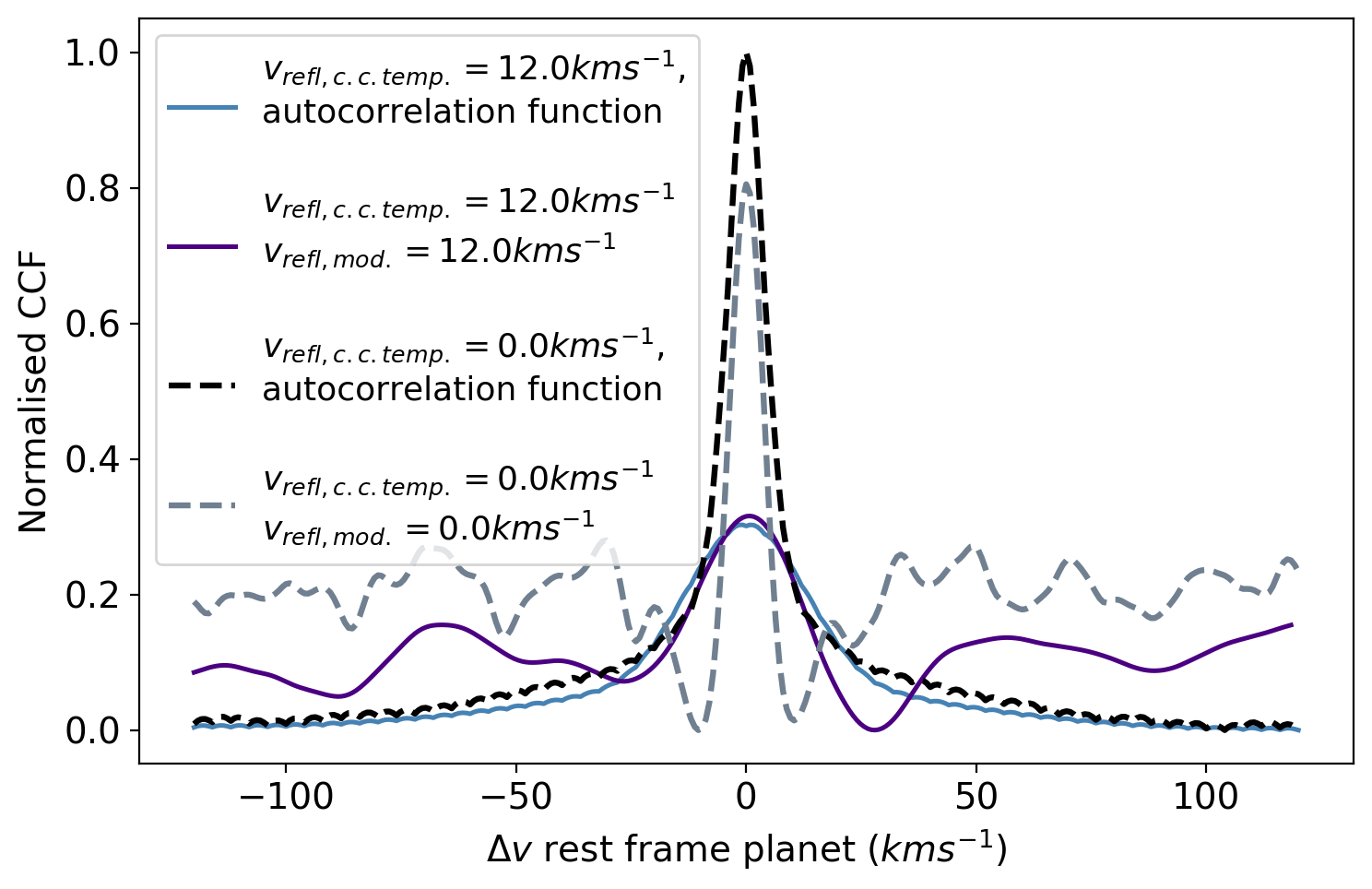}%
        } 
   \caption{Comparison of the autocorrelation function of the model, with (solid) and without (dashed) broadening, to the CCFs obtained after injecting models into the observed data (purple and grey lines), to assess if post-processing affects the FWHM of the signal. All are normalised with respect to the black dashed line for visual purposes. In the autocorrelation functions, one spectrum has the residuals mask from Section~\ref{stel_remove} applied beforehand to better match the steps in the analysis that exclude the more saturated stellar features. The FWHM values of the autocorrelation functions are 34.4 km\,s$^{-1}$ and 10.4 km\,s$^{-1}$, respectively with (blue solid line) and without (black dashed line) broadening. For the injected models, the FWHM of the CCFs are 29.6 km\,s$^{-1}$ (purple solid line), and 7.2 km\,s$^{-1}$ (grey dashed line), respectively with and without broadening. The grey and purple lines correspond to their un-normalised counterparts in Figure~\ref{fig:cc_extract}.}

    \label{fig:cc_auto}        
    \end{figure}
%End Figure  

M15 showed with model injections that their processing techniques acted to broaden the CCF of signals in the data. To confirm that broadening in our analysis is purely astrophysical in origin and not introduced erroneously by any of the steps in our pipeline, we compared the full width at half maximum (FWHM) of the autocorrelation function of our normalised synthetic stellar spectrum ($F_{mod,\,norm}$), with the CCFs obtained after injecting a model with and without rotational broadening (cases 2 and 4). The autocorrelation function for the broadened $F_{mod,\,norm}$ ($v_{refl,\,mod} = 12.0\,\rm km\,s^{-1}$) had FWHM = $34.4$ km\,s$^{-1}$, and without broadening $F_{mod,\,norm}$ ($v_{refl,\,mod} = 0.0\,\rm km\,s^{-1}$) had FWHM = $10.4$ km\,s$^{-1}$. These relatively large FWHMs are in part due to broadening of spectral features in the star due to, for example, pressure or natural broadening, and saturation of certain lines in the optical range of 51 Pegasi's spectrum (e.g. the Ca II H and K, Mg I, and H$\alpha$ lines). We compared these FWHM of the autocorrelation functions to the FWHM of the CCFs obtained from the observed spectra, when injecting a model with and without broadening and using matching cross-correlation templates. The results are shown in Figure~\ref{fig:cc_auto}. For both cases 2 and 4, the FWHM of the CCF is narrower than the model autocorrelation function. We propose that this is because some of the wings of the broadest and shallowest lines of the injected model are removed by the post-processing and stellar line removal. For neither case 2 or 4 does our pipeline broaden the recovered model. Therefore, we can be confident that any broadening is due to astrophysical broadening of the planet signal, not due to our pipeline.\\

\noindent We note that our pipeline was able to recover the real signal of HD 189733b at the same FWHM as found by~\cite{Birkby2013} and ~\cite{Cabot19} (see Appendix~\ref{appen_check_pipeline}). 

\section{Discussion} \label{discuss}

\subsection{Rotational broadening and comparison with previous studies}\label{discuss_interp}
Our results indicate that very low albedo is entirely plausible for 51 Pegasi b, in line with the large body of literature that predicts that hot Jupiters will have extremely low optical albedo, because their high temperatures mean that they lack a reflective cloud deck \citep[see e.g.][]{Rowe2008, Cowan_2011, Heng2013}. However, we cannot confirm that 51 Pegasi b has a low albedo, because rotational broadening due to $v_{\star,p}$ considerably reduces the sensitivity of our data (see Figure~\ref{fig:contour_plots}).

As discussed in Section~\ref{sec:intro_compare}, four previous works sought reflected light from 51 Pegasi b, with conflicting conclusion on the radius and albedo of the planet (M15; \citealp{Borra2018ACF, Marcantonio2019ICA}; S21). While each of these studies acknowledged that the host star rotation period is very long in comparison to orbital period (21.9 days vs 4.23 days respectively), each neglected the effect of the apparent rotational broadening due to the orbital motion of the planet, $v_{\star,p}$ (see Equation~\ref{eq_srefp} and \citealt{Strachan2020}), and instead assumed that the star would not appear broadened from the planet's perspective. In contrast, we have shown here that $v_{\star,p}$ has a significant impact on the rotational broadening of the planet's reflected light spectrum ($v_{refl}=12.0\, \rm km\,s^{-1}$) and the strength with which it can be detected. As a result, Equation~\ref{eq_SN} for the S/N of the detection does not hold true for broadened spectra, requiring modification to account for the reducing effect of broadening on the depth of the spectral lines and hence the number of detectable lines in the spectrum. Only in the special case of a fully synchronised system (i.e. $P_{rot,\star}=P_{orb}$, like for $\tau$ Bo\"o b), can this effect be ignored. An immediate consequence is that the ratio of stellar and planet CCF amplitudes cannot be used directly to calculate contrast and thus $R_p$ and $A_g$ as they have been muted by the broadening. Instead, the CCF would need to be deconvolved to remove the muting effect of the rotational broadening, implying that the parameters proposed by M15 are underestimated, and thus the albedo and radius for 51 Pegasi b would be even brighter or larger, despite their already high values when compared to the radius of a typical hot Jupiter with the same mass and temperature as 51 Pegasi b. 

In the case of a detection, an alternative approach to determining $R_p$ and $A_g$ would be to inject negative versions of the appropriately broadened model into the data until the signals cancelled out (as similarly done with thermal emission detections in the infrared, see e.g. \citealt{Brogi2012, Birkby2017}).

We can attempt a comparison of our work with these previous studies if we temporarily ignore the broadening and muting effects of $v_{\star,p}$, in order to assess the efficacy of the different processing techniques. For this, we consider the results shown in our Figures with $v_{refl} = 0.0\,\rm km\,s^{-1}$ . This is close to, but not exactly the scenarios presented in the other four studies, as our cross-correlation templates are generated in different ways. As discussed in Section~\ref{discuss_interp}, we find that if we inject a model with $v_{refl,\,mod} = 0.0$ km\,s$^{-1}$ with $A_g$ and $R_p$ as proposed by M15 then it is recovered with a $\rm S/N = 18.82$ (see right plot, Figure~\ref{fig:ripple_ccsig_mod}). Our CCF template has matched broadening unlike M15, and thus recovers slightly improved CCF peak values (see Figure~\ref{fig:cc_extract}). However, even if we inject a model with $v_{refl,\,mod} = 12.0$ km\,s$^{-1}$, a weak signal is recovered with $\rm S/N = 5.50$(see left plot, Figure~\ref{fig:ripple_ccsig_mod}). Consequently, we can rule out the high $A_g$ and $R_p$ proposed by M15 for 51 Pegasi b. This corroborates the results of S21, who rule out contrast ratios $\gtrsim 10^{-5}$ for cases without broadening, including the $1.2\times10^{-4}$ contrast ratio suggested by M15. Given that the upper limits reached by S21 and our work are derived from a dataset that contains over five times the amount of spectra (of comparable S/N) used in M15, it is unsurprising they yield a superior sensitivity.

It is interesting then to compare our results solely to S21 given that our datasets are the same, but our model templates and cleaning processes are different, particularly the process of removing the stellar line contamination. Our work operates directly on the spectra, while S21 remove stellar contamination from the CCFs instead. Our closest comparable case to S21 is our case 3 (with no rotational broadening and thus the most sensitive) where we obtained an upper limit on the contrast ratio of $\frac{F_p}{F_\star}<2.40\times10^{-5}$ which is comparable to their $\frac{F_p}{F_\star}\lesssim10^{-5}$. Our recovery of the M15 proposed model yielded $\rm S/N \sim 19$, while S21 recover $\rm S/N \sim 22$ for a similar contrast ratio. These are both very high S/N detections, but we do not assign significance here to their difference. A strict comparison is not possible as the exact details of the injection and recovery scheme are not fully available in S21, hence there may be differences in, for example, the exact S/N metric used, and the noise distribution of the S/N metric. But, given we inject our similar models at the same location within $K_p\pm1$ \rm km\,s$^{-1}$ and $i\pm1^{\circ}$, and we both incorporated the phase function, we conclude broadly that both approaches to removing stellar lines are effective. We do however note some pertinent differences between our spectra-focused approach and their CCF approach to removing stellar lines. S21 do not use 88 frames near superior conjunction to avoid stellar contamination in the CCF, whereas we only need to remove 4 frames for this, allowing more data to be used. We are also able to use more spectral orders than the CCF approach which has to exclude those with broadened features in the host star spectrum. This ability to use extra data may be useful for less abundant datasets, enabling greater phase coverage. We cannot clearly state if our use of synthetic templates is superior to observed templates. While the observed templates in the CCF approach of S21 allows for excellent alignment of the stellar lines, we note that synthetic templates have the advantage of being i) completely free of telluric lines such that they outperform observed templates in this respect, even when great care is taken to correct the tellurics, as done by S21 with \textsc{MOLECFIT} \citep{Smette2015, Kausch2015}, ii) are more straight forward to appropriately broaden without also broadening the planet's albedo function, and iii) are easily combined with models of planet's high resolution albedo function. The latter is particularly important if, for example, abundances are to be derived from high resolution exoplanet reflection spectra. Nonetheless, there remains scope for improvement in both approaches to removing stellar contamination.

Our post-processing and cross-correlation approach differed greatly from that presented in S21, and is most similar to \citet{Hoeijmakers2018} who studied $\tau$ Bo\"o b in reflected light. Our model injections without rotational broadening are most directly comparable to this work, as $\tau$ Bo\"o is a fully synchronised system. Both our studies operate on the spectra directly to remove the host star, including PCA-like systematics removal, and we both use a model from a stellar library to perform the cross-correlation. \citet{Hoeijmakers2018}, however, used a very large repository of archival spectra from multiple spectrographs, obtained relatively sporadically over 15 years, containing a mix of exposure times and phase coverage. While HARPS-N contributes some of their dataset, other spectrographs with lower radial velocity precision and stability are also included. They used a total of 2160 frames with an average S/N of 500 - 1000 per pixel to obtain an upper limit on the contrast of $\frac{F_p}{F_\star} < 1.5\times10^{-5}$. In contrast, the smaller collection of 484 frames used in this work, which were comprised of eight more consistent observing sequences, led to an upper limit on the contrast ratio in case 1 (without rotational broadening) of $\frac{F_p}{F_\star} < 2.40\times10^{-5}$. This is comparable to the result presented in \citet{Hoeijmakers2018}, despite our much smaller dataset. Unlike for the analysis presented in this paper, \citet{Hoeijmakers2018} additionally apply a mask to their CCFs to remove remaining stellar residuals. We also see these residuals in Figure~\ref{fig:cc_plot} (especially in the top right panels), but do not mask them, to avoid interfering with the noise distribution of the $K_p$ - $\Delta v$ maps. A one-to-one comparison is not fully possible due to the different orbital parameters of the system (i.e. $K_p$ is different), thus the full impact of stellar residuals is not the same between our studies. However, we note that due to the different instruments used in \citet{Hoeijmakers2018}, their alignment to the stellar rest frame is based on fitting the positions of common strong stellar features across the dataset, unlike the work presented here where we have relied on the precision radial velocity and stability of HARPS-N to precisely determine the true rest frame of the star and align the data accordingly. It is possible that this instrumental precision and stability has therefore enabled a better alignment and therefore subtraction of the host star spectrum and hence a comparable upper limit, despite the much smaller dataset in our study.

\subsection{Stellar residuals and stellar variability}
The removal of stellar lines, however, is not perfect in either study. One potential reason for this is that neither study accounts for stellar variability. Within the timescale of a single observing night, stellar surface processes including surface granulation, coronal flares and mass ejections could introduce variability~\citep[see e.g.][]{Cegla2013,Meunier2017}. On a longer timescale, stellar spots and faculae could introduce variability between observing nights \citep[see e.g.][]{Meunier2010, Aigrain2012, Dumusque2014, Haywood2014, Haywood2016, Milbourne2019}. Stellar variability is recognised as a serious challenge to the success of Extreme Precision Radial Velocity (EPRV) studies \citep[e.g.][]{Fischer2016, Crass2021}, and has been more recently recognised as a challenge to transmission and thermal day side HRCCS~\citep{vanSluijs2019, Guilluy2020, Nugroho2020}. For reflection HRCCS, small changes in line shape or position due to stellar variability may present as small flux variations at $<10^{-4}$ levels, similar to that expected for reflected light, which are revealed when the time-averaged spectrum is divided out from the aligned spectra, and when nights are combined. Its impact appears greatest for observations taken close to times of superior or inferior conjunction. Future studies could aim to mitigate the effect with modelling or observational strategies.\\

\subsection{Strengths and limits of the S/N analogue} \label{discuss_S/N_analogue}
The S/N analogue used in this work and throughout the literature provides a straightforward and intuitive approach to understanding the extent to which a recovered signal, if any, stands above the noise, and allows us to set a threshold for detection or upper limit. However, the method is based upon the supposition that, for a cross-correlation array that has been co-added in time, a recovered signal will be concentrated at one $\Delta v$ value. This situation is often the case for thermal emission signals in the infrared regime from tidally-locked systems, where any exoplanet or model signal is typically recovered almost entirely at $\Delta v = 0 \,\rm km\,s^{-1}$ \citep[see e.g.][]{Birkby2017}. In this case, the resulting S/N at $\Delta v = 0 \,\rm km\,s^{-1}$ will be representative of the strength of the exoplanet or model signal. However, in the case where a signal is broadened, the S/N analogue rapidly becomes unrepresentative of the total signal because it is spread across a range of $\Delta v$ values (see Figure~\ref{fig:cc_extract}). This impacts the CCFs from spectra of fast rotating exoplanets, for example, $\beta$ Pic b \citep{Snellen2014}, as well as reflected light spectra due to $v_{\star,p}$. We also note that in the case of 51 Pegasi b, its host star spectrum contains near-saturated spectral features that inherently broaden the CCFs too (see Figure~\ref{fig:cc_auto}). This renders the S/N analogue a less suitable metric for determining the significance of a signal when using HRCCS for reflected light and other broadened signals.

An additional issue is that the S/N analogue in principle requires one to know where the planet signal will be so that it can be excluded in the calculation of the standard deviation. This makes it less suitable for a blind search for the planet signal, which is pertinent for non-transiting planets where $K_p$ is not known a priori. On a similar note, the standard deviation depends on the choice of $\Delta v$ range to explore. Finally, in some cases the standard deviation is calculated per $K_p$ value (as was done for this work), but sometimes it is calculated only once across the entire $K_p$ - $\Delta v$ map of values \citep[see e.g.][]{Nugroho2021}. This inconsistency in the literature makes comparison of results difficult. Alternative metrics have been used, for example, Welch's T-test \citep[see e.g.][]{Birkby2017} or log-likelihood \citep[see e.g.][]{BrogiLine2019}, but these come with other complicating issues. In short, the S/N analogue approach has the benefit of simplicity, but it is not ideally suited to reflected light detections, especially where there is significant broadening. Thus, we acknowledge the need to evaluate the precise meaning of S/N, $\sigma$, and other significance metrics in relation to HRCCS, including those that can be reasonably adapted in the case of significant rotational broadening{, which will be investigated in future work.

\subsection{Considerations on the optimisation of \textsc{Sysrem}}\label{discuss_sys}
As stated in Section~\ref{sec:sysrem}, there is precedent for optimising the \textsc{Sysrem} algorithm per spectral order and/or per night because each can suffer from different conditions and systematics \citep[see e.g.][]{Birkby2013, Birkby2017, Nugroho2017, Cabot19, Gibson2020, Kesseli2020}. However, the methods used for determining this optimisation vary, making the outcomes challenging to compare. Typically, a model is injected and recovered per order, with the iteration that gave the highest S/N recovery of the model being used for that order when assessing the real data \citep[see e.g.][]{deKok2013, Birkby2013, Birkby2017, Nugroho2017}. These previous works used a small number of spectral orders compared to HARPS-N, which has 69 orders. Using this approach, we found that the large number of orders acted to give a false positive due to optimised constructive addition of noise at the planet's expected velocity. Furthermore, \citet{Cabot19} stressed the need for caution when optimising the number of \textsc{Sysrem} iterations, and found that the significance of detections could be over-estimated if the algorithm was over-fitted. Therefore, in this work we followed other papers \citep[e.g.][]{Gibson2020, Merritt2020, Nugroho2021}, opting to use a uniform numbers of \textsc{Sysrem} iterations across all orders and nights, as detailed in Section~\ref{stel_remove}, to reduce this effect. A detailed assessment of how to optimally apply \textsc{Sysrem}, including with consideration of rotational broadening, is beyond the scope of this paper, but will be further explored in future work to reach the deepest contrasts.

      \begin{figure*}
        \centering
            \subfloat{%
              \includegraphics[width=18.2cm]{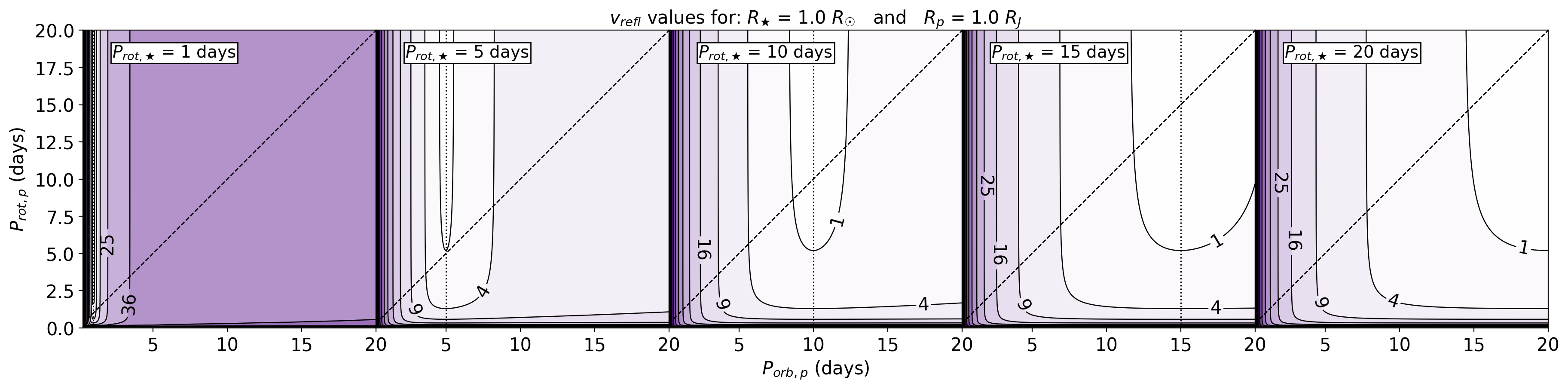}%
            }\hspace{1mm}
          \subfloat{%
              \includegraphics[width=18.2cm]{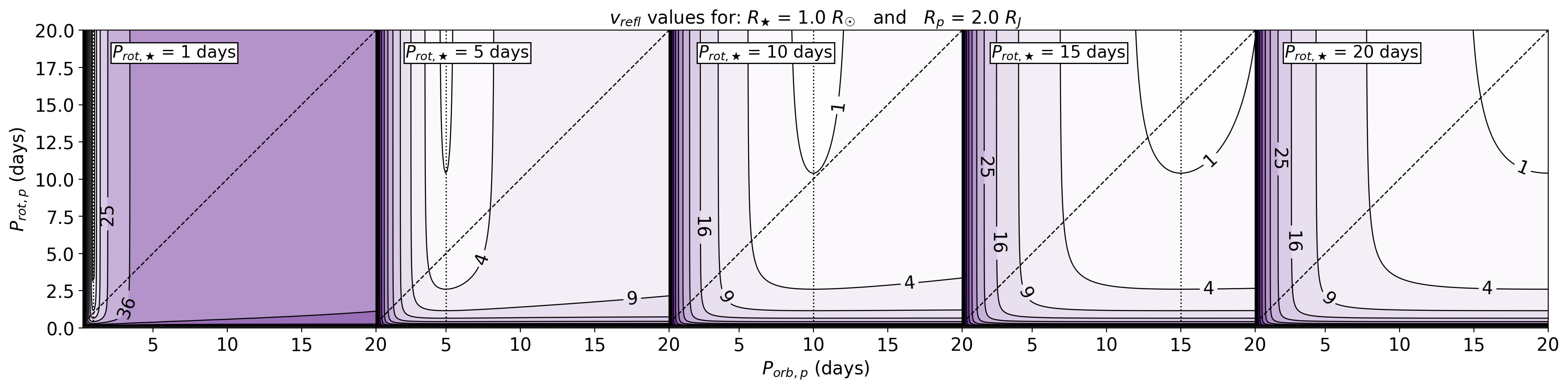}%
         }\hspace{1mm}
        \subfloat{%
              \includegraphics[width=18.2cm]{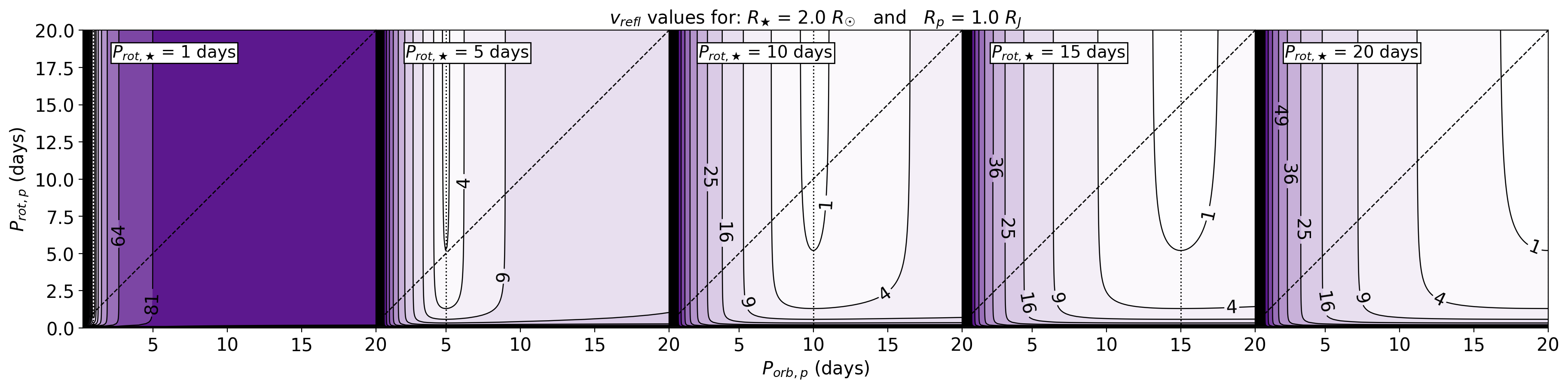}%
            }\hspace{1mm}
          \subfloat{%
              \includegraphics[width=18.2cm]{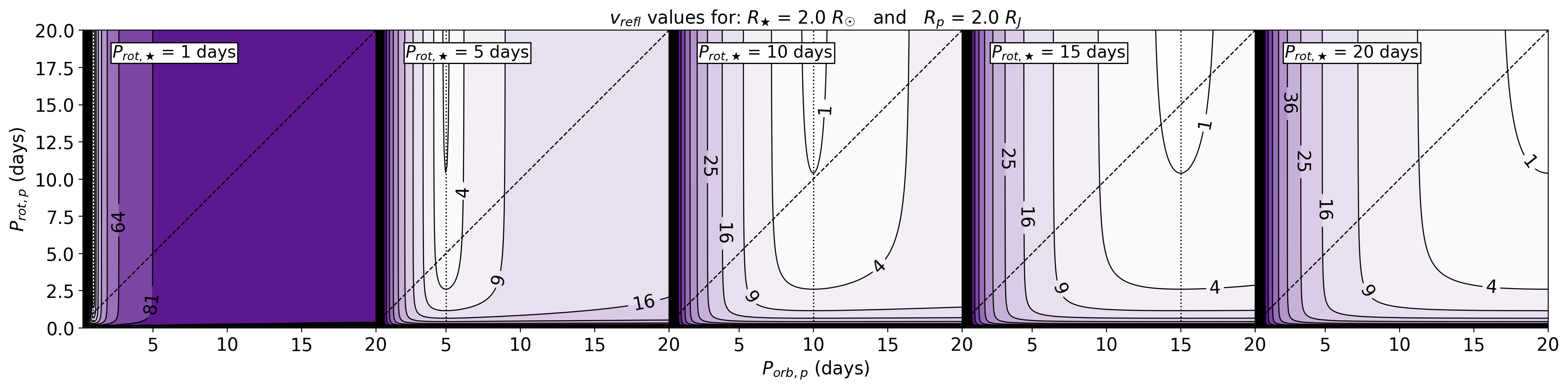}%
         }
        \caption{$v_{refl}$ values, in km\,s$^{-1}$ calculated using the equations defined in \cite{Rodler2010} (see Equations \ref{eq_srefp}, \ref{eq_pproj} and \ref{eq_vrefl}), for a range of exoplanet and stellar parameters. Each row of plots shows $v_{refl}$ values for different values of stellar radii ($R_\star$) and exoplanet radii ($R_p$). The y-axes correspond to the period of the exoplanet's rotation on its own axis ($P_{rot,p}$). The x-axes correspond to the period of the exoplanet's orbit around its host star and extends down to $P_{orb}=0.1$ days). The dashed black diagonal lines show where $P_{rot,p} = P_{orb}$, i.e. the exoplanet is always showing the same face to its host star (as is the case for 51 Pegasi b). Each individual plot corresponds to a different period of the star's rotation on its own axis ($P_{rot,\star}$). The dotted black vertical lines show where $P_{rot,\star} = P_{orb}$, i.e. the star is always showing the same face to the exoplanet. The point where the dashed and dotted black lines cross mark where $P_{rot,p} = P_{rot,\star} = P_{orb}$,  i.e. the star-exoplanet system is in a full tidal lock. }
        \label{fig:roddler_plots}
        \end{figure*}
        
\subsection{Rotational broadening in different systems} 

\begin{table}
    \centering
    \caption{Rotational broadening, $v_{refl}$, of different systems, calculated using Equations~\ref{eq_pproj},~\ref{eq_srefp}, and~\ref{eq_vrefl}. }
    \label{tab:systems}
    \begin{tabular}{l|c|c|c}
          & \textbf{Scenario A \tablefootmark{$\ast$}} & \textbf{Scenario B \tablefootmark{$\dag$}} & \textbf{Scenario C \tablefootmark{$\S$}}\\
          & (Proxima) & & \\
          \hline
          \hline
         $R_\star\, (R_\odot)$ & 0.154 & 0.4 & 0.2 \\
         $M_\star\, (M_\odot)$ & 0.122 & 0.2 & 0.15 \\
         $T_\star\, (K)$ & 3042 & 3000 & 2500 \\
         $P_{rot,\star}$ (days) & 82.6 & 60 & 80 \\
         $R_p\, (R_\oplus)$ & 1.08 & 1 & 1 \\
         $M_p\, (M_\oplus)$ & 1.27 & 1 & 1 \\
         $a$ (AU) & 0.0485 & 0.1 & 0.04 \\
         $P_{orb}$ (days) & 11.2 & 5.2 & 1.1 \\
         $v_{refl}\, ($km\,s$^{-1})$ & 0.6 & 3.6 & 9.0 \\

    \end{tabular}
    \tablefoot{
    \tablefoottext{$\ast$}{Scenario A represents the star Proxima Centauri and the exoplanet Proxima Centauri b. The radius estimate of 1.08$R_\oplus$ is taken from the NASA Exoplanet Archive. }
    \tablefoottext{$\dag$}{Scenario B represents an Earth-like exoplanet in the habitable zone, orbiting a large, fast-rotating M-dwarf.}
    \tablefoottext{$\S$}{Scenario C represents an Earth-like exoplanet in the habitable zone, orbiting a slow-rotating M-dwarf.}
}
\end{table}

The rotational broadening of the reflected light from 51 Pegasi b made a significant impact on the ability of HRCCS to extract its spectrum. We visualise the equations used to calculated $v_{refl}$ for gas giant planets (see Equations~\ref{eq_pproj},~\ref{eq_srefp}, and~\ref{eq_vrefl}) in Figure~\ref{fig:roddler_plots} to highlight the effect of different parameters. It is evident that $v_{\star,p}$ contributes considerably to the value of $v_{refl}$. The magnitude of $v_{\star,p}$ is determined by the rotation of the star on its own axis ($P_{rot,\star}$) and the orbit of the planet ($P_{orb,p}$), and increases as these periods differ. Due to $v_{proj,p}$, which is determined by the ratio between the planet's $R_p$ and $P_{rot,p}$, $v_{refl}$ never equals zero, even when the star and planet system is fully synchronised (see Equation~\ref{eq_pproj}). $v_{refl}$ is greatest for large, rapidly rotating planets where there is a large difference between $P_{rot,\star}$ and $P_{orb}$. However, $v_{refl}$ can still be significant for combinations in between, and we can use these equations to predict its impact on the rotational broadening on different star-planet systems. This is of particular interest to Earth-like exoplanets that orbit M-dwarfs, as these are considered to be amongst the most likely targets for the successful detection of biosignatures in reflected light \citep[e.g. ][]{Snellen2015, Lovis2017}. As examples, we consider Proxima b, and two different hypothetical scenarios for an Earth-like exoplanet in orbit in the habitable zone of two different M-dwarfs. We use the scaling relation $a(HZ,~$au$) = \left(T_\star/5770\right)^2 \left(R_\star/R_\odot\right)$ from \citet{Traub2008} to approximate a semi-major axis of a planet in the habitable zone, and a scaled version of Kepler's $3^{rd}$ law to find the orbital period of the planet in the habitable zone, $P(HZ,$ days$) = 365.25\, a^{3/2} M_\star^{1/2}$. For all scenarios, we assume tidal locking, so that $P_{rot,p} = P_{orb}$. The results are presented in Table~\ref{tab:systems}. 

Consequently, for Proxima b, rotational broadening of its reflected light spectrum is unlikely to have significant impact on HRCCS in extracting is spectrum as the broadening is below the resolution limit of most spectrographs. However, for other scenarios (e.g. scenarios B and C in Table~\ref{tab:systems}) the broadening does exceed the typical velocity per resolution element of $\Delta v = 3\,\rm km\,s^{-1}$  for $R=100,000$ spectrographs, and thus must be carefully considered when inferring the retrieved properties of the planet, and the S/N required to detect the planet's reflected light spectrum.

\subsection{Considering the impact of rotational broadening on the optimal choice of instrument}

\noindent When considering observational plans for HRCCS, one could broadly approximate the impact of rotational broadening to a reduction in resolving power. In the case of 51 Pegasi b, this would be $R\sim c/\Delta v\sim3\times10^5/12=25,000$. This raises the question of whether it is necessary or beneficial to use very high resolution spectrometers for HRCCS for systems where we expect a significant amount of rotational broadening. There are advantages to being able to use lower resolution spectrometers: higher S/N due to a greater number of photons per wavelength channel; a greater range of available instruments; the possibility of using space telescopes. However, using a lower resolution spectrometer for reflection HRCCS would lose one of the scant advantages that the rotational broadening of the reflected light offers, namely that that the planet spectrum is subject to additional broadening (see Equation~\ref{eq_vrefl}), which helps distinguish it from the stellar spectrum.  This advantage would be lost at lower resolution. Nonetheless, it is worth bearing in mind the possibility of employing a wider range of instruments when searching for reflected light from a system with considerable rotational broadening.\\

\section{Conclusions} \label{conclude}
We have presented a search for the reflected light spectrum of 51 Pegasi b using optical high resolution cross-correlation spectroscopy obtained during a dedicated observing programme with the stable, precision radial velocity spectrograph HARPS-N at the TNG, in combination with archival data collected from HARPS-N and HARPS. In contrast to previous studies of 51 Pegasi b, we removed the contaminating host star spectrum directly from the spectra, rather than in cross-correlation space. We found that the precise stellar RVs from HARPS-N allowed better alignment of the spectra to the true stellar rest frame which subsequently allowed us to remove the stellar lines more effectively using \textsc{Sysrem}. However, stellar residuals remain, possibly due to stellar variability during the course of the observations that need to be mitigated in future studies. We also highlight a lack of consistency in the application of \textsc{Sysrem} in the literature. There remains space for improvement in both the removal of stellar lines and telluric lines in our approach and others. Importantly, we found that apparent rotational broadening, due to the difference between the long host star rotation period and the short orbital period of 51 Pegasi b, is significant and strongly impacts the sensitivity of HRCCS in extracting reflected light spectra. We do not make a significant detection of reflected light from 51 Pegasi b, and for the physical scenario with rotational broadening (case 1), we find a S/N $=3$ upper limit on the contrast ratio $\frac{F_p}{F_\star}<7.60\times10^{-5}$, which corresponds to a weakly constrained $A_g < 0.62$ for a typical hot Jupiter with $R_p\sim1.2R_J$. Without broadening (case 3), the sensitivity of our data and pipeline results in a S/N $=3$ upper limit of $\frac{F_p}{F_\star}<2.40\times10^{-5}$}, which would corresponds to a much more tightly constrained $A_g < 0.20$ for $R_p\sim1.2R_J$ if the 51 Pegasi system was fully synchronised ($P_{rot,\star}=P_{orb}$). In both cases, we are able to reject the proposed $A_g$ and $R_p$ in M15, and we stress that our demonstration of the impact of $v_{refl}$ shows it must be accounted for when inferring $R_p$ and $A_g$ from contrast ratios measured with HRCCS, else they will be underestimated. For rocky planets, we find that Proxima b-like systems experience broadening below the velocity resolution of most spectrographs, but other possible M-dwarf habitable zones planets can experience detectable broadening and should be carefully considered when aiming to detect the spectra of these planets. Finally, we find that the S/N analogue used here and in the literature is not an ideal metric for assessing the significance of rotationally broadened signal recovered with HRCCS, because S/N only considers the height of the CCF peak and not its breadth. Other metrics, such as the log-Likelihood \citep[e.g.][]{BrogiLine2019} which account for the shape of the signal, may be more successful and warrant further separate study.

\begin{acknowledgements}
      
 The authors would like to thank P. Uttley for valuable discussions on statistical methods, and I. Snellen and G. Anglada-Escud\'e for helpful insights on cleaning the host star spectrum from the data. We further thank J. Hoeijmakers, M. Brogi, F. Rodler, L. Buchhave, J-M. D\'esert, and D. Charbonneau for helpful discussions that improved the analysis in this paper. We thank Eike Gunther, Hristo Stoev and the TNG operators for all their help in obtaining the data in this work, as well as the wonderful staff at the Observatorio del Roque de Los Muchachos, La Palma. JLB and MEY acknowledge funding from the European Research Council (ERC) under the European Union's Horizon 2020 research and innovation programme under grant agreement No 805445. This work was performed in part under contract with the Jet Propulsion Laboratory (JPL) funded by NASA through the Sagan Fellowship Program executed by the NASA Exoplanet Science Institute. RA acknowledges support from the Spanish MINECO grant ESP2014-57495-C2-1-R. SH acknowledges CNES funding through the grant 837319. PC acknowledges support from Conselho Nacional de Desenvolvimento Cient\'ifico e Tecnol\'ogico (CNPq) under grant 310041/2018-0. This article is based on observations made with the Italian Telescopio Nazionale Galileo (TNG) operated on the island of La Palma by the Fundación Galileo Galilei of the INAF (Istituto Nazionale di Astrofisica) at the Spanish Observatorio del Roque de los Muchachos of the Instituto de Astrofisica de Canarias. This research used the facilities of the Italian Center for Astronomical Archive (IA2) operated by INAF at the Astronomical Observatory of Trieste. This research has made use of the NASA Exoplanet Archive, which is operated by the California Institute of Technology, under contract with the National Aeronautics and Space Administration under the Exoplanet Exploration Program, and NASA’s Astrophysics Data System Bibliographic Services. The following software and packages were used in this work: the HARPS-N Data Reduction Software (DRS, ~\cite{LovisPepe2007}); \texttt{Python v3.7}; \texttt{Python} packages \texttt{Astropy}~\citep{Astropy}, \texttt{NumPy}~\citep{Numpy, NumpyArray}, and \texttt{Matplotlib}~\citep{Matplotlib}.

\end{acknowledgements}

\bibliographystyle{aa}
\bibliography{ref}

\clearpage

\begin{appendix}

\section{Confirming the efficacy of the pipeline} \label{appen_check_pipeline}

\begin{figure}
    \centering
          \includegraphics[width=8.9cm]{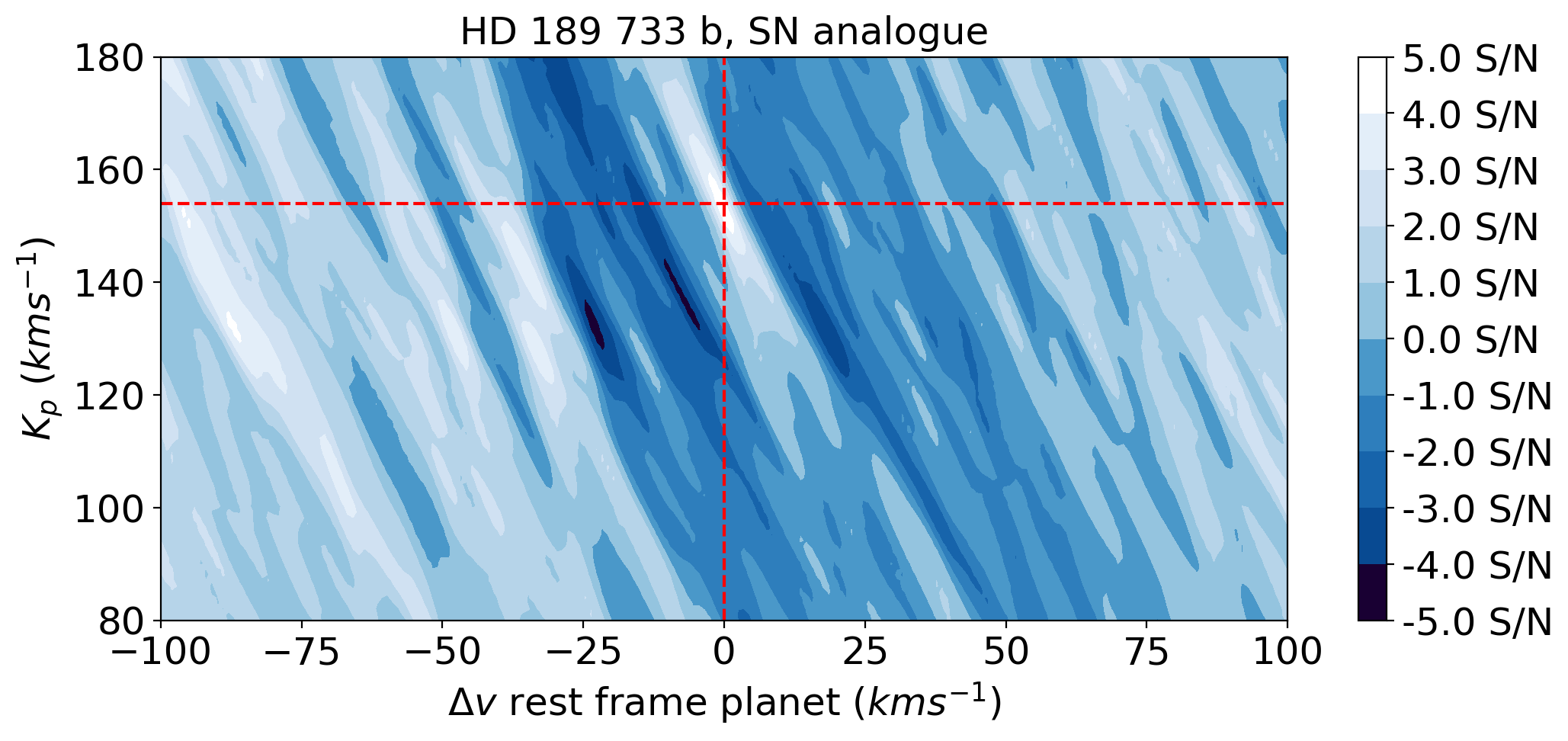}%
        \label{fig:HD189_SN}
    \caption{Detection of water vapour in the atmosphere of HD 189733b found using the pipeline described in this paper. The data are from CRIRES/VLT at $3.2~\mu$m. This serves to verify the efficacy of our pipeline in recovering known detections \citep{Birkby2013, Cabot19}. Axes as for Figures~\ref{fig:ripple_ccsig_nomod} and~\ref{fig:ripple_ccsig_mod}. }
\label{fig:HD189_ripple}
\end{figure}

The pipeline we present in this work is an adaptation of that used for infrared high resolution spectroscopy of exoplanet atmospheres where the telluric lines are the dominant contaminant (e.g. \citealt{Birkby2013, Birkby2017, Brogi2012}). In this work, we make adjustments such as aligning to the true stellar rest frame to improve removal of the stellar lines. To confirm that our pipeline was able to recover previously confirmed signals, we ran it on $3.2~\mu$m CRIRES/VLT observations of HD 189773 b as presented in \citet{Birkby2013} and \citet{Cabot19} who both detected water in the planet's atmosphere. The contaminating tellurics (and few stellar) lines were removed following the method laid out in Section~\ref{stel_remove}, and \textsc{Sysrem} was iterated eight times.
Using the same cross-correlation template presented in \citet{Birkby2013}, we detect a signal of water absorption at $K_p = 152\,\rm km\,s^{-1}$, $\Delta v = 0 \,\rm km\,s^{-1}$ (where $v_{sys} = 2.361\,\rm km\,s^{-1}$) using the pipeline presented in this paper. The detection is shown in Figure~\ref{fig:HD189_ripple}.  The total S/N=4.0 of our detection is slightly lower than \citet{Birkby2013} and \citet{Cabot19} (S/N=4.8), likely due to the difference in the number of \textsc{Sysrem} iterations used to clean the data.

%begin Figure
%end Figure

\section{Additional figures} \label{appen_additional_plots}

\begin{figure*}
    \centering
          \includegraphics[width=18.3cm]{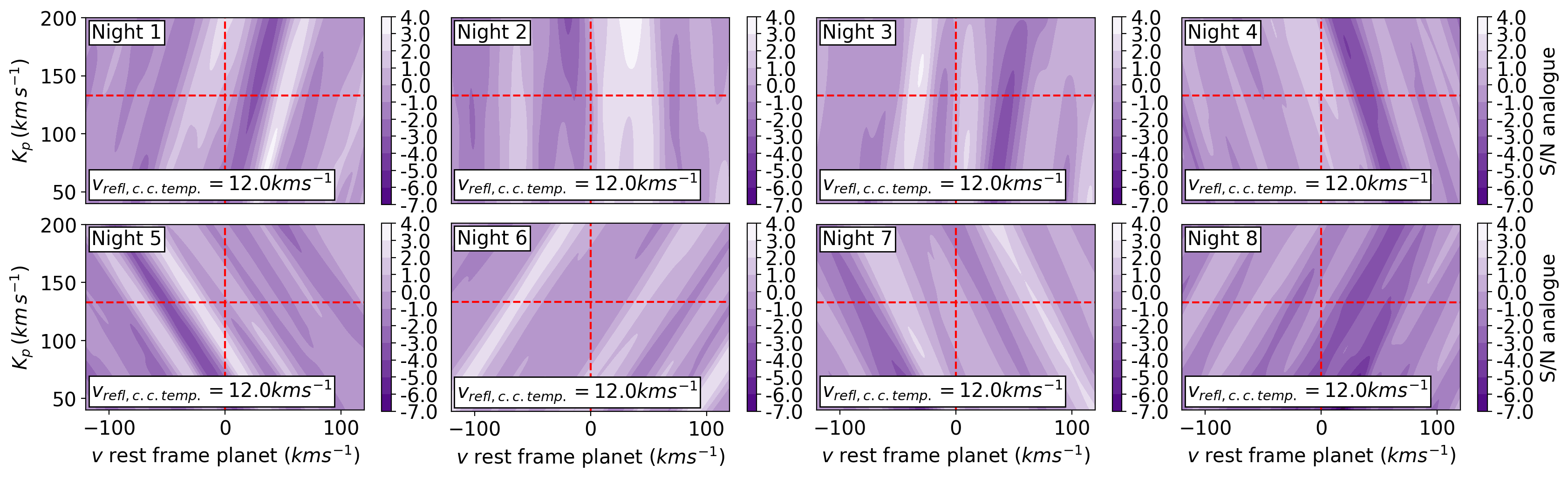}%
    \caption{As for the left plot in Figure~\ref{fig:ripple_ccsig_nomod}, but displaying the individual results from Nights 1, 2, 3, 5, 6, 7 and 8, as opposed to the combined results. Some features appear at the S/N=-5 level, but do not propagate to the all-nights matrix, indicating that these are probably due to residual stellar contamination.}
\label{fig:ripple_appen}
\end{figure*}

We present the equivalent plots displayed in Figure~\ref{fig:waterfall}, for the remaining individuals nights of data (see Figure~\ref{fig:waterfall_appen}), and the equivalent plot as displayed in the left plot of Figure~\ref{fig:ripple_ccsig_nomod}, but for the individual nights, as opposed to all the nights combined (see Figure~\ref{fig:ripple_appen}).

    \begin{figure*}
    \centering
        \vspace{-1.5mm}
        \subfloat{%
          \includegraphics[width=18.4cm]{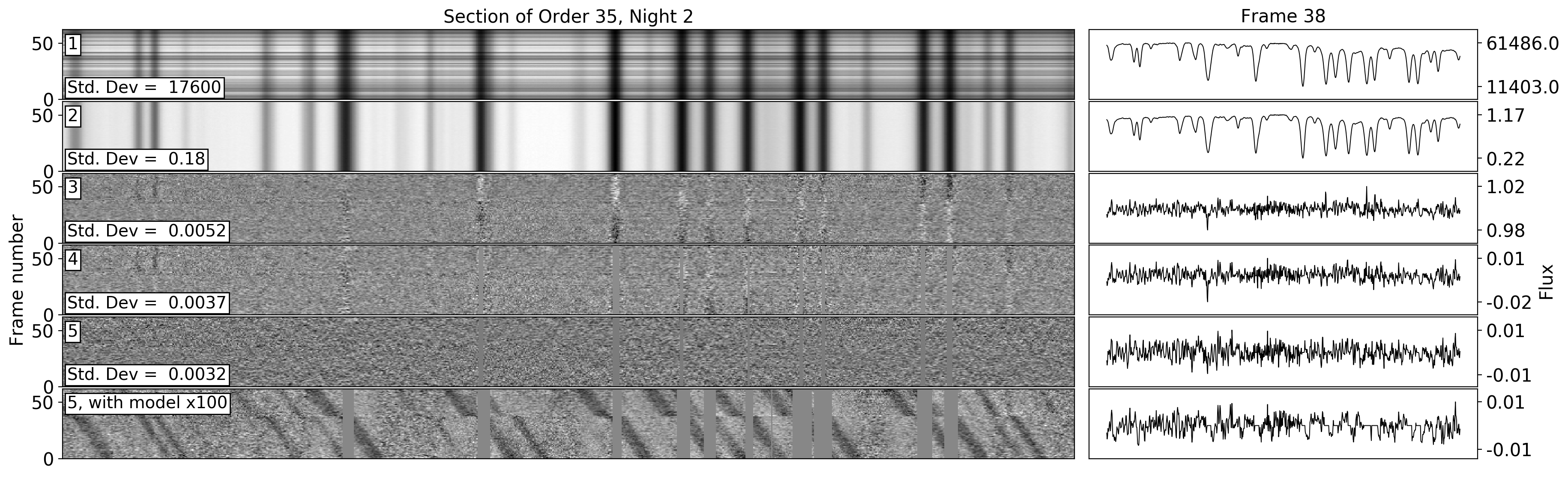}%
        }\vspace{-5mm}

        \subfloat{%
          \includegraphics[width=18.4cm]{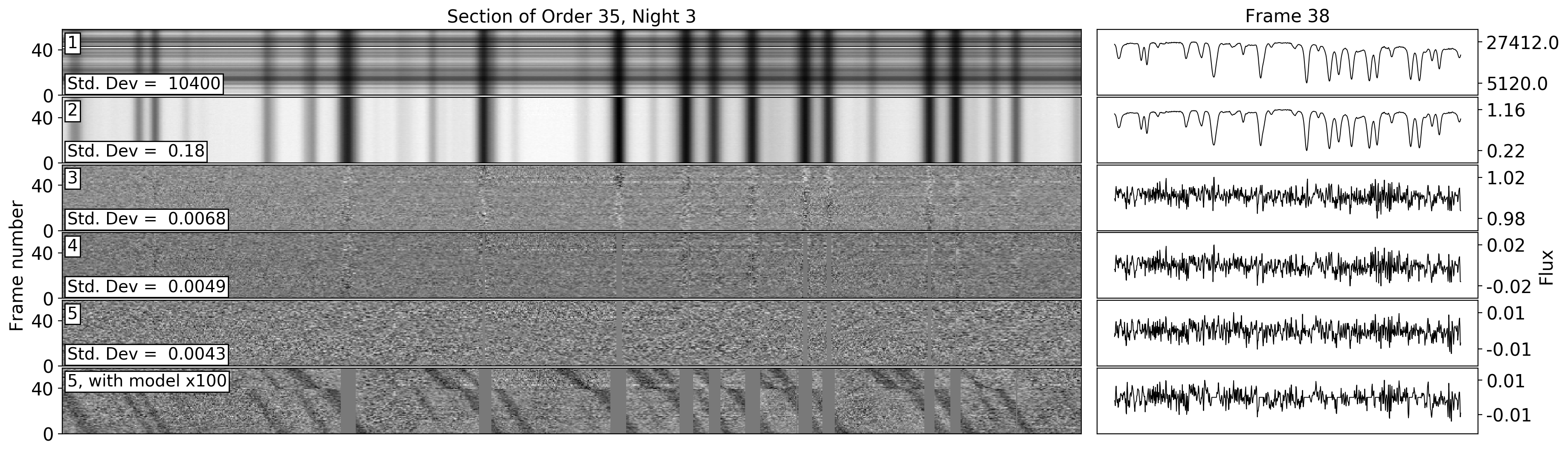}%
        }\vspace{-5mm}
        
        \subfloat{%
          \includegraphics[width=18.4cm]{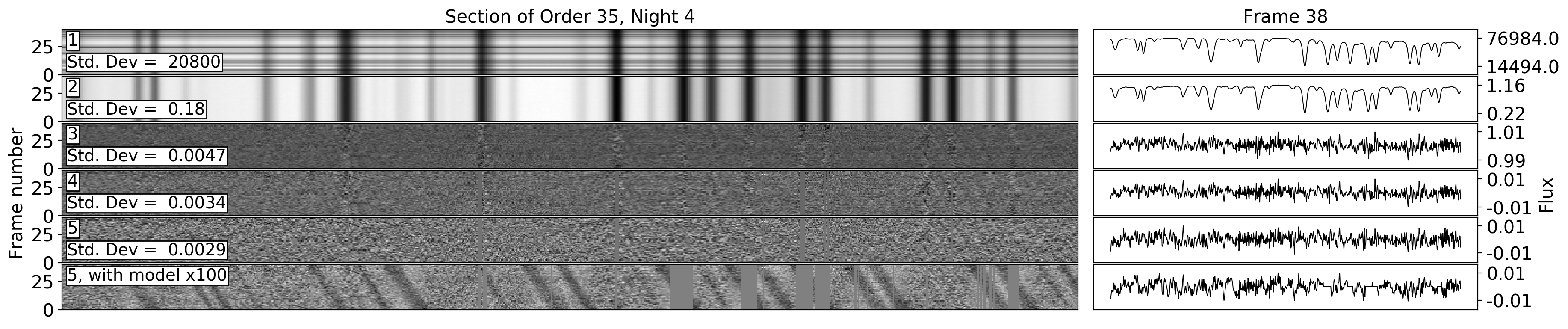}%
        }\vspace{-5mm}

        \subfloat{%
          \includegraphics[width=18.4cm]{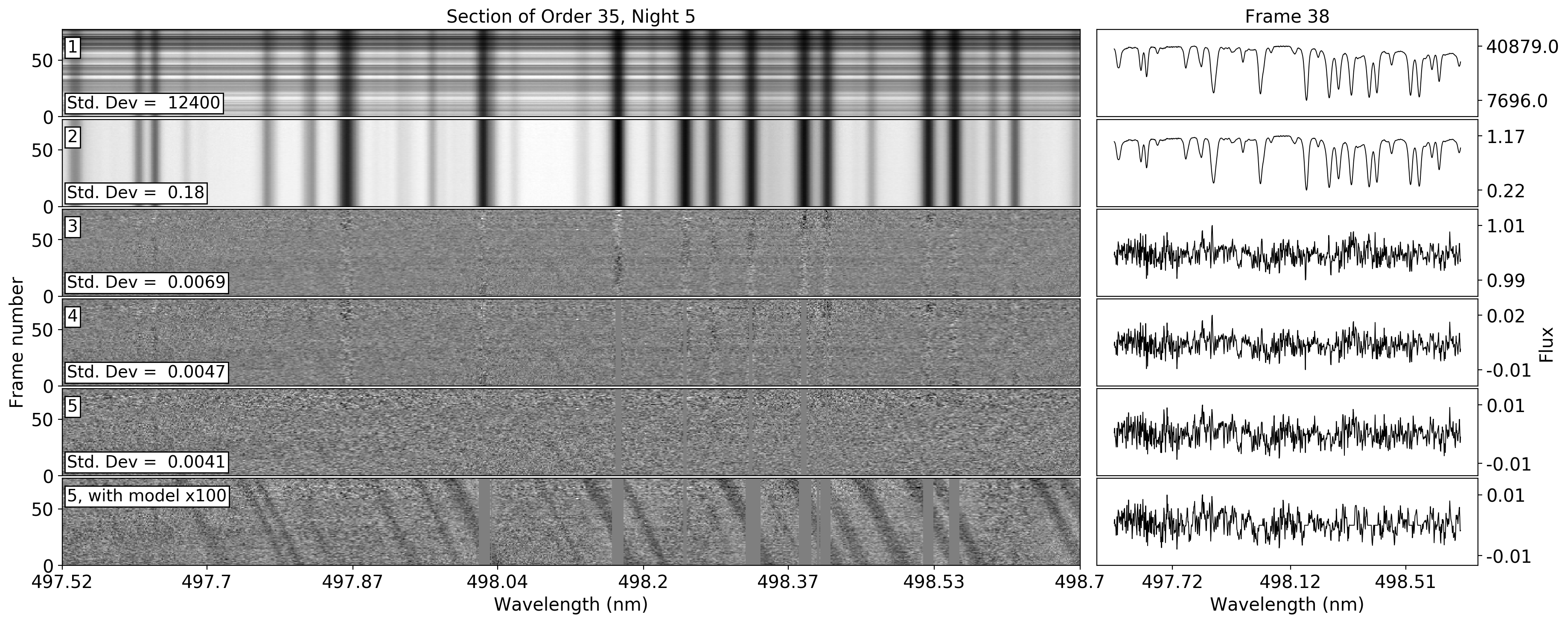}%
        }
    \caption{As for Figure~\ref{fig:waterfall}, but for Nights 2, 3, 4, 5, 6, 7 and 8. Continued on the next page. }
    \label{fig:waterfall_appen}
    \end{figure*}

    \begin{figure*}
    \ContinuedFloat
    \centering
        \vspace{-1.5mm}

        \subfloat{%
          \includegraphics[width=18.4cm]{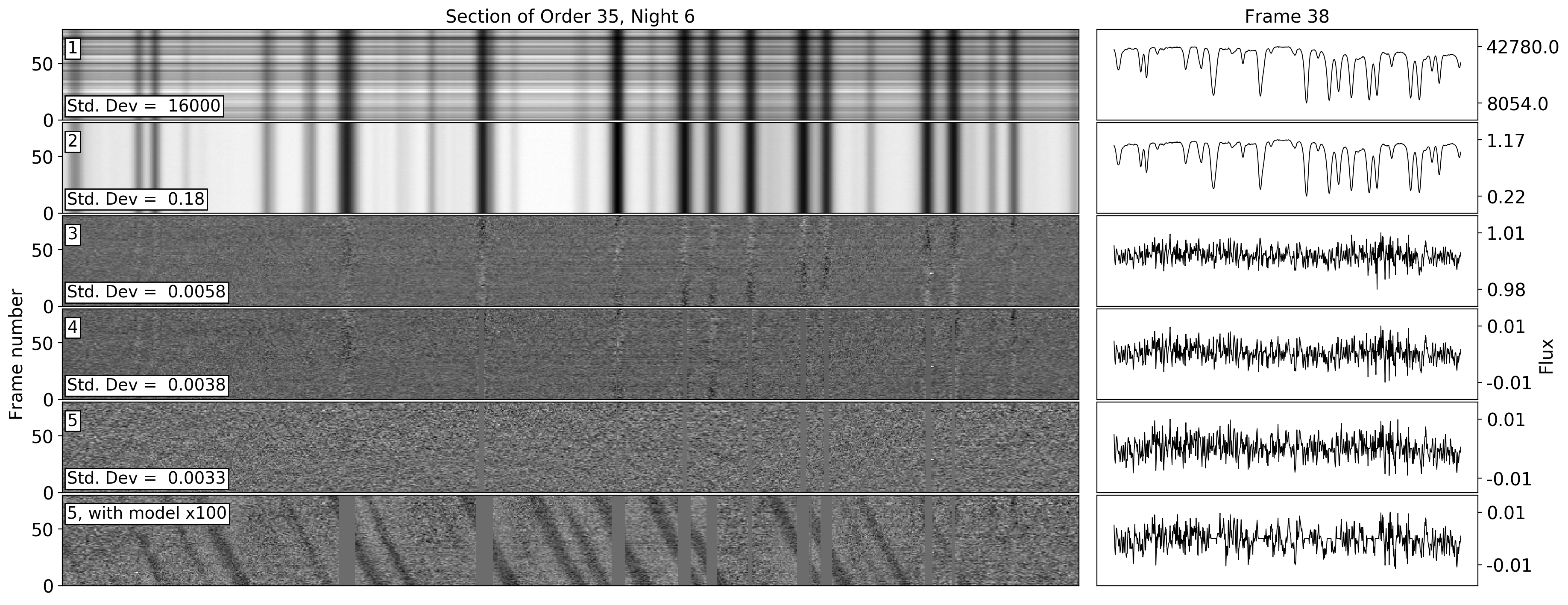}%
        }\vspace{-4mm}
        
        \subfloat{%
          \includegraphics[width=18.4cm]{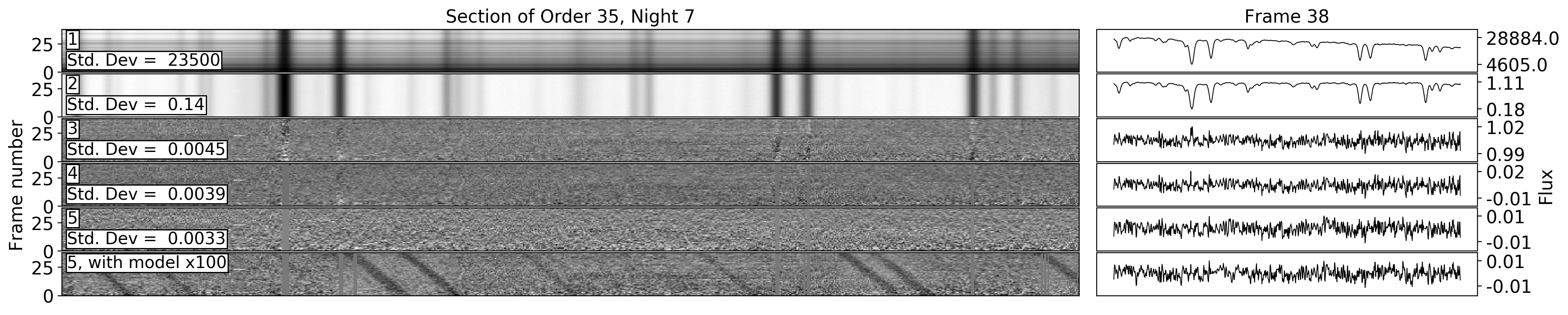}%
        }\vspace{-4mm}
        
        \subfloat{%
          \includegraphics[width=18.4cm]{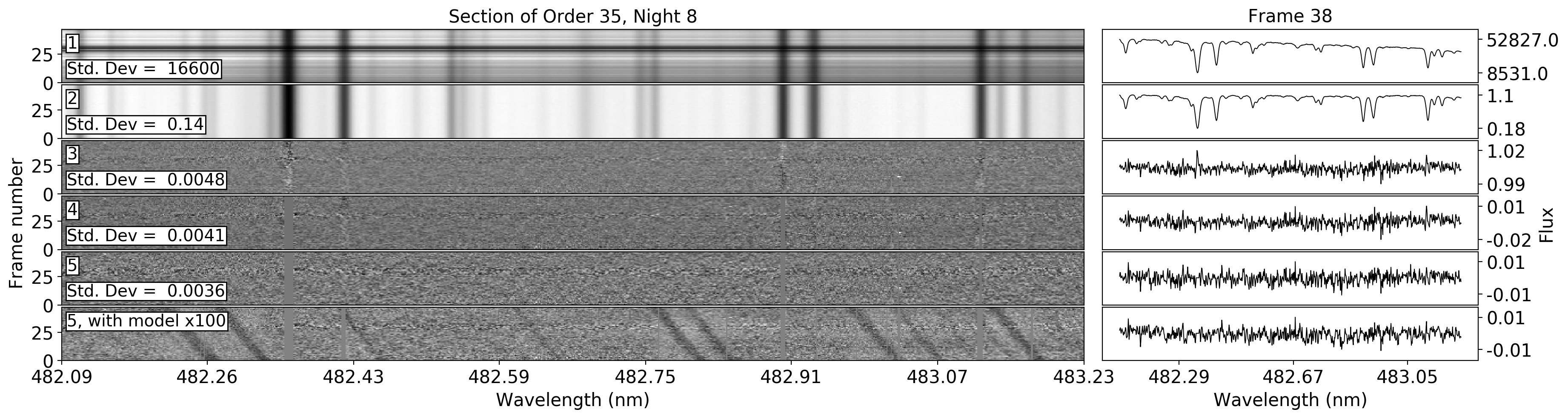}%
        }
    \caption{Figure~\ref{fig:waterfall_appen} continued.}

    \end{figure*}

%begin Figure
%end Figure

\end{appendix}

\end{document}